\documentclass[aps,prd,preprint,superscriptaddress,nofootinbib]{revtex4-2}
\usepackage{graphicx} 
\usepackage[caption=false]{subfig}
\usepackage{amsmath}
\usepackage{amssymb}
\usepackage{amsfonts}
\usepackage{bbm}
\usepackage{dsfont}
\usepackage{color}
\usepackage{float}
\usepackage{hyperref}
\usepackage{array}
\usepackage{physics}
\usepackage{tikz,simpler-wick}
\usetikzlibrary{positioning,calc}
\usetikzlibrary{shapes.misc,shapes.geometric}
\usetikzlibrary{decorations.markings}
\usepackage{bbold}

\usetikzlibrary{arrows.meta, decorations.markings}

\tikzset{
  arrow in middle/.style={
    postaction={decorate},
    decoration={
      markings,
      mark=at position 0.5 with {\arrow{Stealth[red,width=1.5mm,length=2mm]}}
    }
  }
}

\hypersetup{
    colorlinks=true,
    linkcolor=blue,
    filecolor=magenta,      
    urlcolor=cyan,
    pdftitle={Overleaf Example},
    pdfpagemode=FullScreen,
    }
\usepackage[vcentermath]{youngtab}

\newcommand{\bdelta}{\bar{\delta}}
\DeclareMathOperator{\sgn}{sgn}
\newcommand{\blobtwo}{ \begin{tikzpicture}[baseline={([yshift=-.5ex]current bounding box.center)}]
\coordinate (P) at (0,0);

\draw (P) circle (10pt); 

\node[below] at (0,-10pt) {};

\coordinate (P1) at (-0.15,0);
\coordinate (P2) at (0.15,0);

\draw[black] (P1) circle (1.7pt);
\filldraw (P2) circle (1.7pt);

\draw[-, thick] (P1) to (-0.7,0.7);
\draw[dotted, thick] (P1) to (-0.7,-0.7);
\draw[-, thick] (P2) to (0.7,0.7);
\draw[dotted, thick] (P2) to (0.7,-0.7);

\end{tikzpicture}}

\newcommand{\blobleft}{\begin{tikzpicture}[baseline={([yshift=-.5ex]current bounding box.center)}]
\coordinate (P) at (0,0);

\draw (P) circle (10pt); 

\node[below] at (0,-10pt) {};

\coordinate (P1) at (-0.15,0);
\coordinate (P2) at (0.15,0);

\draw[black] (P1) circle (1.7pt);
\filldraw (P2) circle (1.7pt);

\draw[-, thick] (P1) to (-0.7,0.7);
\draw[dotted, thick] (P1) to (-0.7,-0.7);
\end{tikzpicture}}

\newcommand{\blobright}{\begin{tikzpicture}[baseline={([yshift=-.5ex]current bounding box.center)}]
\coordinate (P) at (0,0);

\draw (P) circle (10pt); 

\node[below] at (0,-10pt) {};

\coordinate (P1) at (-0.15,0);
\coordinate (P2) at (0.15,0);

\draw[black] (P1) circle (1.7pt);
\filldraw (P2) circle (1.7pt);

\draw[-, thick] (P2) to (0.7,0.7);
\draw[dotted, thick] (P2) to (0.7,-0.7);
\end{tikzpicture}}

\newcommand{\beq}{\begin{equation}}
\newcommand{\eeq}{\end{equation}}
\newcommand{\beqn}{\begin{eqnarray}}
\newcommand{\eeqn}{\end{eqnarray}}

\newcommand{\cN}{\mathcal{N}}

\renewcommand{\d}{\mathrm{d}} 
\newcommand{\eps}{\epsilon}

\newcommand{\nt}{\addtocounter{equation}{1}\tag{\theequation}}

\usepackage{enumitem}
\setlist{leftmargin=6mm}

\begin{document}

\title{On the stabilizer complexity of Hawking radiation}
\author{Ritam Basu}
\email{ritam.basu@tifr.res.in}
\affiliation{Department of Theoretical Physics, 
Tata Institute of Fundamental Research, 1 Homi Bhabha Road, Mumbai 400005, India}
\author{Onkar Parrikar}
\email{parrikar@theory.tifr.res.in}
\affiliation{Department of Theoretical Physics, 
Tata Institute of Fundamental Research, 1 Homi Bhabha Road, Mumbai 400005, India}
\author{Suprakash Paul}
\email{suprakash.paul@tifr.res.in}
\affiliation{Department of Theoretical Physics, 
Tata Institute of Fundamental Research, 1 Homi Bhabha Road, Mumbai 400005, India}
\author{Harshit Rajgadia}
\email{harshit.rajgadia@tifr.res.in}
\affiliation{Department of Theoretical Physics, 
Tata Institute of Fundamental Research, 1 Homi Bhabha Road, Mumbai 400005, India}
\date{\today}

\begin{abstract}
We study the complexity of Hawking radiation for an evaporating black hole from the perspective of the stabilizer theory of quantum computation. Specifically, we calculate Wigner negativity -- a magic monotone which can be interpreted as a measure of the stabilizer complexity, or equivalently, the complexity of classical simulation -- in various toy models for evaporating black holes. We first calculate the Wigner negativity of Hawking radiation in the PSSY model directly using the gravitational path integral, and show that the negativity is $O(1)$ before the Page transition, but becomes exponentially large past the Page transition. We also derive a universal, information theoretic formula for the negativity which interpolates between the two extremes. We then study the Wigner negativity of radiation in a dynamical model of black hole evaporation. In this case, the negativity shows a sharp spike at early times resulting from the coupling between the black hole and the radiation system, but at late times when the system settles down, we find that the negativity satisfies the same universal formula as in the PSSY model. Finally, we also propose a geometric formula for Wigner negativity in general holographic states using intuition from fixed area states and random tensor networks, and argue that a python's lunch in the entanglement wedge implies a stabilizer complexity which is exponentially large in $\frac{1}{8G_N}$ times the difference between the areas corresponding to the outermost and minimal extremal surfaces.
\end{abstract}

\maketitle

\section{Introduction}

The black hole information paradox is a sharp puzzle that highlights the mysterious nature of black holes. According to Hawking's seminal work \cite{Hawking}, a black hole emits thermal radiation, and so a black hole formed from the gravitational collapse of matter in a pure state apparently evaporates away leaving radiation in an approximately thermal state. This process seemingly violates the unitarity of quantum mechanics. In more technical terms, Hawking's calculation suggests that the von Neumann entropy of the state of the radiation keeps growing with time until the end of evaporation, violating the basic fact that entropy of a bi-partite quantum state is upper bounded by the dimension of the Hilbert space of the smaller subsystem, and should thus satisfy the Page curve \cite{Page}. {Recent progress on the black hole information problem \cite{Penington:2019npb, Almheiri:2019psf, Almheiri:2019hni, Almheiri:2019yqk, Penington:2019kki, Almheiri:2019qdq} (see also \cite{Almheiri:2020cfm} for a review and further references) has clarified how a unitary Page curve emerges in the calculation of the radiation entropy using generalizations of ideas developed in the AdS/CFT context.} A heuristic way to understand why the radiation entropy keeps rising in Hawking's picture is that the black hole emits more and more Hawking quanta with time, and each of these quanta is entangled with its Hawking partner behind the black hole horizon. So, the increasing radiation entropy stems from a continuous build-up of entanglement between the early radiation and the degrees of freedom in the black hole interior. A crucial insight in the recent resolution of the information paradox was that after the Page time, a new quantum extremal surface close to the black hole horizon becomes dominant. From the entanglement wedge reconstruction paradigm \cite{Headrick:2014cta, Czech:2012bh, Almheiri:2014lwa, Dong:2016eik, Harlow:2016vwg, Faulkner:2017vdd, Petzmap}, this implies that the black hole interior degrees of freedom are no longer independent degrees of freedom entangled with the radiation, but are in fact encoded in the early radiation. A crucial role in this analysis is played by replica wormholes \cite{Penington:2019kki, Almheiri:2019qdq} -- non-trivial Euclidean wormhole saddle points of gravity which dominate the R\'enyi entropy calculation after Page time and stop the growth of the radiation entropy predicted by Hawking's calculation. 

While the above resolution to the black hole information paradox is extremely elegant and satisfying in many ways, it highlights the essentially non-local nature of quantum gravity: degrees of freedom behind the black hole horizon, which are causally inaccessible to an observer at infinity, can nevertheless be encoded in -- and thus manipulated via -- the degrees of freedom at infinity. In local quantum field theory without gravity, this would be impossible; what makes it possible (at least, in principle) in gravity is the existence of gravitational wormholes.\footnote{Also, as emphasized in \cite{Balasubramanian:2006iw, Raju:2019qjq, Chowdhury:2020hse}, the ADM Hamiltonian in gravity is a boundary term at infinity and well-defined observables in gravity are non-local on account of the need for gravitational dressing. These features already point to an essential non-locality in gravity.} {Thus, if the semiclassical picture of the black hole interior makes sense, then it appears that quantum gravity must be fundamentally non-local, in that an observer with access to all the early radiation from a sufficiently old black hole can perform a unitary that manipulates a qubit deep inside the causally inaccessible black hole interior. }Nevertheless, following the pioneering work of Harlow and Hayden \cite{Harlow:2013tf}, several authors have suggested that observers at infinity with \emph{computationally bounded resources} should not be able to perform such causality violating operations \cite{Brown:2019rox, Kim:2020cds, Akers:2022qdl, Yang:2023zic} (see also \cite{Balasubramanian:2022fiy, Balasubramanian:2023xdp}). Thus, the semi-classical notion of spacetime is protected by computational complexity. 

A very similar situation arises in a more general context in AdS/CFT, where a fixed boundary subregion $B$ in some states can have multiple corresponding homologous extremal surfaces in the bulk spacetime. If the outermost extremal surface (on the side of the asymptotic subregion $B$) does not have the smallest area, then the entanglement wedge is said to have a python's lunch \cite{Brown:2019rox}. The python's lunch region -- i.e., the portion of the entanglement wedge behind the outermost extremal surface, is causally inaccessible from the asymptotic boundary, and in fact cannot be brought in causal contact with the asymptotic boundary by simple boundary operations \cite{Engelhardt:2021mue}. Nevertheless, semi-classical degrees of freedom in the python's lunch region are encoded in and can be manipulated from the boundary subregion $B$, much like in the case of the black hole interior in our previous discussion. Taking inspiration from the analogy between holography and tensor networks, it was conjectured in \cite{Brown:2019rox} (see also \cite{Engelhardt:2021mue, Engelhardt:2021qjs, May:2024epy, Arora:2024edk, Engelhardt:2023bpv}) that for degrees of freedom in the python's lunch region, the complexity of bulk reconstruction should scale as:
\beq 
C \sim \exp\left[\frac{1}{2}\frac{\left(A(\gamma_{\text{bulge}})-A(\gamma_{\text{out}})\right)}{4G_N} \right],
\eeq
where $\gamma_{\text{out}}$ is the outermost extremal surface and $\gamma_{\text{bulge}}$ is the ``bulge'' surface (roughly corresponding to the maximal cross section of the python's lunch, but see \cite{Brown:2019rox} for a more precise definition). The python's lunch conjecture tries to make precise the idea that the semi-classical notion of spacetime in AdS/CFT is computationally (but not information theoretically) protected. It would be useful to have some concrete boundary calculations which shed light on this conjecture; part of the difficulty with this is the lack of a computable notion of complexity.

In this paper, we study the complexity of the Hawking radiation from the perspective of the stabilizer theory of quantum computation. The Gottesman-Knill theorem \cite{gottesman1998heisenberg, Aaronson:2004xuh, Mari_2012} identifies a class of circuits -- called stabilizer circuits -- that can generate arbitrary amounts of entanglement, and yet be efficiently simulated on classical computers. In the latter sense, stabilizer circuits are essentially classical, and do not have inherent quantum computational advantage. As one might expect, the stabilizer model of quantum computation is not universal. It was proposed by Bravyi and Kitaev \cite{Bravyi_2005} that one can get universal quantum computation by supplementing stabilizer circuits with some number of external ``resource'' states that provide the non-stabilizerness required for universal quantum computation. These authors called this quantum resource ``magic''. In \cite{Veitch_2012, Veitch_2014}, a resource theory of magic was developed which centered around the construction of magic monotones -- measures of magic that do not increase under stabilizer operations. One can think of a magic monotone as a measure of the ``stabilizer complexity'', or the non-stabilizer content of a quantum circuit. Intuitively, we can also regard it as the complexity of classically simulating the quantum circuit on a classical computer. Remarkably, it was shown in \cite{Veitch_2012} (building on the work of \cite{Mari_2012}) that a simple and computable magic monotone can be constructed out of the \emph{Wigner function}. The Wigner function is a quasi-probability distribution which seeks to represent a quantum state as a distribution in phase space. For finite (prime) dimensional Hilbert spaces, the phase space is taken to be a lattice, and the corresponding Wigner function is called the discrete Wigner function \cite{WOOTTERS19871, Leonard, sphere, quantumcomp, Galois, Gross_2006, classicality}. The discrete Wigner function satisfies many of the properties of a standard probability distribution, except for one -- it can take negative values. It was shown in \cite{Veitch_2012} that the amount of negativity in the Wigner function (see \eqref{negdef} for the definition) is a magic monotone. In \cite{Pashayan_2015}, more detailed arguments were given to show that the Wigner negativity in a circuit provides a measure of the complexity of classical simulation using quasi-probability methods. 

Our goal in this paper is to study the stabilizer complexity of the state of the Hawking radiation in an evaporating black hole. More precisely, we will work with Wigner negativity as our magic monotone, and calculate the negativity of the radiation in toy models of black hole evaporation. Our main results are as follows:
\begin{itemize}
\item In section \ref{sec:PSSY}, we consider the Penington-Shenker-Stanford-Yang (PSSY) model \cite{Penington:2019kki} for black hole evaporation in JT gravity coupled to an external bath. In this model, we compute the Wigner negativity directly by using the gravitational path integral and show that the negativity is $O(1)$ before the Page transition, but grows exponentially large after the Page transition and is given by
\beq \label{eq:main_neg}
\mathcal{N} \sim \sqrt{\frac{2}{\pi}}\,\exp\left[\frac{1}{2}\left(S_{\text{max}} - S_2\right)\right],
\eeq 
where $S_{\text{max}} = \log D$ is the entropy of the maximally mixed state on the radiation system, while $S_2$ is the second R\'enyi entropy of the radiation after the Page point. We also derive a more general formula which interpolates between the two extreme regimes. Importantly, gravity gives a universal, information theoretic formula for the Wigner negativity, essentially independent of the choice of computational basis for the bath subsystem.{\footnote{This essentially happens because the gravitational path integral computes an ensemble average over Hamiltonians and pseudorandom boundary conditions in the dual theory.} In this calculation, an important role is played by two-boundary Euclidean wormholes (the $n=2$ replica wormholes), which end up dominating past the Page point. }

\item In section \ref{sec:dynamical}, we then turn to a dynamical model of black hole evaporation where the black hole is coupled to a bath and evolved in time by a joint Hamiltonian involving interaction terms between the black hole and the bath. With the simplifying assumption that the bath operators which couple to the black hole look like random matrices (picked from some unitarily invariant ensemble) with respect to the computational basis, we once again derive a universal formula for the Wigner negativity of the radiation system. This negativity shows a sharp spike (sort of like a ``negativity shock'') immediately after the bath is coupled to the black hole, but then decays due to decoherence and settles down at late times to a time-independent value which precisely matches the gravity formula we obtained in the PSSY model. 

\item Finally, in section \ref{sec:python}, we consider Wigner negativity in general holographic states satisfying the Ryu-Takayanagi formula for entanglement entropy. Here one must define the negativity with some care, because the boundary Hilbert space in AdS/CFT is infinite dimensional to begin with. Using intuition from fixed-area states and random tensor networks, we propose a geometric formula for the Wigner negativity: whenever there is a large gap between $\frac{1}{4G_N}$ times the areas of the outermost extremal surface $\gamma_{\text{out}}$ and the true extremal surface $\gamma_{\text{min}}$ (with minimal area), the Wigner negativity is given by
\beq 
\mathcal{N}\sim \sqrt{\frac{2}{\pi}}\exp\left[\frac{1}{2}\frac{\left(A(\gamma_{\text{out}}) - A(\gamma_{\text{min}})\right)}{4G_N}\right],
\eeq 
while the negativity is $O(1)$ when the gap vanishes. As a consequence of this formula, a python's lunch region in the entanglement wedge corresponds to an exponentially large stabilizer complexity. Our formula for the negativity has a similar flavor to the python's lunch conjecture, but also different -- firstly it does not seem to involve the bulge surface, and secondly the Wigner negativity is not directly the same as the reconstruction complexity relevant for the python's lunch conjecture.  
\end{itemize}

\section{Preliminaries}
\label{sec:prelim}

\subsection{The discrete Wigner function}

For any finite dimensional quantum system of odd, prime dimension $D$,\footnote{The discrete Wigner function formalism is nicest when $D$ is either an odd prime or a power of an odd prime, although most aspects of the formalism work more generally for odd $D$, and generalizations exist for even $D$ as well. When $D$ is an odd prime power, the Wigner function can be shown to be the unique quasi-probability distribution which ``geometrizes'' the action of the Clifford group in terms of symplectic transformations on phase space \cite{Gross_2006}. For simplicity, we will take $D$ to be an odd prime number in this work. Most of our results should also work when $D$ is a power of an odd prime.} there exists a natural quasi-probability representation called the \emph{discrete Wigner function} \cite{WOOTTERS19871, Leonard, sphere, quantumcomp, Galois, Gross_2006, classicality}. Let $\left\{|k\rangle\right\}_{k=0}^{D-1}$ be an ordered, orthonormal basis for the Hilbert space. The essential idea is to interpret this as the ``position'' basis, and correspondingly construct a ``phase space''. We define the \emph{discrete phase space} $\mathcal{P}$ as the lattice $\mathbb{Z}_D \times \mathbb{Z}_D$ of size $D^2$. 
With respect to the chosen basis, we define a set of $D^2$ operators $A(q,p)$ called phase-point operators, each labelled by a phase space point :
\begin{equation}\label{eq:A_Wooters}
A(q,p) = \sum_{k,\ell=0}^{D-1} \widehat{\delta}_{2q,k+l}e^{\frac{2\pi i}{D} (k-\ell)p } \ketbra{k}{\ell},
\end{equation}
where the hatted Kronecker delta $\widehat{\delta}$ is the $\text{mod}\,D$ version, i.e., it is one when $(k+\ell) = 2q\,\text{mod}\,D$, and zero otherwise. The discrete Wigner function for a density matrix $\rho$ is now defined (in analogy with the continuous case \cite{Wigner}) as:
\beq 
W_{\rho}(q,p) = \frac{1}{D} \mathrm{Tr}\left(\rho A(q,p)\right).
\eeq 
The Wigner function attempts to represent the quantum state $\rho$ as a probability distribution in phase space, much like is the case in classical mechanics. Indeed, it satisfies the following properties:
\begin{enumerate}
\item The discrete Wigner function is real and is unit normalized, i.e., $\sum_{q,p=0}^{D-1} W_{\rho}(q,p) = 1$, assuming $\text{Tr}\,(\rho)=1.$
\item Summing over one of the directions, say either $p$ or $q$, reduces the Wigner function to the probability density along the other direction:
\beq 
\sum_{p=0}^{D-1} W_{\rho}(q,p) = \langle q|\rho|q\rangle,\;\;\;\sum_{q=0}^{D-1} W_{\rho}(q,p) = \langle p|\rho|p\rangle,
\eeq 
where we have defined the ``momentum eigenstates'' as $|p\rangle = \frac{1}{\sqrt{D}}\sum_{q=0}^{D-1}e^{\frac{2\pi i pq}{D}}|q\rangle$. 
\item The time evolution of the discrete Wigner function follows a discrete version of the Moyal equation \cite{WOOTTERS19871}, which can be thought of as the quantum analog of the Liouville equation from classical mechanics.  
\end{enumerate}
While the above properties seem to suggest that we should regard the Wigner function as a probability distribution in phase space, this interpretation fails for an important reason -- the Wigner function can take negative values at some points in phase space. Indeed, had this not been the case, then quantum mechanics would reduce to classical probabilistic dynamics. The negativity of the Wigner function will play a central role in this work. 

So far we have discussed Wigner functions for states, but Wigner functions can also be defined for quantum operations/channels \cite{Mari_2012, Wang_2019}. For a quantum channel $\mathcal{E}(\rho) = \sum_m E_m\,\rho\,E_m^{\dagger}$ given in terms of a Krauss operator representation $\{E_m\}$, the Wigner function is defined as 
\beq 
W_{\mathcal{E}}(q,p|q',p')  = \frac{1}{D}\sum_m\mathrm{Tr}\left(A(q,p)\,E_m\,A(q',p')\,E_m^{\dagger}\right).
\eeq 
This can be thought of as a representation of the quantum channel as a quasi-stochastic matrix on phase space. Indeed, given an input state $\rho$ with the Wigner function $W_{\rho}(q,p)$, the Wigner function of the output state $\mathcal{E}(\rho)$ satisfies
\beq 
W_{\mathcal{E}(\rho)}(q,p) = \sum_{q',p'} W_{\mathcal{E}}(q,p|q',p') \,W_{\rho}(q',p').
\eeq 
Similarly, for a positive operator valued measure $\{M_n\}$, the corresponding Wigner function is defined as:
\beq 
W(n | q,p) = \mathrm{Tr}\left(M_n\,A(q,p)\right).
\eeq 
This definition is designed to satisfy:
\beq 
\text{Tr}\,(M_n\rho) = \sum_{q,p} W(n| q,p)W_{\rho}(q,p),
\eeq 
and turns the POVM into a quasi conditional probability distribution. In this way, the outcome of a general quantum circuit:
\beq
P_n = \mathrm{Tr}\left(M_n\,\mathcal{E}_m \cdots \mathcal{E}_1(\rho)\right),
\eeq 
translates to quasi-stochastic evolution in terms of the Wigner function representations:
\beq \label{outcome}
P_n = \sum_{q,p}\sum_{q_1,p_1,\cdots q_m,p_m}\sum_{q',p'}W(n|q,p)\, W_{\mathcal{E}_m}(q_m,p_m|q_{m-1},p_{m-1})\cdots W_{\mathcal{E}_1}(q_1,p_1|q',p')\,W_{\rho}(q',p'),
\eeq
with the important caveat that while $P_n$ is always non-negative, the intermediate Wigner functions are not always everywhere non-negative. From the above discussion, it is natural to regard the set of states, operations and measurements with non-negative Wigner functions as being \emph{classical}, and the amount of negativity in the Wigner function as a measure of non-classicality. One can make several arguments for this (see \cite{Basu:2025mmm} for further discussion); here we will focus on one line of reasoning motivated by complexity considerations.

\subsection{Wigner negativity as stabilizer complexity}

A direct link between positivity of the Wigner function and stabilizer quantum computation is provided by the \emph{Gottesman-Knill theorem} \cite{gottesman1998heisenberg, Aaronson:2004xuh}. At a heuristic level, any quantum circuit which involves initializing to a Wigner positive state, quantum operations with positive Wigner functions, and measurements defined by POVMs with positive Wigner functions can be efficiently simulated on a classical computer \cite{Mari_2012, Veitch_2012}. The underlying intuition is that any quantum circuit constructed out of Wigner positive elements 
 reduces to classical stochastic  (Markovian) evolution in terms of the Wigner function representation -- see equation \eqref{outcome}, where now all the Wigner functions are genuine probability distributions. As long as each of these Wigner probability distributions can be efficiently sampled, then we can also efficiently sample the output probability distribution of the quantum circuit on a classical computer.\footnote{Everywhere positive Wigner functions tend to be severely constrained  and have very few free parameters, see for instance the characterization in \cite{Gross_2006} for pure Wigner positive states and Clifford unitaries. So, we expect that sampling from such distributions should be efficient, although this is not always guaranteed, especially for mixed states and quantum channels.} Interestingly, one can generalize this argument to the case where circuit elements are not necessarily Wigner positive, and one finds that the amount of negativity in the circuit is directly connected to the classical simulation complexity \cite{Pashayan_2015, Wang_2019}.

From the above discussion, it is natural to interpret the negativity of the Wigner function as a measure of the complexity of classical simulation. One way to make this more concrete is in terms of the resource theory of stabilizer computation \cite{Veitch_2012, Veitch_2014}. Pure states with positive Wigner functions and convex combinations thereof are called \emph{stabilizer states}.\footnote{Usually stabilizer states are introduced as $+1$ eigenstates of a subgroup of Pauli operators, but the above definition turns out to be equivalent \cite{Gross_2006}.} In the resource theory of stabilizer computation, stabilizer states are regarded as being ``free of cost''. \emph{Clifford unitaries}, which constitute a special subgroup of the unitary group which map stabilizer states to stabilizer states, are also regarded as being free.\footnote{In the discrete phase space formalism, Clifford unitaries correspond to symplectic affine transformations on the discrete phase space. Furthermore, for a system with $n$ qudits, the most general Clifford unitary can be constructed out of $O(n \log d)$ elementary one and two-qudit gates \cite{Hostens:2005svl}. } In addition, one is also allowed to bring in ancillas initialized to stabilizer states, perform joint Clifford unitaries, make measurements on the ancilla in the computational basis, and use conditioning on previous measurements or classical randomness; all these operations constitute ``allowed'' operations in stabilizer protocols. However, stabilizer protocols are \emph{not} universal -- one can think of them as generating a class of circuits that can be efficiently simulated classically. In order to achieve universal quantum computation, one needs access to additional \emph{resource} states, sometimes referred to as ``magic states'' \cite{Bravyi_2005}. Much like entanglement plays the role of a resource in the theory of quantum communication with local operations and classical communication being treated as ``free'', \emph{magic} states play the role of a resource in the theory of stabilizer quantum computation, with stabilizer states and operations being treated as cheap \cite{Veitch_2014}. Magic has found many applications in theoretical physics, such as in the context of conformal field theories \cite{White:2020zoz}, many-body systems \cite{Liu:2020yso, Turkeshi:2023lqu, tirrito2025anticoncentrationnonstabilizernessspreadingergodic, Fliss:2020yrd}, chaos and random circuits \cite{White:2020hgn, Goto_2022, Leone2021quantumchaosis, Chen:2022yza, Niroula:2023meg, PhysRevLett.132.140401, Zhang:2024fyp, Oliviero:2022url} and to some extent also in quantum gravity \cite{Cao:2023mzo, Cao:2024nrx, 2024JHEP...05..264B, Basu:2025mmm}.     

In order to quantify the notion of magic, one defines a \emph{magic monotone} $\mathcal{M}$ as any real-valued, non-negative function on density matrices which vanishes on stabilizer states, and is, on average, non-increasing under any stabilizer protocol \cite{Veitch_2014}. By on average, we mean that if $\Lambda$ is a stabilizer protocol potentially involving a measurement, and if under $\Lambda$ we have $\rho \stackrel{\Lambda}{\to} \{p_i,\sigma_i\}$ (i.e., the outcome is the state $\sigma_i$ with probability $p_i$), then 
\beq 
\mathcal{M}(\rho) \geq \sum_i p_i \mathcal{M}(\sigma_i).
\eeq 
Importantly, it may be the case that for some $i$, $\mathcal{M}(\rho) < \mathcal{M}(\sigma_i)$, as long as the corresponding probability is sufficiently small. It turns out that the \emph{sum negativity} of the Wigner function, defined as:
\begin{equation}
    \label{negdef}
    \cN_s(\rho)= \frac{1}{2}\sum_{q,p} (|W_{\rho}(q,p)|-1),
\end{equation}
is a magic monotone \cite{Veitch_2014}. This gives Wigner negativity a direct operational meaning: if there exists a stabilizer protocol which converts an initial state $\rho_0$ to a final state $\rho$ with probability $p$, then 
\beq 
p \leq \frac{\mathcal{N}_s(\rho_0)}{\mathcal{N}_s(\rho)}.
\eeq 
Thus, in order to obtain the desired state, one would need to repeat \emph{any} such protocol at least $\frac{\mathcal{N}_s(\rho)}{\mathcal{N}_s(\rho_0)}$ number of times. Equivalently, one must start with a $\rho_0$ which comprises of order $\log \mathcal{N}_s(\rho)$ number of copies of a resource state that has some $O(1)$ negativity in order to have a significant success probability for conversion to $\rho$. In this sense, one may regard Wigner negativity as a measure of ``stabilizer complexity'', i.e., the minimal, unavoidable complexity in preparing $\rho$ through a stabilizer protocol. Alternatively, one could think of Wigner negativity as a measure of the non-stabilizer content of the state, or ``how far'' the state is from being a stabilizer state. Since stabilizer circuits can be simulated efficiently on a classical computer, a yet another interpretation of Wigner negativity is as a measure of the complexity of classical simulation using the Wigner quasi-probability representation (see \cite{Pashayan_2015} for more details). This last point requires some care, especially in the presence of a local tensor product structure -- in that case, one can have states which have large Wigner negativity and yet are classically simulable using other (i.e., non-quasi-probability) methods, on account of low entanglement \cite{Vidal:2003pmm}.\footnote{We thank Alex May for bringing this to our attention.} Thus, Wigner negativity can be thought of as an operationally meaningful notion of complexity from these various points of view. We should emphasize that by complexity, we do not mean circuit complexity \cite{nielsen2005geometric, Nielsen_2006, Susskind:2014moa, susskind2014computational, Jefferson_2017, Chapman_2018, Brown_2018,  Balasubramanian:2019wgd, Brandao:2019sgy, Haferkamp:2021uxo, Balasubramanian:2021mxo}, although it would be interesting to study the connection to circuit complexity further.

In this paper we will focus on the quantity
\beq \label{eq:conv}
\mathcal{N}(\rho) := 1+ 2\mathcal{N}_s(\rho)= \sum_{q,p}|W_{\rho}(q,p)|,
\eeq 
and we will refer to this quantity as the Wigner negativity. The negativity has several nice properties:
\begin{enumerate}
\item It is evident from the definition that $\mathcal{N} \geq 1$, where the lower bound is saturated if and only if the Wigner function is nowhere negative. 

\item The negativity of a state $\rho$ is also bounded above by $
\sqrt{\frac{D}{e^{S_2}}}$, where $S_2$ is the second R\'enyi entropy of $\rho$. To see this, we note that
\begin{equation}
\sum_{q,p} W^2(q,p) = \frac{1}{D}\Tr(\rho^2). 
\end{equation}
Together with Jensen's inequality and the fact that $\mathrm{Tr}\,\rho^2 =e^{-S_2}$, this gives us the desired bound:
\beq
\sum_{q,p} |W(q,p)| \leq D \sqrt{\sum_{q,p} W^2(q,p)} = \sqrt{D} \sqrt{\mathrm{Tr}\,\rho^2}  =  \sqrt{D e^{-S_2}}.
\eeq

\item The negativity is convex, i.e., for some set of probabilities $\{p_i\}$, 
\beq
\mathcal{N}(\sum_i p_i\rho_i ) \leq \sum_i p_i \mathcal{N}(\rho_i).
\eeq 
\item Finally, the negativity gives a lower bound on a natural measure of decoherence \cite{Streltsov_2017} and spreading \cite{Balasubramanian:2022tpr, Nandy:2024htc, Baiguera:2025dkc}, namely the \emph{half-R\'enyi diagonal entropy}: given a density matrix $\rho$ and an orthonormal basis $\mathcal{B} = \left\{|i\rangle\right\}$, the half-R\'enyi diagonal entropy is defined as
\beq 
H^{(1/2)}(\rho) = 2\log\left(\sum_i \lambda^{1/2}_i\right),\;\;\lambda_i = \langle i|\rho|i\rangle.
\eeq 
Then, the Wigner negativity in the basis $\mathcal{B}$ satisfies the lower bound:
\beq
 \log\,\cN(\rho) \leq H^{(1/2)}(\rho).
\eeq
Actually, the Wigner function transforms covariantly under any Clifford unitary, and so the negativity remains invariant under Clifford unitary transformations. This immediately implies the stronger bound:
\beq
 \log\,\cN(\rho) \leq \text{min}_{U\in \text{Cliff}}H^{(1/2)}(U\rho U^{\dagger}),
\eeq
where the minimization is over Clifford unitaties. One can interpret this bound as saying that a state with some amount of negativity in its Wigner function must have the above minimal and unavoidable spreading in the wavefunction along \emph{any} direction in phase space. 
\end{enumerate}

\section{Wigner negativity in the PSSY model}\label{sec:PSSY}
In this section, we will compute the Wigner negativity for Hawking radiation in a toy model of black hole evaporation due to Penington, Shenker, Stanford and Yang (PSSY) \cite{Penington:2019kki}. {The emphasis in this section will be on computing the Wigner negativity using the \emph{gravitational path integral}. As we will see, the answer obtained from the gravitational path integral will be completely information theoretic in nature (i.e., independent of the choice of computational basis for the radiation) and matches with the formula for Wigner negativity in Haar random states obtained in \cite{White:2020hgn}}. 
\subsection{Setup}
The Euclidean action of JT gravity with end of the world (EOW) branes is given by:
\begin{equation}
    I=I_{\mathrm{JT}}+\mu \int_{\text {brane }} \mathrm{d} s,
\end{equation} 
\begin{equation}
    I_{\mathrm{JT}}=-\frac{S_{0}}{2 \pi}\left[\frac{1}{2} \int_{\mathcal{M}} \sqrt{g} R+\int_{\partial \mathcal{M}} \sqrt{h} K\right]-\left[\frac{1}{2} \int_{\mathcal{M}} \sqrt{g} \phi(R+2)+\int_{\partial \mathcal{M}} \sqrt{h} \phi (K-1)\right],
\label{eq:jt-action}
\end{equation}
with the boundary conditions:
\begin{equation}
    \left.\mathrm{d} s^{2}\right|_{\partial \mathcal{M}}=\frac{1}{\epsilon^{2}} \mathrm{~d} \tau^{2}, \quad \left.\phi\right|_{\partial\mathcal{M}}=\frac{\phi_b}{\epsilon}, \quad \epsilon \rightarrow 0.
\end{equation}
The boundary condition on the EOW brane is given by
\begin{equation}
    \partial_{n} \phi=\mu\geq 0, \quad K=0.
\end{equation}
We can regard $S_0$ as the entropy of the higher-dimensional extremal black hole for which the JT theory is the two-dimensional reduction, or it can be thought of as the ground state entropy of the JT system. The EOW brane hosts semi-classical, intrinsic degrees of freedom, which we label as $i,j$ etc.,  running from 1 to $D$. These intrinsic EOW brane states are taken to be orthogonal in the bulk semi-classical description i.e., these indices along the brane are contracted with the propagator $\delta_{ij}$. We may regard these internal states of the EOW brane as a toy model for bulk matter. 

JT gravity in asymptotically $AdS_2$ spacetimes is dual to a quantum mechanical theory $B$ on the asymptotic boundary -- the path integral of JT gravity computes ensemble averaged quantities from the boundary point of view \cite{Saad:2019lba}, where the averaging is over an ensemble of Hamiltonians. More precisely, JT gravity is dual to a double-scaled random matrix theory, with the leading contribution to the density of states given by:
\beq
\rho(s) = \frac{s}{2\pi^2}\sinh(2\pi s),\;\;\; E = \frac{s^2}{2}.
\eeq 

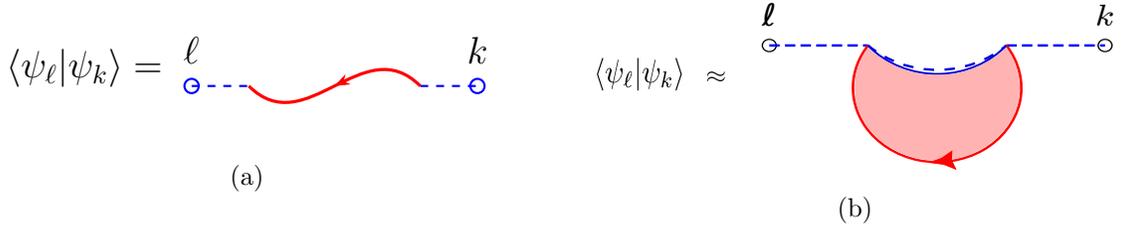
\begin{figure}[t]
  \centering

  \subfloat[]{\label{fig:OverlapBound}%
\begin{minipage}[c]{0.48\textwidth}%
\centering
    \begin{tikzpicture}[thick, scale=0.95, every node/.style={font=\Large}, baseline={(0,-0.1)}]

    \node at (-1.5,0.25) {$\langle \psi_\ell | \psi_k \rangle = $};

    \draw[blue, thick, dashed] (0,0) -- (0.8,0);
    \draw[blue, thick, dashed] (3.2,0) -- (4,0);

    \draw[red, very thick, arrow in middle]
      (3.2,0) .. controls (2.4,0.8) and (1.6,-0.8) .. (0.8,0);

    \draw[blue, thick] (0,0) circle (0.1);
    \draw[blue, thick] (4,0) circle (0.1);

    \node at (0,0.5) {$\ell$};
    \node at (4,0.5) {$k$};

    \end{tikzpicture}
\end{minipage}}
  \hfill
  \subfloat[]{\label{fig:OverlapBulk}%
\begin{minipage}[c]{0.48\textwidth}%
\centering
    \includegraphics[scale=1.8]{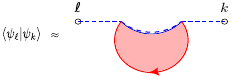}
\end{minipage}}

  \caption{\textbf{(a)}: Boundary condition to evaluate the overlap $\langle \psi_{\ell}|\psi_k\rangle$:The solid line represents the asymptotic boundary of Euclidean $\text{AdS}_2$ with renormalized length $\beta$. The arrow indicates the direction of time from ket to bra. Indices attached to the dashed lines represent the states of the EOW branes. The boundary condition suggests that EOW branes of type $\ell$ or $k$ should intersect the asymptotic $\text{AdS}_2$ boundary at the appropriate endpoints. \textbf{(b)}: The bulk gravity path integral corresponding to the overlap, subject to the boundary conditions illustrated on the left.
}. 
  \label{fig:psi-overlap}
\end{figure}

The main object of interest for us in this paper is a stationary black hole solution in Jackiw-Teitelboim (JT) gravity with an end of the world (EOW) brane capping off the geometry behind the black hole horizon. This black hole solution describes the semi-classical bulk dual to the following excited state in the dual quantum mechanical boundary theory:
\beq \label{eq:coeff}
|\psi_i\rangle = \sum_a2^{\frac{1}{2}-\mu}\Gamma(\mu-\frac{1}{2}+i\sqrt{2E_a})e^{-\frac{\beta}{2}E_a} C_{ia} |E_a\rangle,
\eeq 
where the coefficients $C_{ia}$ are i.i.d complex Gaussian random variables \cite{Penington:2019kki}. The states $|\psi_i\rangle$ are approximately orthogonal. The overlap $\langle \psi_{\ell}|\psi_k\rangle$, or more precisely, its ensemble average, is given by the gravity path integral (see figure \ref{fig:psi-overlap}): 
\beq 
\overline{\langle \psi_i | \psi_j\rangle} = e^{S_0}Z_1\,\delta_{ij},
\eeq 
where $Z_1$ is the disc partition function of JT gravity with one asymptotic boundary and one EOW brane:
\begin{equation}
   Z_1 = \int \rho(s) y(s), \quad \rho(s) = \frac{s}{2 \pi^2}\sinh( 2\pi s),\; \quad y(s) = e^{-\frac{\beta s^2}{2}} 2^{1 - 2 \mu}|\Gamma( \mu - \frac{1}{2} + is)|^2.
\end{equation}

More generally, the JT gravity path integral on the disc topology with $n$ asymptotic boundaries of length $\beta$ and $n$ EOW branes is given by $e^{S_0} Z_n$, where
\begin{equation}
   Z_n = \int \rho(s) y(s)^n, \quad \rho(s) = \frac{s}{2 \pi^2}\sinh( 2\pi s),\; \quad y(s) = e^{-\frac{\beta s^2}{2}} 2^{1 - 2 \mu}|\Gamma( \mu - \frac{1}{2} + is)|^2.
\end{equation}

In order to model an evaporating scenario, one entangles the black hole system $B$ with a quantum mechanical bath/reservoir $R$ \cite{Penington:2019kki}:
\begin{equation}
\label{eq: global states}
    \left|\Psi\right\rangle= \frac{1}{\sqrt{D\, e^{S_0}Z_1}} \sum_{i=1}^{D}\left|\psi_i\right\rangle_B \otimes |i \rangle_R,
\end{equation} 
where the states $\{|i\rangle\}$ in $R$ are orthonormal, and following \cite{Penington:2019kki}, we have chosen the normalization factor to ensure that:
\beq 
\overline{\langle \Psi|\Psi\rangle} = 1.
\eeq 
Equation \eqref{eq: global states} should be thought of as the late-time equilibrium state we get by coupling the black hole to a bath of Hilbert space dimension $D$ through random interaction terms (without any conservation laws), and waiting long enough till the system settles down. The reduced density matrix for the bath is given by:
\begin{equation}
    \rho_R=\frac{1}{D\, e^{S_0} Z_1} \sum_{i,j=1}^{D} \langle \psi_i|\psi_j\rangle_{B}\,|j\rangle \langle i|_R .
\end{equation}
For small $D$, one can treat the states $\left\{|\psi_i\rangle\right\}$ as being approximately orthogonal. This leads to a maximally mixed state on the bath whose entropy grows as $\log D$. However, when $D$ becomes $O(e^{S_0})$, one can no longer treat the states $\left\{|\psi_i\rangle\right\}$ as being orthogonal. Indeed, it was shown in \cite{Penington:2019kki} that in the replica trick computation of the radiation entropy, a new gravitational saddle, i.e., the replica wormhole, makes the dominant contribution past the Page point \cite{Penington:2019kki}. This cuts off the naive growth of the entropy with $D$, thus realizing the expected Page curve.

Our goal in this section is to compute the Wigner negativity of the reduced density matrix $\rho_R$, viewed as a measure of the stabilizer complexity, or equivalently, the complexity of classically simulating the state of the Hawking radiation. In doing so, we will use the rules of the gravitational path integral. In order to compute the Wigner function, we must choose an orthonormal basis for the bath Hilbert space. Let us first consider the radiation basis $\{|i\rangle\}$ which appears in equation \eqref{eq: global states}. The Wigner function with respect to this basis is given by:\footnote{Here we are taking $D$ to be an odd prime for convenience, and no further structure on the radiation Hilbert space is assumed. If the radiation Hilbert space has some locality structure -- for instance, if it consists of a tensor product of $n$ local qudits (for some odd prime $d$) -- then one can also use the Cartesian product phase space and its corresponding Wigner function \cite{WOOTTERS19871, Gross_2006}. Many of our conclusions will remain unaffected even in that case, with $D = d^n$.}
\begin{eqnarray}\label{eq:wig_neg}
    W(q,p)&=& \frac{1}{D^2 e^{S_0} Z_1}\sum_{k,\ell=0}^{D-1} \widehat{\delta}_{2q,k+\ell} \;e^{-\frac{i2\pi P}{D}(k-\ell)}\; \langle \psi_\ell|\psi_k\rangle_B \nonumber \\
    &=& \frac{1}{D^2 e^{S_0} Z_1}\sum_{k,\ell=0}^{D-1} A_{k\ell} \langle \psi_\ell|\psi_k\rangle_B,
\end{eqnarray}
where $A_{k\ell}=\langle \ell| A(q,p)|k\rangle$. Any other basis $\{|\tilde{i}\rangle\}$ can be written in terms of the above radiation basis as $|\tilde{i}\rangle = \sum_{i'}U_{ii'}|i'\rangle$, for some unitary matrix $U$. The corresponding Wigner function is given by
\beq 
W_{U}(q,p) = \frac{1}{D^2 e^{S_0}Z_1}\sum_{k,\ell,k',\ell'} A^{(U)}_{k\ell}\langle \psi_\ell|\psi_k\rangle_B,\;\; A^{(U)}_{k\ell}= U_{kk'}A_{k'\ell'} U^{\dagger}_{\ell'\ell}.
\eeq 
As we will see, the negativity of the above Wigner function computed using the gravitational path integral only involves traces of products of the form $\text{Tr}\left[(A^{(U)})^n\right]$, and so the unitary $U$ cancels out. This means that the answer for the negativity is \emph{independent} of the choice of basis. Two remarks: 
\begin{enumerate}
    \item In general, the Wigner negativity very much depends on the choice of basis. However, in our gravity calculation, the negativity seems to have a universal basis independent form, which can then only depend on information theoretic properties of the boundary state $\rho_R$. This essentially happens because the gravitational path integral computes an ensemble average over Hamiltonians and pseudorandom boundary conditions in the dual theory. As we will see later, this same universal (i.e., basis-independent) information-theoretic form for the Wigner negativity also appears more generally in chaotic systems with random matrix properties.  
    \item We have assumed that the basis for the bath is chosen independently, without prior knowledge of the microscopic details of the black hole. For instance, had we fine-tuned the basis in $R$ to involve microscopic information -- such as BH microstate energies, or the specific coefficients $C_{ia}$ in equation \eqref{eq:coeff} -- then the above analysis would not be valid, because one would need to account for additional Euclidean wormholes emanating from the various $U_{ij}$ matrices.
\end{enumerate}

\begin{figure}[t]
  \centering

  \subfloat[]{\label{fig:W}%
\begin{minipage}[c]{0.48\textwidth}%
\centering
    \includegraphics[scale=1.2]{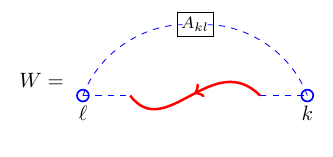}
\end{minipage}}
  \hfill
  \subfloat[]{\label{fig:Wbar}%
\begin{minipage}[c]{0.48\textwidth}%
\centering
    \includegraphics[scale=1.7]{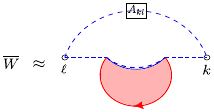}
\end{minipage}}

  \caption{(a) The boundary conditions for computing the Wigner function $W$. (b) The  gravitational path integral with the given boundary conditions: the same EOW brane gives $\delta_{kl}$, and so the diagram evaluates to $e^{S_0} Z_1\; \text{Tr}(A)$}. 
  \label{fig:boundaryCond}
\end{figure}

From equation \eqref{eq:wig_neg}, we see that the boundary conditions required to evaluate the Wigner function can be represented diagrammatically as in figure \ref{fig:W}, where the phase point operators are appended between the two terminal nodes. We can compute the ensemble average of the Wigner function $\overline{W(q,p)}$ by using the gravity path integral with these specified boundary conditions, as shown in figure \ref{fig:Wbar}. This just gives:
\beq\label{eq:avW}
\overline{W(q,p)} = \frac{1}{D^2e^{S_0}Z_1} \sum_{k,\ell=0}^{D-1} A_{k\ell} e^{S_0}Z_1 \delta_{k,\ell}= \frac{1}{D^2 } \text{Tr} (A) =\frac{1}{D^2}.
\eeq
So, on average, the Wigner function looks uniform and positive. However, this does not mean that the state of the Hawking radiation is simple -- what we actually want to compute is the average of the Wigner negativity, and as we will see in the next section, the negativity of the averaged Wigner function is not always the same as the averaged Wigner negativity. Indeed, we will find that after Page time, the Wigner function has large Wigner negativity; it's just that the averaging in equation \eqref{eq:avW} washes out this fine-grained information. 
\vspace{1em} 

\subsection{First look at Wigner negativity}

Building on the above discussion, our goal is to compute the ensemble averaged Wigner negativity of the Hawking radiation using the gravitational path integral:
\beq 
\overline{\mathcal{N}} = \sum_{q,p}\overline{ |W(q,p)|},
\eeq 
where note that the absolute value is inside the average. Indeed, the absolute value in the above formula creates trouble -- while calculating the ensemble average of the Wigner function and its powers is straightforward using the gravity path integral, the absolute value is a non-analytic function. In the next section, we will present a systematic way to deal with this difficulty using an integral representation of the absolute value function. But to develop some rough intuition, we will first employ the following somewhat heuristic \emph{replica trick}: we first evaluate the ensemble average over $W^{2n}$ for integer $n$, and analytically continue the result to $n=\frac{1}{2}$ in order to extract the average of the absolute value of $W$:
\begin{equation}
    \lim_{n\to \frac{1}{2}}\sum_{q,\;p}\overline{W^{2n}(q,p)}. \label{eq:replica}
\end{equation}

\begin{figure}[t]
  \centering
  \includegraphics[scale=1.8]{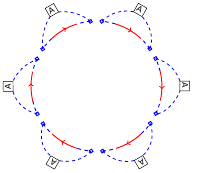} 
  \caption{Schematic diagram representing the boundary conditions used to evaluate $W^{2n}$, shown here for $n=3$. Six copies of figure \ref{fig:W} are arranged in a circular layout, with each $A$ denoting the insertion of a phase point operator. A gravitational path integral with these boundary conditions generally admits multiple saddle point solutions, corresponding to different ways of filling in the bulk region.}
  \label{fig:W2nreplica_bc}
\end{figure}

The point is that it is much easier to calculate $W^{2n}$ using the gravitational path integral. In order to do so, we set up the appropriate boundary conditions corresponding to $W^{2n}$ (see figure \ref{fig:W2nreplica_bc}), and then let gravity fill in the bulk in all possible ways consistent with these boundary conditions. A key point is that for large values of $n$, multiple geometries contribute to $\overline{W^{2n}}$, and resumming the contributions from all such saddles becomes a somewhat non-trivial task. We will come back to this in the next section, but to gain some initial insight, we will begin with a simplified analysis in this subsection. We will consider the two extreme limits $D \ll e^{S_0}$ and $D \gg e^{S_0}$, where the gravity path integral and subsequent analytic continuation can be performed straightforwardly. This approach will help capture the essential features of the Wigner negativity across the Page transition.

\begin{figure}[t]
  \centering

  \subfloat[]{\label{fig:beforePage}%
\begin{minipage}[t]{0.45\textwidth}%
\centering
    \includegraphics[scale=1.6]{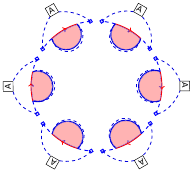}
\end{minipage}}
  \hspace{0.1cm}
  \subfloat[]{\label{fig:afterPage}%
\begin{minipage}[t]{0.45\textwidth}%
\centering
    \includegraphics[scale=1.6]{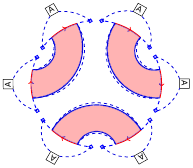}
\end{minipage}}

  \caption{The replica trick for $n=3$. (a) In the regime $D\ll e^{S_0}$, the fully disconnected geometry dominates the gravitational path integral. Each disconnected piece contributes $e^{S_0} Z_1 \text{Tr}(A)$, leading to a total contribution of $(e^{S_0}Z_1)^6$. (b) In contrast, for $D\gg e^{S_0}$, the dominant saddle is a pairwise connected geometry. Each connected component connects two asymptotic boundaries. Each pair gives a factor $\text{Tr}(A^2)=D$ from the sum over EOW brane indices, and the full diagram evaluates to $(D e^{S_0}Z_2)^3$.}
  \label{fig:extreme saddles}
\end{figure}

First, consider the regime $D\ll e^{S_0}$. In this limit, the dominant contribution comes from the completely disconnected diagram, shown in figure \ref{fig:beforePage}. All other diagrams come with lower powers of $e^{S_0}$, and are suppressed when $D \ll e^{S_0}$. This gives:
\begin{equation}
    \overline{W^{2n}(q,p)}\approx \big(\;\overline{W(q,p)}\;\big)^{2n}=\frac{1}{D^{4n}} \;\;\;\;\cdots\;\;\;\; (D\ll e^{S_0}).
\end{equation}
Upon analytic continuation to $n=\frac{1}{2}$, we find that the averaged Wigner negativity is given by 
\beq 
\overline{\mathcal{N}} \approx 1,\;\;\; \cdots \;\;\;(D \ll e^{S_0}).
\eeq 
Thus, the state of the Hawking radiation well before the Page transition has vanishing stabilizer complexity (recall from equation \eqref{eq:conv} that $\overline{\mathcal{N}}_s = \frac{1}{2}(\overline{\mathcal{N}} - 1)$).  

Next, we consider the opposite limit, where $D\gg e^{S_0}$. In this limit, the dominant diagrams are given by pairwise connected geometries, as illustrated in figure 
\ref{fig:afterPage}. The reason such pair-wise connected diagrams become important is as follows: while such diagrams are subleading in $e^{S_0}$ compared to the fully disconnected diagram, the EOW brane index contractions for such diagrams give an enhancement at large $D$ coming from the fact that $\text{Tr}(A^2) = D$ (see equation \eqref{Apowers}). For $2n$ boundaries, such pairwise-connected configurations can be constructed in $\frac{(2n)!}{2^n n!}$ distinct ways. Contributions from other geometries (with three and higher boundary connections) are subleading and can be ignored. This leads to the following expression:
\begin{equation}
    \overline{W^{2n}(q,p)}\approx \frac{(2n)!}{2^n n!}\; \frac{\big( e^{S_0}Z_2\; \text{Tr}(A^2) \big)^n}{\big( e^{S_0}Z_1 \; D^2\big)^{2n}} =\frac{(2n)!}{2^n n!}\;\bigg(\frac{Z_2}{Z_1^2}\bigg)^n \frac{1}{e^{nS_0}D^{3n}} \;\;\;\cdots\;\;\;\; (D\gg e^{S_0}).
\end{equation}
Analytically continuing this expression to $n \to \frac{1}{2}$,  we find that for $D \gg e^{S_0}$, the averaged negativity is given by
\beq 
\overline{\mathcal{N}} \approx \sqrt{\frac{2}{\pi}} \exp\left[\frac{1}{2}(S_{\text{max}} - S_2)\right],\;\;\; \cdots \;\;\; (D\gg e^{S_0}) ,
\eeq 
where $S_2$ is the second R\'enyi entropy of the radiation post the Page transition
\beq \label{eq:secondrenyi}
S_2 = S_0 - \log\left(\frac{Z_2}{Z_1^2}\right),
\eeq 
and $S_{\text{max}} = \log D$ is the coarse-grained entropy, or equivalently the entropy of the maximally mixed state on $R$. Thus, we see that after the Page point, the Wigner negativity becomes exponentially large despite the averaged Wigner function being uniform and positive, as in equation \eqref{eq:avW}. So, the state of the Hawking radiation post the Page transition has an exponentially large stabilizer complexity. It is interesting to note that this complexity is given universally -- i.e., in a basis independent way -- in terms of information theoretic quantities. 

The above simple exercise already reveals the asymptotic behavior of the Wigner negativity as a function of the ratio $D/e^{S_0}$. However, when $D$ and $e^{S_0}$ are of comparable magnitude, one must, in principle, account for other diagrams beyond the fully disconnected and the pairwise connected ones in the gravitational path integral. This makes the analytic continuation non-trivial. To capture the intermediate behavior of the Wigner negativity in this regime, a more careful analysis is required, which we turn to in the next subsection.


\subsection{Integral representation method}

\begin{figure}[t]
  \centering
  \includegraphics[scale=3.7]{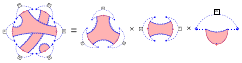} 
  \caption{A generic saddle that arises in the computation of $\overline{W^6}$. It can be decomposed into three irreducible components: a three-boundary contraction, a two-boundary contraction, and a one-boundary contraction.
}
  \label{fig:W6generic}
\end{figure}

A more systematic approach to deal with the absolute value in calculating the ensemble average of the Wigner negativity is to use the following integral representation: 
\begin{align*} \label{eq:wignev_intrep}\nt 
    |W|= W \sgn(W) &= \lim_{\epsilon \rightarrow 0}\int_{-\infty}^{\infty} \frac{\d z}{2\pi i}\; \frac{2z}{z^2+\epsilon^2} W \exp (izW) \\ &= \lim_{\epsilon \rightarrow 0}  \int_{-\infty}^\infty \frac{\d z}{2\pi i} \frac{-2i z}{z^2 + \epsilon^2}  \frac{\d}{\d z} \exp (i z W).
\end{align*}
In order to avoid cumbersome notation, we will simply denote the Wigner function as $W$, leaving the dependence on $(q,p)$ implicit, but it should be understood that the following analysis works point-wise in phase space. The above integral representation is useful because it reduces the calculation of the ensemble average over the non-analytic function $|W|$ to the ensemble average of an analytic function, namely $e^{i z W}$. Expanding out the exponential and using equation \eqref{eq:wig_neg}, 
the ensemble average of $\overline{e^{it W}}$ is given by:
\begin{equation}\label{eq:sumN}
 \overline{  e^{i z W}} = \sum_{n =1}^{\infty}  \frac{(iz)^n}{n!} \overline{W^n}=\sum_{n =1}^{\infty}  \frac{1}{n!} \left(\frac{iz}{D^2e^{S_0} Z_1}\right)^n\overline {\left( \sum_{k,\ell=0}^{D-1} A_{k\ell} \langle \psi_\ell | \psi_k \rangle_B \right)^n}.
\end{equation}
So, we need to compute $\overline{ \prod_{i =1}^{n} \langle \psi_{\ell_i} |\psi_{k_i}\rangle} $. Each overlap $\langle \psi_{\ell}|\psi_k\rangle$ corresponds to one asymptotic boundary. The bulk dual to the ensemble average corresponds to a sum over all possible ways in which the $n$ boundaries can be filled-in by a gravitational bulk. In the present case, it suffices to only consider bulk geometries topologically corresponding to multiple discs; geometries with handles are suppressed by powers of $e^{-S_0}$. Furthermore, since the index contractions are all determined by the structure of contractions of the EOW branes, handles do not modify the invariant we get from contractions of the phase-point operators, and so, do not give any additional enhancements. For example, for $n=3$, we have the following possibilities:
\begin{figure}[t]
  \centering
  \includegraphics[scale=1.15]{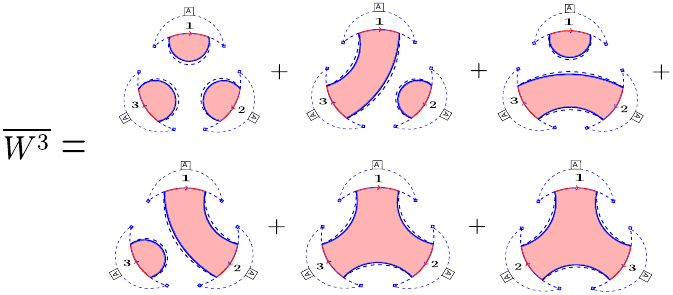} 
  \caption{All possible geometries (at disk level) contributing to the gravitational path integral for $\overline{W^3}$.} 
  \label{fig:W3example}
\end{figure}
For arbitrary $n$, the quantity $\overline{W^n}$ receives contributions from a series of such terms, corresponding to various bulk geometries with multiple discs. A key observation is that \textit{any} such term can be decomposed into a product of irreducible building blocks. We illustrate this structure diagrammatically with a representative example in figure \ref{fig:W6generic}. For a configuration with $n$ boundaries, any generic term can be expressed as a product of $p_1$ one-boundary contractions (i.e., $p_1$ discs, each with one asymptotic boundary and one EOW brane), $p_2$ two-boundary contractions (i.e., $p_2$ discs, each with two asymptotic boundaries and two EOW branes), $p_3$ three-boundary contractions (i.e., $p_3$ discs, each with three asymptotic boundaries and three EOW branes), and so on, up to $p_n$ $n$-boundary contractions, subject to the constraint $\sum_{j=1}^{n} j p_j = n$. Here, each $j$-boundary contraction is assumed to be irreducible, i.e., fully connected.

We can now organize the sum over saddles as a sum over products of ``connected components" where each connected component corresponds to boundaries sharing a common gravitational bulk. For $n = 3$, this is shown in figure \ref{fig:W3example}. After multiplying with the weights $A_{k\ell}$ and summing over all the indices, we get\footnote{The combinatorial factor comes about as follows: there are a total of $\frac{n!}{p_1! (2p_2)!\cdots (n p_n)!}$ ways of partitioning the $n$ boundaries into $p_1$ single components, $p_2$ pairs, $p_3$ triples, and so on. Furthermore, the number of ways of grouping $ jp_j$ boundaries into $p_j$ groups of $j$ boundaries each is given by $\frac{(jp_j)!}{p_j! (j!)^{p_j}}$.} : 
\begin{equation}
   \overline  {W^n} = \left( \frac{1}{D^2 e^{S_0}Z_1}\right)^n \sum_{p_1 \dots p_n}\delta_{\sum_r r p_r, n}{\frac{n!}{p_1! \dots p_n!}} \prod_{j = 1}^{n} \left(\frac{D_j}{j!}\right)^{p_j}. \label{eq:Wn}
\end{equation}
Here $D_j$ is JT gravity path integral corresponding to a fully connected diagram with $j$ asymptotic boundaries and $j$ EOW branes:
\begin{equation}
    D_j = (j-1)! \,  e^{S_0} \, Z_j \, \Tr(A^j),
\end{equation}
where the $(j-1)!$ is a combinatorial factor that counts the number of possible ways to fully contract the $j$ asymptotic boundaries. It is easy to evaluate $\Tr(A^k)$ by repeatedly using the formula \cite{WOOTTERS19871}
\beq
A(q_1,p_1) A(q_2,p_2) = \sum_{q_3,p_3} \Gamma(q_1,p_1,q_2,p_2|q_3,p_3) A(q_3,p_3),
\eeq 
$$\Gamma(q_1,p_1,q_2,p_2 | q_3,p_3) = \frac{1}{D}\exp\left[\frac{4\pi i}{D}\Big(p_1(q_3-q_2) + p_2(q_1-q_3) + p_3(q_2-q_1)\Big)\right],$$
and this gives
\begin{equation}\label{Apowers}
    \Tr(A^k) = \begin{cases}
        1  \quad  \text{if} \; k \; \text{is} \; \text{odd}, \\ 
        D  \quad  \text{if} \; k \; \text{is} \;\text{even}.
    \end{cases}
\end{equation}
Since the only dependence on the phase space variables $(q, p)$ is through $\Tr(A^j)$, we see that $\overline{|W|}$ is independent of $(q,p)$.

\begin{figure}[t]
    \centering
    \begin{tabular}{l r}
    \includegraphics[scale=0.38]{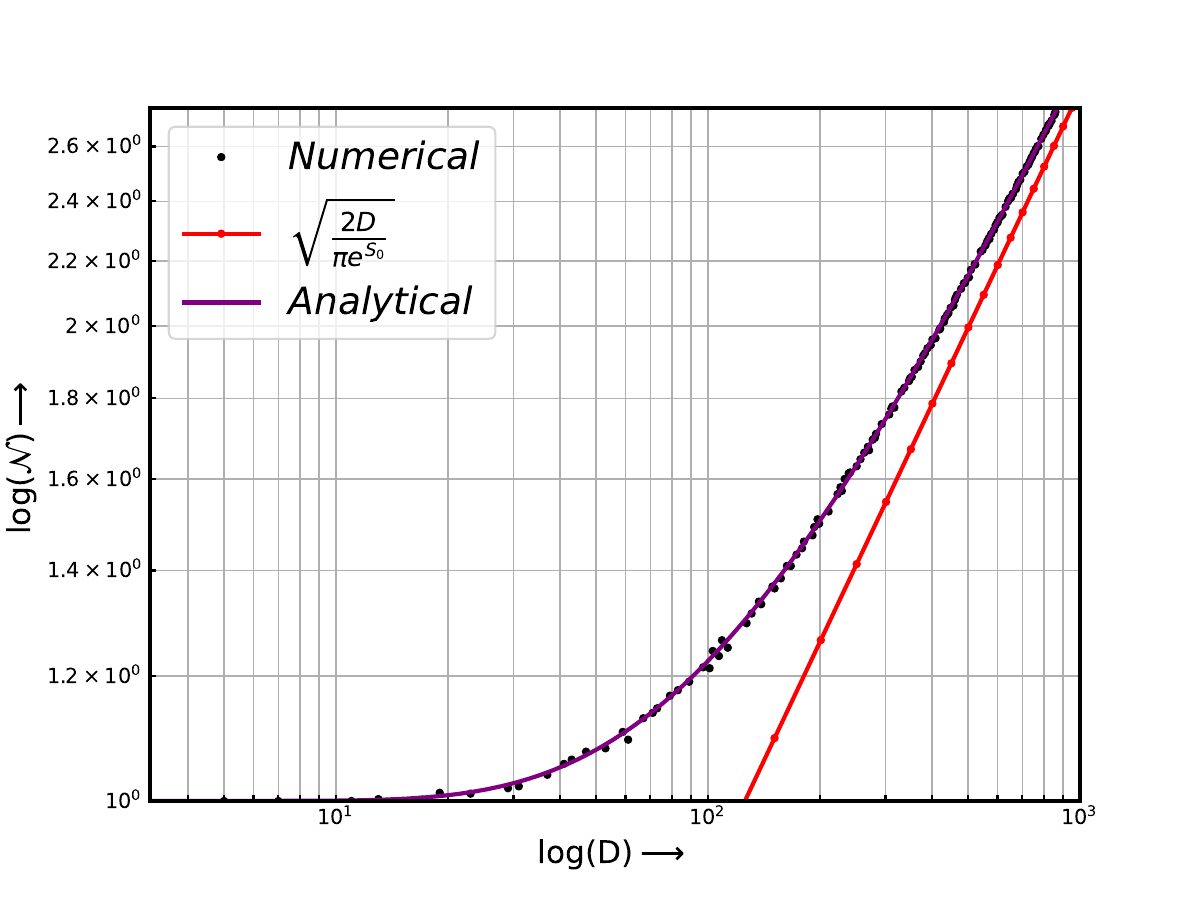} & \includegraphics[scale=0.35]{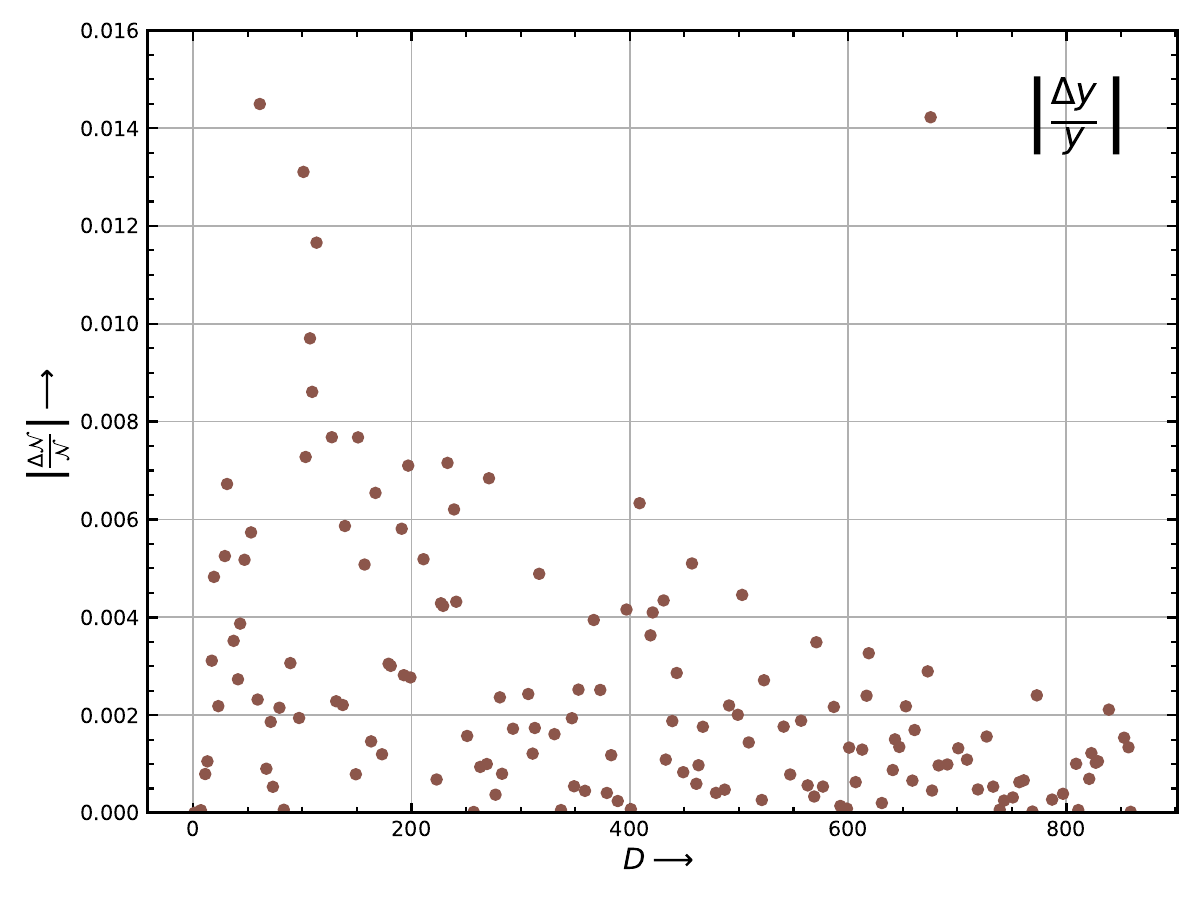}
    \end{tabular}
    \caption{ (\textbf{Left}) Log-log plot of Wigner negativity for the microcanonical ensemble with the black hole Hilbert space dimension $e^{S_{\text{micro}}}=80$. The black dots are numerically calculated values for the state in equation \eqref{eq:coeff} for one particular choice of the $C_{ia}$, and where the energies are taken to be identical. \textbf{(Right)} Relative error between analytical formula and numerical data.}
    \label{fig:micro}
\end{figure}

Going back to equation \eqref{eq:sumN}, we can now perform the sum over $n$ exactly. The sum over $n$ effectively decouples the sums over $p_j$, allowing each to be evaluated independently, and we get
\begin{eqnarray}
    \overline{e^{izW}} &=& \sum_{n=0}^{\infty}\frac{(iz)^n}{n!} \overline{W^n}\nonumber \\
    &=&\sum_{n=0}^{\infty}\left( \frac{iz}{D^2 e^{S_0}Z_1}\right)^n \frac{1}{n!} \sum_{p_1 \dots p_n}\delta_{\sum_r r p_r, n}{\frac{n!}{p_1! \dots p_n!}} \prod_{j = 1}^{n}\left( \frac{D_j}{j!}\right)^{p_j} \nonumber \\
    &=& \prod_{j=1}^{\infty} \left\{\sum_{p_j=0}^{\infty}\frac{1}{p_j!} \left[ \left( \frac{iz}{D^2 e^{S_0}Z_1} \right)^j \frac{D_j}{j!}\right]^{p_j}\right\}\nonumber\\
     &=& \exp \left(\sum_{j=1}^{\infty} \frac{(iz/D^2 e^{S_0} Z_1)^j}{j!} D_j \right) .
     \end{eqnarray}
     Using the explicit form of gravitational amplitudes $D_j$, we get
     \beq \label{eq:exp}
     \overline{e^{izW}}= \exp \left[e^{S_0} \left\{\sum_{m=1}^{\infty} \left(\frac{iz}{D^2 e^{S_0} Z_1} \right)^{2m-1} \frac{Z_{2m-1}}{(2m-1)} + D\sum_{m=1}^{\infty} \left(\frac{iz}{D^2 e^{S_0}Z_1} \right)^{2m} \frac{Z_{2m}}{2m}\right\}\right] .
\eeq
Now, we can insert the above expression in equation (\ref{eq:wignev_intrep}) to calculate the ensemble averaged Wigner negativity:
\begin{equation} 
\overline{|W|} =\lim_{\varepsilon \rightarrow 0} \int_{-\infty}^{\infty} \frac{\d z}{2\pi i} \frac{-2iz}{z^2 + \varepsilon^2} \frac{\d}{\d z}\left( \overline{\exp(i z W)}\right).
\end{equation}
It is convenient to define the rescaled integration variable $t=\frac{z}{D e^{S_2}}$, where $S_2$ is defined in equation \eqref{eq:secondrenyi}. In terms of this rescaled integration variable, we get:
\begin{equation}
\overline{|W|} =\lim_{\eps \rightarrow 0} \frac{1}{D e^{S_2} } \int_{-\infty}^{\infty} \frac{\d t}{2\pi i} \frac{-2it}{t^2 + \eps^2} \frac{\d}{\d t} \left(\overline{\exp(i D e^{S_2} t W)}\right) ,
\end{equation}
where $\epsilon = \frac{\varepsilon}{De^{S_2}}$. 
To calculate the integral exactly, we need to know the partition functions $Z_j$ for all $j>0$. However, the integral simplifies if we take the large $D$ and large $\exp(S_0)$ limit with the ratio $e^{S_0}/D$ fixed. In this limit, we see that the exponent simplifies:
\begin{align*}
\overline{|W|} = \lim_{\eps \rightarrow 0} \frac{1}{De^{S_2} } \int_{-\infty}^{\infty}  \d t \frac{-2it}{t^2 + \eps^2} \frac{\d}{\d t} \exp \left[ \frac{e^{S_2}}{D} \left\{ it - \frac{t^2 }{2} \right\} + O\left(\frac{1}{D^2} \right)\right].
\end{align*}
In this limit, terms in the exponential with powers of $t^3$ and higher are suppressed by a factor of $\frac{1}{D^2}$, so these terms can be dropped at large-$D$. Thus, the integral reduces to:
\begin{equation}\label{eq:integral}
    \overline{|W|} =\lim_{\eps \rightarrow 0}  \frac{1}{D e^{S_2} } \int_{-\infty}^{\infty} \d t \frac{-2 it}{t^2 + \eps^2} \frac{\d}{\d t} \exp \left[ \frac{e^{S_2}}{D} \left( i t - \frac{t^2}{2} \right)\right] .
\end{equation}
The above integral can be evaluated exactly (see Appendix \ref{sec:integral} for details). This gives the following result for the ensemble average over the absolute value of the Wigner function:
\begin{equation}
    \overline{|W|} = \frac{1}{D^2}   \left( \frac{\exp(-r)}{\sqrt{\pi r}} + \erf(\sqrt{r}) \right), \quad r =  \frac{e^{S_2} }{2 D},
\end{equation}
where $S_2$ is defined in equation \eqref{eq:secondrenyi} and 
\begin{equation}
\erf(z) = \frac{2}{\sqrt{\pi}}\int_0^z \d t\, e^{-t^2}.
\end{equation}
Since $|\overline{W}|$ does not depend on $(q,p)$, the ensemble averaged Wigner negativity is therefore given by:
\begin{equation}\label{eq:neg}
    \overline{\mathcal{N}} = \erf(\sqrt{r}) + \frac{\exp(-r)}{\sqrt{\pi r}}.
\end{equation}
Note that when $r \to \infty$ (corresponding to $D \ll e^{S_2}$), $\overline{\mathcal{N}} \sim 1$, while when $r\to 0$ (corresponding to $D \gg e^{S_2}$), then $\overline{\mathcal{N}} \sim \sqrt{\frac{2D}{\pi e^{S_2}}}$, which is the result we obtained from the replica trick calculation in the previous section. Equation \eqref{eq:neg} interpolates between these two extreme limits. Notably, the answer obtained from the gravitational path integral is completely information theoretic in nature, does not depend on the choice of computational basis, and is closely related to the formula for Wigner negativity in Haar random states \cite{White:2020hgn}. In appendix \ref{sec:numerics}, we argue that fluctuations around the ensemble average are suppressed in $\frac{1}{De^{S_0}}$. In figure \ref{fig:micro}, we compare our analytical formula for the Wigner negativity with direct numerical calculations for the microcanonical ensemble and find good agreement. 


So, to summarize, the state of the Hawking radiation in the PSSY model has an $O(1)$ Wigner negativity before the Page transition, corresponding to low stabilizer complexity. Thus, before the Page transition, the state of the Hawking radiation is classically simulable. On the other hand, after the Page transition, the Hawking radiation has an exponentially large Wigner negativity, and thus an exponentially large stabilizer complexity. In this case, the state of the Hawking radiation is hard to simulate classically. Interestingly, the radiation system contains the black hole interior in its entanglement wedge after the Page transition, so one interpretation is that the large complexity is associated with the black hole interior. We will return to this point in section \ref{sec:python}.

\section{A dynamical model of black hole evaporation}\label{sec:dynamical}

In this section, we study Wigner negativity in a dynamical model of black hole evaporation. We consider two quantum systems $B$ (for black hole) and $R$ (for radiation) coupled to each other. The Hilbert space dimensions of $B$ and $R$ are $e^{S}$ and $D$ respectively. The subsystem $B$ should be thought of as the quantum system dual to the black hole and the subsystem $R$ should be identified with radiation bath. Our goal is to couple the black hole (in some pure state) to a ``simple'' pure state of the bath and study the growth of Wigner negativity of the bath system in real time. {The calculation in this section can be regarded as a time-dependent generalization of the calculations in \cite{White:2020hgn} (see also \cite{Basu:2025mmm}). }

\subsection{Ensemble averaged negativity}
 We choose a computational basis $\mathcal{B} = \left\{|i_0\rangle\right\}$ for the bath. We start with a product state of the total system of the form $|\chi_0\rangle_B \otimes |\psi_0\rangle_R$, where $\psi_0$ is a pure, stabilizer (i.e., Wigner positive) state with respect to the computational basis, and $\chi_0$ is a state of the $B$ system which is dual to a black hole on the gravity side. The combined $BR$ system is then evolved in time with some joint Hamiltonian $H_{BR}$. We will consider a joint Hamiltonian of the form:
 \beq \label{eq:intterms}
 H_{BR} = O_B\otimes O_R,
 \eeq 
with the further assumption that $O_R$ is a random matrix with respect to the computational basis on $R$, drawn from some unitarily invariant ensembles. We are interested in computing the growth of Wigner negativity of the time evolved density matrix of subsystem $R$:
\begin{equation}
    \rho_R(t)= \text{Tr}_B( e^{-iH_{BR}t} |\chi_0\rangle \langle\chi_0|_B \otimes |\psi_0\rangle\langle\psi_0|_R e^{iH_{BR}t}),
\end{equation}
 with respect to the computational basis. In general, it is hard to compute the growth of negativity for one particular choice of Hamiltonian. Our assumption that the operator $O_R$ is a random matrix allows us to get some analytical handle on the problem using tools from random matrix theory. This way, instead of computing the Wigner negativity for one specific choice of Hamiltonian, we can compute the ensemble average:
\begin{equation}
    \overline{\mathcal{N}(\rho_R(t))}=\int \; dH_{BR} \; \mu(H_{BR})\; \mathcal{N}_{H_{BR}}(\rho_R(t)),
\end{equation}
where by the average over $H_{BR}$, we really mean the average over $O_R$ with some unitarily invariant measure $\mu$.

In order to compute the negativity, we will follow the strategy explained in \cite{Basu:2025mmm}. Let $\mathcal{B}=\{ |i_0\rangle\}_{i=0}^{D-1}$ be the chosen computational basis on the bath. Without loss of generality, we can take the initial state $\psi_0$ of the bath to be the first basis vector $|0\rangle$ from the computational basis. The Wigner function corresponding to $\rho_R(t)$ in the computational basis is given by:
\begin{equation}
    W_{\rho_R, \mathcal{B}}(q,p)=\frac{1}{D} \sum_{k,\ell=0}^{D-1} e^{\frac{i2\pi }{D}(\ell-k) \;p}\; \widehat{\delta}_{2q,k+\ell} \; \langle k_0|\rho_R(t)|\ell_0\rangle.
\end{equation}
Now consider another basis: $U\cdot\mathcal{B}=\{ |i\rangle\}_{i=0}^{D-1}$, where $|i\rangle=U|i\rangle_0$ with $U|\psi_0\rangle=|\psi_0\rangle$. This new basis is just a unitary rotation of the computational basis $\mathcal{B}$ which leaves the first basis vector $|\psi_0\rangle$ invariant. So, the unitary $U$ acts on a $D-1$ dimensional subspace. We can write:
\begin{equation}
    |i\rangle = \begin{cases}|\psi_0\rangle\;\;\; \cdots\;\;\; (i=0)\\\sum_{j=1}^{D-1}U_{ij}|j_0\rangle\;\;\; \cdots\;\;\; (i\neq 0),\end{cases} \label{eq:new_basis}
\end{equation}
where $U_{ij} = \langle j_0| U |i_0\rangle$. Equivalently,
\begin{equation}
    |i\rangle = \delta_{i,0}|\psi_0\rangle + \bdelta_{i,0}\,\sum_{j=1}^{D-1}U_{ij}|j_0\rangle, \label{eq:new_basis_2}
\end{equation}
with the definition $\bdelta_{m,n} = (1-\delta_{m,n})$. The reason to introduce this rotated basis will become clear below. In the rotated basis, the Wigner function can be written as,
\begin{equation}
    W_{\rho_R,\;U\cdot\mathcal{B}}(q,p)=\frac{1}{D} \sum_{k,\ell} e^{\frac{i2\pi }{D}(\ell-k) \;p}\; \widehat{\delta}_{2q,k+\ell} \; \langle k_0|U^{\dagger}\rho_R(t)U|\ell_0\rangle.
\end{equation}
Since $U$ acts only on $R$ and the partial trace is performed over subsystem $B$, we can write,
\begin{eqnarray}
    U^{\dagger}\rho_R(t)U&=&\text{Tr}_B(U^{\dagger} e^{-iH_{BR}t} |\chi_0\rangle \langle\chi_0|_B \otimes |\psi_0\rangle\langle\psi_0|_R e^{iH_{BR}t}\; U) 
 \nonumber \\
    &=& \text{Tr}_B(U^{\dagger} e^{-iH_{BR}t} |\chi_0\rangle \langle\chi_0|_B \otimes U|\psi_0\rangle\langle\psi_0|_RU^{\dagger} e^{iH_{BR}t}\; U) \nonumber \\
    &=&\text{Tr}_B(U^{\dagger} e^{-iH_{BR}t} \;U|\chi_0\rangle \langle\chi_0|_B \otimes |\psi_0\rangle\langle\psi_0|_R\;U^{\dagger} e^{iH_{BR}t}\; U) ,\label{inv}
\end{eqnarray}
where in the second line we have used the fact $U|\psi_0\rangle=|\psi_0\rangle$. Consider now the ensemble averaged negativity with respect to the rotated basis $U\cdot \mathcal{B}$:
\begin{equation}
    \overline{\mathcal{N}_{\mathcal{U.B}}(\rho_R(t))}=\int \;dH_{BR} \; \mu(H_{BR}) \; \sum_{q,p}|W_{U\cdot \mathcal{B}}(q,p)|. \label{neg_equi}
\end{equation} 
From equation \eqref{inv}, together with unitary invariance of the measure $\mu$ under $H_{BR} \to (1_B\otimes U) \; H_{BR}\; (1_B\otimes U^{\dagger})$, 
we get,
\begin{equation} \label{eq:neg_inv}
    \overline{\mathcal{N}_{U\cdot \mathcal{B}}(\rho_R(t))}=\overline{\mathcal{N}_{\mathcal{B}}(\rho_R(t))}.
\end{equation}
So, the ensemble averaged negativity is invariant under unitary rotations of the computational basis that leave the initial state $\psi_0$ invariant. We can thus write the negativity in the computational basis in the following way: 
\begin{equation}
    \overline{\mathcal{N}_{\mathcal{B}}(\rho_R(t))}=\frac{1}{\text{Vol}_{U(D-1)}}\int \;dU \int  \;dH_{BR} \; \mu(H_{BR}) \; \sum_{q,p}|W_{U\cdot\mathcal{B}}(q,p)|,
\end{equation}
where we have by hand introduced an integral over $U$ which cancels with the measure factor, given equation \eqref{eq:neg_inv}. We now switch the order of the integration and perform the Haar integration first:
\begin{equation}
    \overline{\mathcal{N}_{\mathcal{B}}(\rho_R(t))}=\int  \;dH_{BR} \; \mu(H_{BR}) \; \mathcal{N}_{\text{Haar}}, \label{negativity}
\end{equation}
where $\mathcal{N}_{\text{Haar}}$ is given by:
\begin{equation}
    \mathcal{N}_{\text{Haar}}=\frac{1}{\text{Vol}_{U(D-1)}}\int \;dU \; \sum_{q,p}|W_{U\cdot \mathcal{B}}(q,p)|. \label{Haar1}
\end{equation}
The quantity $\mathcal{N}_{\text{Haar}}$ can be interpreted as the Wigner negativity averaged over all possible basis transformations leaving the initial state fixed, and therefore it is an inherently information theoretic quantity. Our goal now is to compute $\mathcal{N}_{\text{Haar}}$ using the integral representation method discussed in the previous section.

\subsection{Integral representation method and Haar integration}

As before, we will use the following integral representation to compute the Haar average over the absolute value of the Wigner function:
\begin{equation}
    |W|=\lim_{\epsilon \to 0}\frac{1}{2 \pi i}\int_{-\infty}^{\infty} dz\;\frac{2z}{z^2+\epsilon^2}\; We^{izW}. \label{resol2}
\end{equation}
From equations \eqref{Haar1} and \eqref{resol2}, we get:
\begin{eqnarray}
    \mathcal{N}_{\text{Haar}}&=&\lim_{\epsilon \to 0}\frac{1}{2 \pi i}\int_{-\infty}^{\infty} dz\;\frac{2z}{z^2+\epsilon^2}\;\sum_{q,p}\;\frac{1}{\text{Vol}_{U(D-1)}}\int \;dU \; W_{U\cdot\mathcal{B}}(q,p)\;e^{iz\;W_{U\cdot\mathcal{B}}(q,p)} \nonumber \\
    &=&\sum_{q,p}\;\lim_{\epsilon \to 0}\frac{1}{2 \pi i}\int_{-\infty}^{\infty} dz\;\frac{2z}{z^2+\epsilon^2}\; \mathcal{I}'(q,p;z), \label{eval_I}
\end{eqnarray}
where 
\beq \label{eq:haar_int}
\mathcal{I}'(q,p;z) = \frac{1}{\text{Vol}_{U(D-1)}}\int \;dU \; W_{U\cdot\mathcal{B}}(q,p)\;e^{iz\;W_{U\cdot\mathcal{B}}(q,p)} .
\eeq 
Our goal now is to evaluate $\mathcal{I}'(q,p;z)$ using the standard tools of Haar integration. To this end, we will expand out the exponential in powers of $W_{U\cdot\mathcal{B}}$, perform the average term by term, and re-sum the result in an appropriate limit. 

In order to do this systematically, we will use the diagrammatic notation explained in \cite{Basu:2025mmm}. We first write down the expression for the Wigner function in the basis $U\cdot \mathcal{B}$ (see equation \eqref{eq:new_basis_2}):
\begin{multline}
    W_{U\cdot \mathcal{B}\;}(q,p)=\frac{1}{D} \sum_{i,i'=0}^{D-1} e^{\frac{i2\pi }{D}(i'-i) \;p}\; \widehat{\delta}_{2q,i+i'} \; \langle i'|\rho_R(t)|i\rangle \\
    =\frac{1}{D}\sum_{i,i',j,j'=0}^{D-1}e^{\frac{i2\pi }{D}(i'-i) \;p}\; \widehat{\delta}_{2q,i+i'} \; \big[\delta_{i',0} \delta_{i,0} \; S_0+\delta_{i',0}\bdelta_{i,0}\; U_{ij}S_j\\
    +\bdelta_{i',0}\delta_{i,0}\; U_{i'j'}^{*}S_{j'}^{*}+\bdelta_{i',0}\bdelta_{i,0}\; U^{*}_{i'j'}U_{ij}S_{j'j}  \big], \label{eq:Wig_newbasis}
\end{multline}
where we have defined the following quantities,
\begin{eqnarray}
    S_0 &:=& \langle \psi_0|\rho_R(t)|\psi_0\rangle ,
 \label{eq:S0}\\
    S_j &:=& \langle \psi_0|\rho_R(t)|j_0\rangle,  \label{eq:Sj} \\
    S_{j'j} &:=& \langle j'_0|\rho_R(t)|j_0\rangle.  \label{eq:Sj1j2}
\end{eqnarray}
It is convenient to diagrammatically represent each of the terms in equation \eqref{eq:Wig_newbasis} in order to facilitate the Haar integration over $U$. We will use the following notation:
\begin{eqnarray}
    \begin{tikzpicture}[baseline={([yshift=-.5ex]current bounding box.center)}]
\coordinate (P) at (0,0);

\draw (P) circle (10pt); 

\node[below] at (0,-10pt) {};

\coordinate (P1) at (-0.15,0);
\coordinate (P2) at (0.15,0);

\draw[black] (P1) circle (1.7pt);
\filldraw (P2) circle (1.7pt);


\end{tikzpicture} &=& \frac{1}{D}\sum_{i,i'} e^{\frac{i2\pi }{D}(i'-i) \;p}\; \widehat{\delta}_{2q,i+i'} \; \delta_{i',0} \delta_{i,0} \; S_0 =\frac{1}{D}\delta_{q,0}S_0, \nonumber \\
\begin{tikzpicture}[baseline={([yshift=-.5ex]current bounding box.center)}]
\coordinate (P) at (0,0);

\draw (P) circle (10pt); 

\node[below] at (0,-10pt) {};

\coordinate (P1) at (-0.15,0);
\coordinate (P2) at (0.15,0);

\draw[black] (P1) circle (1.7pt);
\filldraw (P2) circle (1.7pt);

\draw[-, thick] (P1) to (-0.7,0.7);
\draw[dotted, thick] (P1) to (-0.7,-0.7);
\end{tikzpicture} &=& \frac{1}{D} \sum_{i,i',j}e^{\frac{i2\pi }{D}(i'-i) \;p}\; \widehat{\delta}_{2q,i+i'} \;  \delta_{i',0}\bdelta_{i,0}\; U_{ij}S_j,  \nonumber \\
\begin{tikzpicture}[baseline={([yshift=-.5ex]current bounding box.center)}]
\coordinate (P) at (0,0);

\draw (P) circle (10pt); 

\node[below] at (0,-10pt) {};

\coordinate (P1) at (-0.15,0);
\coordinate (P2) at (0.15,0);

\draw[black] (P1) circle (1.7pt);
\filldraw (P2) circle (1.7pt);

\draw[-, thick] (P2) to (0.7,0.7);
\draw[dotted, thick] (P2) to (0.7,-0.7);
\end{tikzpicture} &=& \frac{1}{D} \sum_{i,i',j'}e^{\frac{i2\pi }{D}(i'-i) \;p}\; \widehat{\delta}_{2q,i+i'} \;  \bdelta_{i',0}\delta_{i,0}\; U_{i'j'}^{*}S_{j'}^{*}, \nonumber \\
\begin{tikzpicture}[baseline={([yshift=-.5ex]current bounding box.center)}]
\coordinate (P) at (0,0);

\draw (P) circle (10pt); 

\node[below] at (0,-10pt) {};

\coordinate (P1) at (-0.15,0);
\coordinate (P2) at (0.15,0);

\draw[black] (P1) circle (1.7pt);
\filldraw (P2) circle (1.7pt);

\draw[-, thick] (P1) to (-0.7,0.7);
\draw[dotted, thick] (P1) to (-0.7,-0.7);
\draw[-, thick] (P2) to (0.7,0.7);
\draw[dotted, thick] (P2) to (0.7,-0.7);

\end{tikzpicture} &=&\frac{1}{D}\sum_{i,i',j,j'} e^{\frac{i2\pi }{D}(i'-i) \;p}\; \widehat{\delta}_{2q,i+i'} \;  \bdelta_{i',0}\bdelta_{i,0}\; U^{*}_{i'j'}U_{ij}S_{j'j}. 
\end{eqnarray}
The convention is as follows: each copy of the Wigner function corresponds to a blob. For each blob, the empty dot corresponds to the un-primed index $i$ (corresponding to the ket) and the shaded dot corresponds to the primed index $i'$ (corresponding to the bra). No legs attached refers to the case when that particular index has been set to zero. The matrix elements $U_{ij}$ are denoted by a couple of legs coming out of the (empty/shaded) dot. The solid line corresponds to the $i$ index of the unitary matrix and the dotted line corresponds to the $j$ index. Note that the dotted indices are appropriately contracted with either $S_j$, $S_{j'}^*$ or $S_{j'j}$. With this notation, equation \eqref{eq:Wig_newbasis} can be written as,
\begin{equation}
    W_{U\cdot \mathcal{B}\;}(q,p)= 
     \begin{tikzpicture}[baseline={([yshift=-.5ex]current bounding box.center)}]
\coordinate (P) at (0,0);

\draw (P) circle (10pt); 

\node[below] at (0,-10pt) {};

\coordinate (P1) at (-0.15,0);
\coordinate (P2) at (0.15,0);

\draw[black] (P1) circle (1.7pt);
\filldraw (P2) circle (1.7pt);


\end{tikzpicture}\;  +
\begin{tikzpicture}[baseline={([yshift=-.5ex]current bounding box.center)}]
\coordinate (P) at (0,0);

\draw (P) circle (10pt); 

\node[below] at (0,-10pt) {};

\coordinate (P1) at (-0.15,0);
\coordinate (P2) at (0.15,0);

\draw[black] (P1) circle (1.7pt);
\filldraw (P2) circle (1.7pt);

\draw[-, thick] (P1) to (-0.7,0.7);
\draw[dotted, thick] (P1) to (-0.7,-0.7);
\end{tikzpicture} \;+
 \begin{tikzpicture}[baseline={([yshift=-.5ex]current bounding box.center)}]
\coordinate (P) at (0,0);

\draw (P) circle (10pt); 

\node[below] at (0,-10pt) {};

\coordinate (P1) at (-0.15,0);
\coordinate (P2) at (0.15,0);

\draw[black] (P1) circle (1.7pt);
\filldraw (P2) circle (1.7pt);

\draw[-, thick] (P2) to (0.7,0.7);
\draw[dotted, thick] (P2) to (0.7,-0.7);
\end{tikzpicture} \; +   
\begin{tikzpicture}[baseline={([yshift=-.5ex]current bounding box.center)}]
\coordinate (P) at (0,0);

\draw (P) circle (10pt); 

\node[below] at (0,-10pt) {};

\coordinate (P1) at (-0.15,0);
\coordinate (P2) at (0.15,0);

\draw[black] (P1) circle (1.7pt);
\filldraw (P2) circle (1.7pt);

\draw[-, thick] (P1) to (-0.7,0.7);
\draw[dotted, thick] (P1) to (-0.7,-0.7);
\draw[-, thick] (P2) to (0.7,0.7);
\draw[dotted, thick] (P2) to (0.7,-0.7);

\end{tikzpicture}.\label{eq:Wigner_diag}
\end{equation} 

The utility of the diagrammatic notation is that it allows us to visualize Haar integration over $W^n$ in terms of contractions of the various legs in these diagrams, As a quick warm-up, let us first compute the average over the Wigner function $W_{U\cdot \mathcal{B}}$. Using the formula:
\beq 
\frac{1}{\text{Vol}_{U(D-1)}}\int dU\; U_{ij}U_{i'j'}^{*}=\frac{1}{D-1}\delta_{ii'}\delta_{jj'},
\eeq 
this can be represented as:
\begin{eqnarray}\label{eq:av_wig}
    \frac{1}{\text{Vol}_{U(D-1)}}\int dU\,W_{U\cdot \mathcal{B}} &=&
\begin{tikzpicture}[baseline={([yshift=-.5ex]current bounding box.center)}]
  \coordinate (P) at (0,0);
  \draw (P) circle (10pt); 
  \node[below] at (0,-10pt) {};

  \coordinate (P1) at (-0.15,0);
  \coordinate (P2) at (0.15,0);
  \draw[black] (P1) circle (1.7pt);
  \filldraw (P2) circle (1.7pt);
\end{tikzpicture}
\; + \;
\begin{tikzpicture}[baseline={([yshift=-.5ex]current bounding box.center)}]
  \coordinate (P) at (0,0);
  \draw (P) circle (10pt);

  \coordinate (P1) at (-0.15,0);
  \coordinate (P2) at (0.15,0);
  \draw[black] (P1) circle (1.7pt);
  \filldraw (P2) circle (1.7pt);

  \draw[black, out=120, in=60, min distance=1.5cm] (P1) to (P2);
  \draw[dotted, out=-120, in=-60, min distance=1.5cm] (P1) to (P2);
\end{tikzpicture} =\frac{1}{D} \delta_{q,0} S_0+\frac{1}{D(D-1)}\bar{\delta}_{q,0}(1-S_0),
\end{eqnarray}
where in the second equality, we have used $\sum_{j}S_{jj} = (1- S_0)$ . One can easily check that the averaged Wigner function is normalized i.e. $\sum_{q,p} \overline{W_{U.\mathcal{B}}}=1$. Note that the quantity $S_0$ is the square of the \emph{fidelity}\footnote{Recall that the fidelity between two density matrices $\rho$ and $\sigma$ is defined as $F(\rho,\sigma) = \text{Tr}\left(\sqrt{\sqrt{\rho}\sigma \sqrt{\rho}}\right)$. When one of the states is pure, this formula reduces to $F(\psi_0, \rho) = \sqrt{\langle \psi_0|\rho|\psi_0\rangle}$.} between $\rho_R(t)$ and the initial state $\psi_0$:
\beq 
S_0(t) = \Big(F(\psi_0, \rho_R(t))\Big)^2.
\eeq 
The fidelity is a measure of the distance between between the two states, and in particular satisfies $0 \leq  F(\rho,\sigma) \leq 1$. Thus, $0\leq S_0\leq 1$. From equation \eqref{eq:av_wig}, we see that the Haar-averaged Wigner function, and consequently also the ensemble averaged Wigner function $\overline{W}_{\mathcal{B}}$ is non-negative everywhere. However, this does not mean that the Haar/ensemble averaged Wigner negativity vanishes, as we will now see.

\subsection{Diagrammatics and resummation}

In order to compute the negativity, we need to calculate the Haar integral $\mathcal{I}'(q,p;z)$ in equation \eqref{eq:haar_int}. First note that,
\begin{eqnarray}
    \mathcal{I}'(q,p;z) =\frac{1}{\text{Vol}_{U(D-1)}}\int dU\; W_{U.\mathcal{B}}\;e^{izW_{U.\mathcal{B}}}=-i\frac{\partial}{\partial z} \mathcal{I}(q,p;z),
\end{eqnarray}
where 
\beq 
\mathcal{I}(q,p;z) = \frac{1}{\text{Vol}_{U(D-1)}}\int dU \;e^{izW_{U.\mathcal{B}}}.
\eeq 
So, it will be sufficient for us to compute $\mathcal{I}(q,p;z)$ as a function of $z$. Using the diagrammatic expression for $W_{U.\mathcal{B}}$ in equation \eqref{eq:Wigner_diag}, we can write,
\begin{eqnarray}
   \mathcal{I}(q,p;z)&=&\frac{1}{\text{Vol}_{U(D-1)}}\int dU \exp \left\{iz 
    \bigg[ \; 
     \begin{tikzpicture}[baseline={([yshift=-.5ex]current bounding box.center)}]
\coordinate (P) at (0,0);

\draw (P) circle (10pt); 

\node[below] at (0,-10pt) {};

\coordinate (P1) at (-0.15,0);
\coordinate (P2) at (0.15,0);

\draw[black] (P1) circle (1.7pt);
\filldraw (P2) circle (1.7pt);


\end{tikzpicture}\;  +
\begin{tikzpicture}[baseline={([yshift=-.5ex]current bounding box.center)}]
\coordinate (P) at (0,0);

\draw (P) circle (10pt); 

\node[below] at (0,-10pt) {};

\coordinate (P1) at (-0.15,0);
\coordinate (P2) at (0.15,0);

\draw[black] (P1) circle (1.7pt);
\filldraw (P2) circle (1.7pt);

\draw[-, thick] (P1) to (-0.7,0.7);
\draw[dotted, thick] (P1) to (-0.7,-0.7);
\end{tikzpicture} \;+
 \begin{tikzpicture}[baseline={([yshift=-.5ex]current bounding box.center)}]
\coordinate (P) at (0,0);

\draw (P) circle (10pt); 

\node[below] at (0,-10pt) {};

\coordinate (P1) at (-0.15,0);
\coordinate (P2) at (0.15,0);

\draw[black] (P1) circle (1.7pt);
\filldraw (P2) circle (1.7pt);

\draw[-, thick] (P2) to (0.7,0.7);
\draw[dotted, thick] (P2) to (0.7,-0.7);
\end{tikzpicture} \; +   
\begin{tikzpicture}[baseline={([yshift=-.5ex]current bounding box.center)}]
\coordinate (P) at (0,0);

\draw (P) circle (10pt); 

\node[below] at (0,-10pt) {};

\coordinate (P1) at (-0.15,0);
\coordinate (P2) at (0.15,0);

\draw[black] (P1) circle (1.7pt);
\filldraw (P2) circle (1.7pt);

\draw[-, thick] (P1) to (-0.7,0.7);
\draw[dotted, thick] (P1) to (-0.7,-0.7);
\draw[-, thick] (P2) to (0.7,0.7);
\draw[dotted, thick] (P2) to (0.7,-0.7);

\end{tikzpicture} \; \bigg]\right\}  \\
&=& \exp \bigg[ iz \;
\begin{tikzpicture}[baseline={([yshift=-.5ex]current bounding box.center)}]
\coordinate (P) at (0,0);

\draw (P) circle (10pt); 

\node[below] at (0,-10pt) {};

\coordinate (P1) at (-0.15,0);
\coordinate (P2) at (0.15,0);

\draw[black] (P1) circle (1.7pt);
\filldraw (P2) circle (1.7pt);


\end{tikzpicture}\; \bigg] \;
\underbrace{\frac{1}{\text{Vol}_{U(D-1)}}\int dU  \exp \left\{iz 
    \bigg[ \; 
\begin{tikzpicture}[baseline={([yshift=-.5ex]current bounding box.center)}]
\coordinate (P) at (0,0);

\draw (P) circle (10pt); 

\node[below] at (0,-10pt) {};

\coordinate (P1) at (-0.15,0);
\coordinate (P2) at (0.15,0);

\draw[black] (P1) circle (1.7pt);
\filldraw (P2) circle (1.7pt);

\draw[-, thick] (P1) to (-0.7,0.7);
\draw[dotted, thick] (P1) to (-0.7,-0.7);
\end{tikzpicture} \;+
 \begin{tikzpicture}[baseline={([yshift=-.5ex]current bounding box.center)}]
\coordinate (P) at (0,0);

\draw (P) circle (10pt); 

\node[below] at (0,-10pt) {};

\coordinate (P1) at (-0.15,0);
\coordinate (P2) at (0.15,0);

\draw[black] (P1) circle (1.7pt);
\filldraw (P2) circle (1.7pt);

\draw[-, thick] (P2) to (0.7,0.7);
\draw[dotted, thick] (P2) to (0.7,-0.7);
\end{tikzpicture} \; +   
\begin{tikzpicture}[baseline={([yshift=-.5ex]current bounding box.center)}]
\coordinate (P) at (0,0);

\draw (P) circle (10pt); 

\node[below] at (0,-10pt) {};

\coordinate (P1) at (-0.15,0);
\coordinate (P2) at (0.15,0);

\draw[black] (P1) circle (1.7pt);
\filldraw (P2) circle (1.7pt);

\draw[-, thick] (P1) to (-0.7,0.7);
\draw[dotted, thick] (P1) to (-0.7,-0.7);
\draw[-, thick] (P2) to (0.7,0.7);
\draw[dotted, thick] (P2) to (0.7,-0.7);

\end{tikzpicture} \; \bigg]\right\}}_{\mathcal{J}}\nonumber. \label{eq:resum'}
\end{eqnarray}
The first blob with no legs comes out of the integral since this piece is independent of $U$. To evaluate the rest of the integral $\mathcal{J}$, we can expand out the part of the exponential containing blobs with one pair of legs attached. Thus we get,
\begin{eqnarray}
   \mathcal{J}&=&\frac{1}{\text{Vol}_{U(D-1)}} \int dU \exp \bigg[ iz  
   \begin{tikzpicture}[baseline={([yshift=-.5ex]current bounding box.center)}]
\coordinate (P) at (0,0);

\draw (P) circle (10pt); 

\node[below] at (0,-10pt) {};

\coordinate (P1) at (-0.15,0);
\coordinate (P2) at (0.15,0);

\draw[black] (P1) circle (1.7pt);
\filldraw (P2) circle (1.7pt);

\draw[-, thick] (P1) to (-0.7,0.7);
\draw[dotted, thick] (P1) to (-0.7,-0.7);
\draw[-, thick] (P2) to (0.7,0.7);
\draw[dotted, thick] (P2) to (0.7,-0.7);

\end{tikzpicture}
   \bigg]\times \; \sum_{n=0}^{\infty} \frac{(iz)^n}{n!} \;
   \bigg[
   \begin{tikzpicture}[baseline={([yshift=-.5ex]current bounding box.center)}]
\coordinate (P) at (0,0);

\draw (P) circle (10pt); 

\node[below] at (0,-10pt) {};

\coordinate (P1) at (-0.15,0);
\coordinate (P2) at (0.15,0);

\draw[black] (P1) circle (1.7pt);
\filldraw (P2) circle (1.7pt);

\draw[-, thick] (P1) to (-0.7,0.7);
\draw[dotted, thick] (P1) to (-0.7,-0.7);
\end{tikzpicture} \;+
   \begin{tikzpicture}[baseline={([yshift=-.5ex]current bounding box.center)}]
\coordinate (P) at (0,0);

\draw (P) circle (10pt); 

\node[below] at (0,-10pt) {};

\coordinate (P1) at (-0.15,0);
\coordinate (P2) at (0.15,0);

\draw[black] (P1) circle (1.7pt);
\filldraw (P2) circle (1.7pt);

\draw[-, thick] (P2) to (0.7,0.7);
\draw[dotted, thick] (P2) to (0.7,-0.7);
\end{tikzpicture} 
   \bigg]^n \nonumber \\
&=&  \sum_{n=0}^{\infty} \frac{(iz)^n}{n!} \;
   \frac{1}{\text{Vol}_{U(D-1)}} \int dU \underbrace{\exp \bigg[ iz 
   \begin{tikzpicture}[baseline={([yshift=-.5ex]current bounding box.center)}]
\coordinate (P) at (0,0);

\draw (P) circle (10pt); 

\node[below] at (0,-10pt) {};

\coordinate (P1) at (-0.15,0);
\coordinate (P2) at (0.15,0);

\draw[black] (P1) circle (1.7pt);
\filldraw (P2) circle (1.7pt);

\draw[-, thick] (P1) to (-0.7,0.7);
\draw[dotted, thick] (P1) to (-0.7,-0.7);
\draw[-, thick] (P2) to (0.7,0.7);
\draw[dotted, thick] (P2) to (0.7,-0.7);

\end{tikzpicture}
    \bigg]}_{A} \; 
   \underbrace{\bigg[
   \begin{tikzpicture}[baseline={([yshift=-.5ex]current bounding box.center)}]
\coordinate (P) at (0,0);

\draw (P) circle (10pt); 

\node[below] at (0,-10pt) {};

\coordinate (P1) at (-0.15,0);
\coordinate (P2) at (0.15,0);

\draw[black] (P1) circle (1.7pt);
\filldraw (P2) circle (1.7pt);

\draw[-, thick] (P1) to (-0.7,0.7);
\draw[dotted, thick] (P1) to (-0.7,-0.7);
\end{tikzpicture} \;+
   \begin{tikzpicture}[baseline={([yshift=-.5ex]current bounding box.center)}]
\coordinate (P) at (0,0);

\draw (P) circle (10pt); 

\node[below] at (0,-10pt) {};

\coordinate (P1) at (-0.15,0);
\coordinate (P2) at (0.15,0);

\draw[black] (P1) circle (1.7pt);
\filldraw (P2) circle (1.7pt);

\draw[-, thick] (P2) to (0.7,0.7);
\draw[dotted, thick] (P2) to (0.7,-0.7);
\end{tikzpicture} 
   \bigg]^n }_{B} . \label{eq:HaarInt1'}
\end{eqnarray}
{Note that we have not yet performed the Haar integration. We can now expand the
exponential in $A$, after which the Haar integral is evaluated by combining the
blobs in $A$ and $B$ in all possible ways and contracting the corresponding
open legs. To keep the presentation concise, we quote only the final result of the Haar integration here. The details of the computation are deferred to
Appendix \ref{sec:Iqpz}. As explained in the Appendix, the final result can be expressed in terms of
fully connected $n$-blob contractions. There are two classes of such contractions. The first class consists of irreducible $n$-body contractions constructed from
$n$ blobs, where blobs of type-$B$ are always present, while blobs of type-$A$
may or may not be included. The open legs are  then contracted in all possible ways
such that the resulting diagram is fully connected. We denote the set of all
such contractions by $\mathcal{D}_{n}'$. The second class consists of irreducible $n$-body contractions formed entirely
from type-$A$ blobs. Again, the open legs of the $A$ blobs are contracted among
themselves to produce a fully connected diagram. We denote this set by $\mathcal{D}_{n,0}$. We show a few examples below.}

\begin{eqnarray}
        \mathcal{D}_{2}'&=& 
        \begin{tikzpicture}[baseline={([yshift=-.5ex]current bounding box.center)}]
\coordinate (P) at (0,0);
\coordinate (Q) at (40pt,0);

\draw (P) circle (10pt);
\draw (Q) circle (10pt); 

\coordinate (P1) at (-4pt,0);
\coordinate (P2) at (4pt,0);
\coordinate (Q1) at (36pt,0);
\coordinate (Q2) at (44pt,0);

\draw[black] (P1) circle (1.7pt);
\filldraw (P2) circle (1.7pt);
\draw[black] (Q1) circle (1.7pt);
\filldraw (Q2) circle (1.7pt);

\draw[black,out = 60,in = 120, looseness = 1.5] (P2) to (Q1);
\draw[dotted,out = -60,in = -120, looseness = 1.5] (P2) to (Q1);
\end{tikzpicture} \\
\mathcal{D}_{3}' &=&
\begin{tikzpicture}[baseline={([yshift=-.5ex]current bounding box.center)}]
\coordinate (P) at (0,0);
\coordinate (Q) at (30pt,0);
\coordinate (R) at (60pt,0);

\draw (P) circle (10pt);
\draw (Q) circle (10pt); 
\draw (R) circle (10pt);

\coordinate (P1) at (-4pt,0);
\coordinate (P2) at (4pt,0);
\coordinate (Q1) at (26pt,0);
\coordinate (Q2) at (34pt,0);
\coordinate (R1) at (56pt,0);
\coordinate (R2) at (64pt,0);

\draw[black] (P1) circle (1.7pt);
\filldraw (P2) circle (1.7pt);
\draw[black] (Q1) circle (1.7pt);
\filldraw (Q2) circle (1.7pt);
\draw[black] (R1) circle (1.7pt);
\filldraw (R2) circle (1.7pt);

\draw[black,out = 60,in = 120, looseness = 2.0] (P2) to (R1);
\draw[black,out = 120,in = 60, min distance = 1.2cm] (Q1) to (Q2);

\draw[dotted,out = -120,in = -60, min distance = 1.0cm] (P2) to (Q1);
\draw[dotted,out = -120,in = -60, min distance = 1.0cm] (Q2) to (R1); 
\end{tikzpicture}  \label{eq:D3prime'}
\end{eqnarray}
\begin{eqnarray}
    \mathcal{D}_{1,0}&= &
    \begin{tikzpicture}[baseline={([yshift=-.5ex]current bounding box.center)}]
  \coordinate (P) at (0,0);
  \draw (P) circle (10pt);

  \coordinate (P1) at (-0.15,0);
  \coordinate (P2) at (0.15,0);
  \draw[black] (P1) circle (1.7pt);
  \filldraw (P2) circle (1.7pt);

  \draw[black, out=120, in=60, min distance=1.5cm] (P1) to (P2);
  \draw[dotted, out=-120, in=-60, min distance=1.5cm] (P1) to (P2);
\end{tikzpicture}  \\
\mathcal{D}_{2,0}&=&
\begin{tikzpicture}[baseline={([yshift=-.5ex]current bounding box.center)}]
\coordinate (P) at (0,0);
\coordinate (Q) at (40pt,0);

\draw (P) circle (10pt);
\draw (Q) circle (10pt);

\coordinate (P1) at (-4pt,0);
\coordinate (P2) at (4pt,0);
\coordinate (Q1) at (36pt,0);
\coordinate (Q2) at (44pt,0);

\draw[black] (P1) circle (1.7pt);
\filldraw (P2) circle (1.7pt);
\draw[black] (Q1) circle (1.7pt);
\filldraw (Q2) circle (1.7pt);

\draw[black,out = 60,in = 120, looseness = 1.5] (P1) to (Q2);
\draw[black,out = 60,in = 120, looseness =1.5] (P2) to (Q1);

\draw[dotted,out = -60,in = -120, looseness = 1.5] (P1) to (Q2);
\draw[dotted,out = -60,in = -120, looseness =1.5] (P2) to (Q1); 
\end{tikzpicture} \; +
\begin{tikzpicture}[baseline={([yshift=-.5ex]current bounding box.center)}]
\coordinate (P) at (0,0);
\coordinate (Q) at (40pt,0);

\draw (P) circle (10pt);
\draw (Q) circle (10pt);

\coordinate (P1) at (-4pt,0);
\coordinate (P2) at (4pt,0);
\coordinate (Q1) at (36pt,0);
\coordinate (Q2) at (44pt,0);

\draw[black] (P1) circle (1.7pt);
\filldraw (P2) circle (1.7pt);
\draw[black] (Q1) circle (1.7pt);
\filldraw (Q2) circle (1.7pt);

\draw[black,out = 60,in = 120, looseness = 1.5] (P1) to (Q2);
\draw[black,out = 60,in = 120, looseness =1.5] (P2) to (Q1);

 \draw[dotted, out=-120, in=-60, min distance=1.5cm] (P1) to (P2);
  \draw[dotted, out=-120, in=-60, min distance=1.5cm] (Q1) to (Q2);
\end{tikzpicture} \;+
\begin{tikzpicture}[baseline={([yshift=-.5ex]current bounding box.center)}]
\coordinate (P) at (0,0);
\coordinate (Q) at (40pt,0);

\draw (P) circle (10pt);
\draw (Q) circle (10pt);

\coordinate (P1) at (-4pt,0);
\coordinate (P2) at (4pt,0);
\coordinate (Q1) at (36pt,0);
\coordinate (Q2) at (44pt,0);

\draw[black] (P1) circle (1.7pt);
\filldraw (P2) circle (1.7pt);
\draw[black] (Q1) circle (1.7pt);
\filldraw (Q2) circle (1.7pt);

\draw[dotted,out = -60,in = -120, looseness = 1.5] (P1) to (Q2);
\draw[dotted,out = -60,in = -120, looseness =1.5] (P2) to (Q1);

 \draw[black, out=120, in=60, min distance=1.5cm] (P1) to (P2);
  \draw[black, out=120, in=60, min distance=1.5cm] (Q1) to (Q2);
\end{tikzpicture}
\end{eqnarray}

\noindent We can now write down the final expression for $\mathcal{I}(q,p;z)$:
\begin{equation} 
   \mathcal{I}(q,p;z) =  \exp \bigg( \sum_{n=1}^{\infty} \;(iz)^n \; \mathcal{D}_{n}\bigg), \label{eq:Haarfinal'}
\end{equation}
where 
\begin{eqnarray}
    \mathcal{D}_{1}&=& \mathcal{D}_{1,0}\; + 
    \begin{tikzpicture}[baseline={([yshift=-.5ex]current bounding box.center)}]
\coordinate (P) at (0,0);

\draw (P) circle (10pt); 

\node[below] at (0,-10pt) {};

\coordinate (P1) at (-0.15,0);
\coordinate (P2) at (0.15,0);

\draw[black] (P1) circle (1.7pt);
\filldraw (P2) circle (1.7pt);


\end{tikzpicture}\; \\
\mathcal{D}_{n}&=&\frac{\mathcal{D}_{n,0}}{n!}+\mathcal{D}'_{n} \;\;\;\; n=2,3,\cdots. \label{eq:HigherDia}
\end{eqnarray}

\begin{figure}[t]
    \centering
    \begin{tabular}{ccc}
    \includegraphics[height=3.7cm]{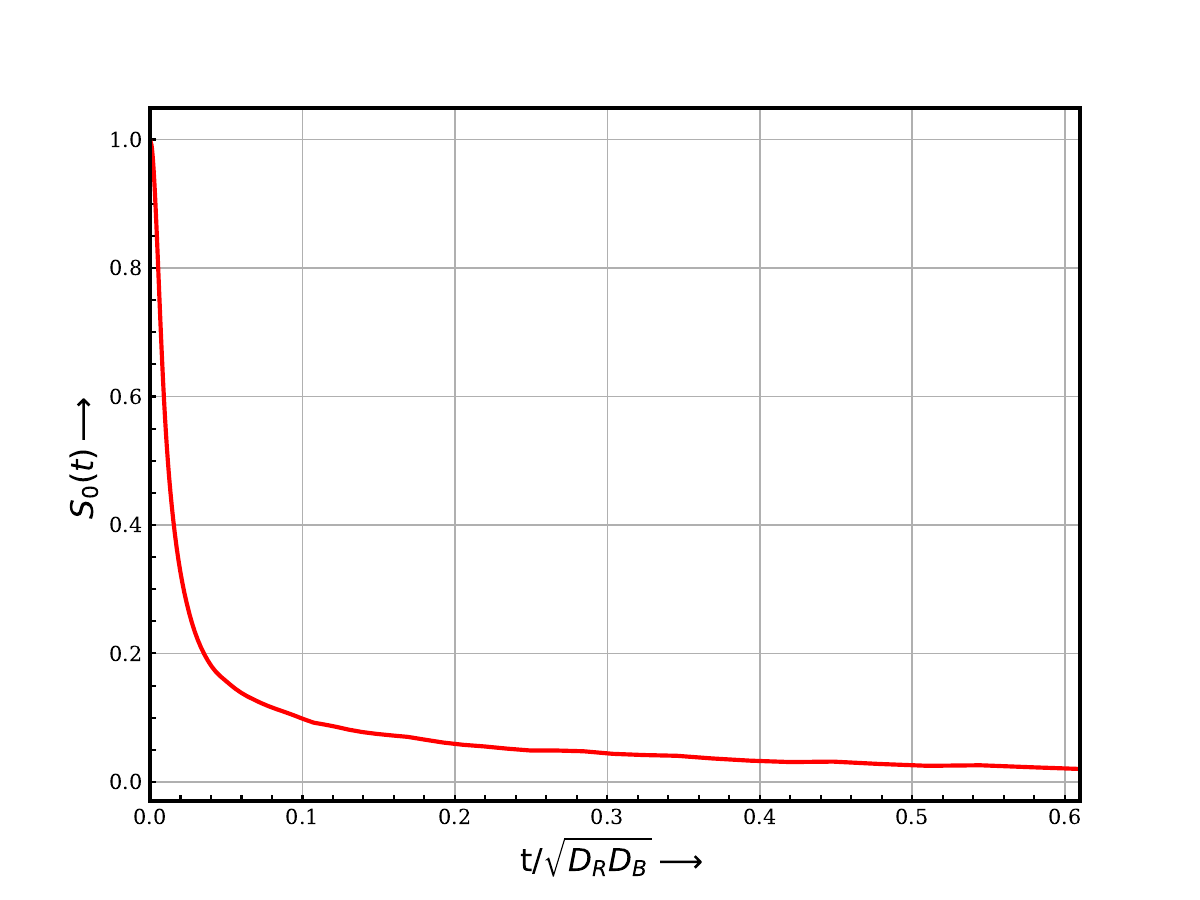} & \includegraphics[height=3.7cm]{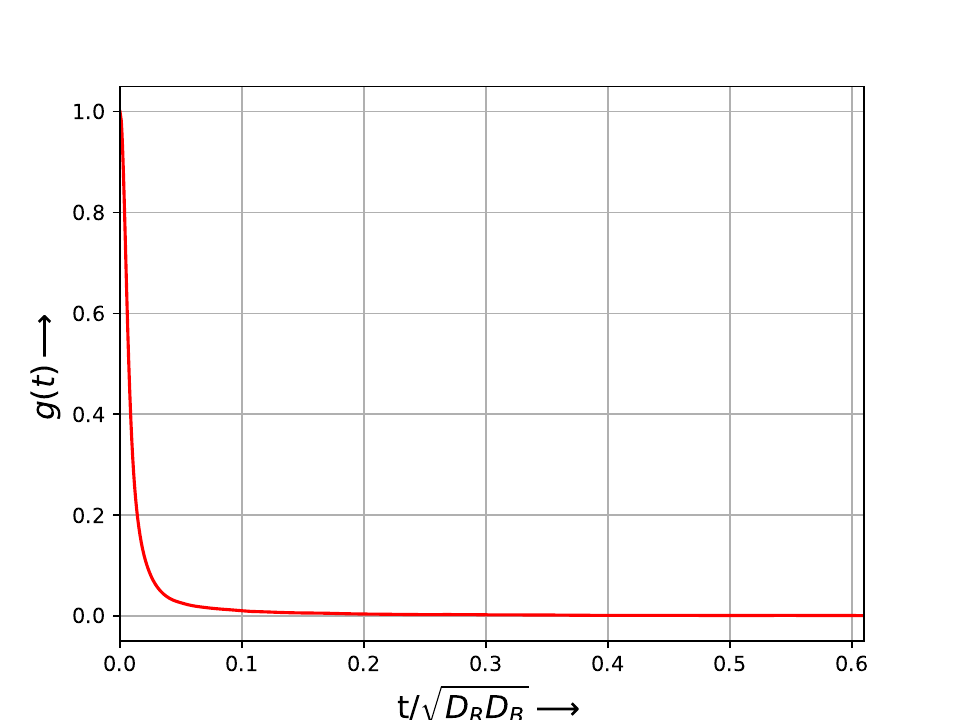}&
    \includegraphics[height=3.7cm]{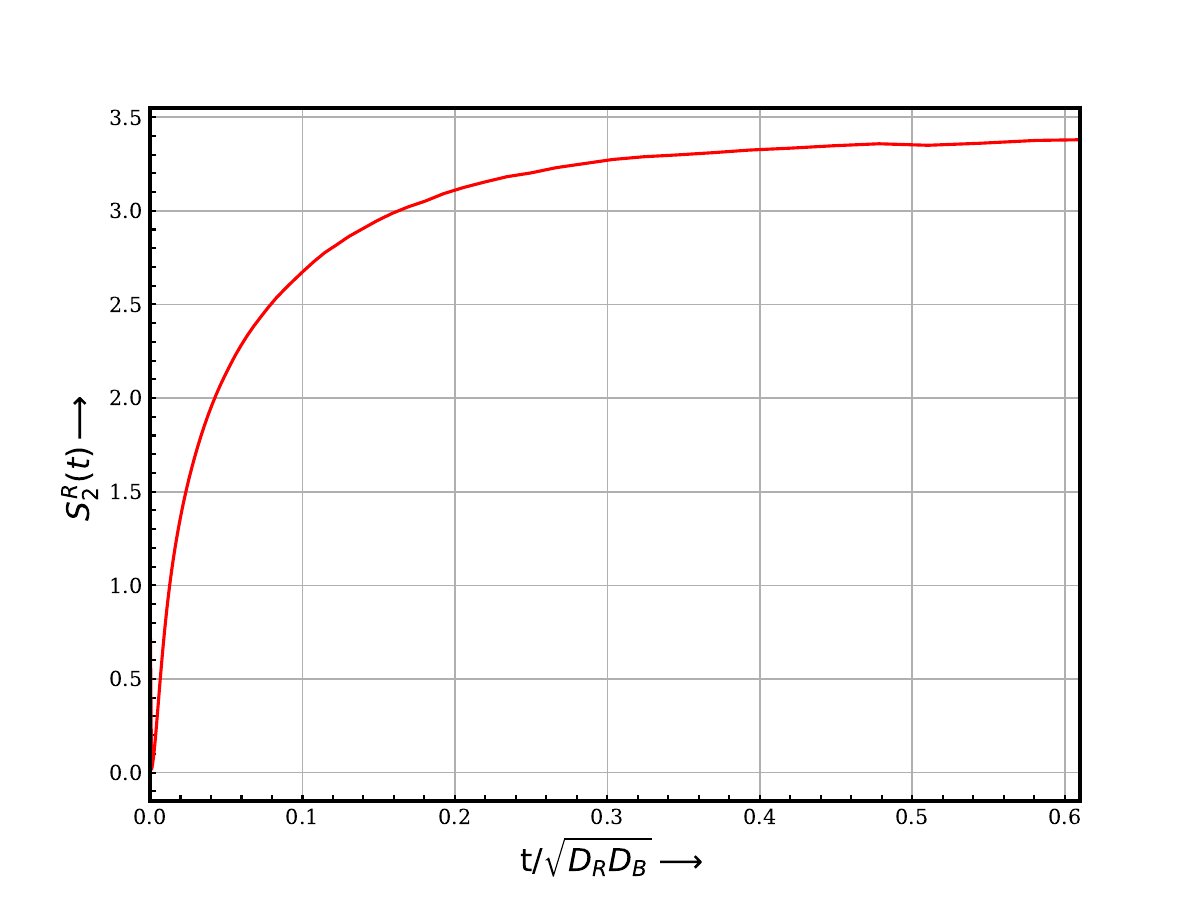}
    \end{tabular}
    \caption{The invariants $S_0$, $g$ and $S_2^R$ as functions of time for the situation where the operators $O_B$ and $O_R$ in the interaction term \eqref{eq:intterms} are taken to be random matrices from the GUE.}
    \label{fig:canno2}
\end{figure}

From equations \eqref{eval_I} and \eqref{eq:Haarfinal}, the Haar averaged Wigner negativity is given by:
\begin{equation}
    \mathcal{N}_{\text{Haar}}=\lim_{\epsilon \to 0} \sum_{q,p}\frac{1}{2\pi i }\int_{-\infty}^{\infty} dz \; \frac{2z}{z^2+\epsilon^2} \bigg( -i\frac{\partial}{\partial z} \bigg)\;  e^ {\bigg( \sum_{n=1}^{\infty} \;(iz)^n \; \mathcal{D}_{n}\bigg)}. \label{eq:wignerabs0}
\end{equation}
By redefining $s = \frac{2\mathcal{D}_2}{\mathcal{D}_1} z$, we can write this integral as
\begin{eqnarray}
    \mathcal{N}_{\text{Haar}} &=& \lim_{\epsilon \to 0} \sum_{q,p}\frac{2\mathcal{D}_2}{2\pi i\mathcal{D}_1 }\int_{-\infty}^{\infty} ds \; \frac{2s}{s^2+(\frac{2\mathcal{D}_2\epsilon}{\mathcal{D}_1})^2} \bigg( -i\frac{\partial}{\partial s} \bigg)\;  e^ { \sum_{n=1}^{\infty} \;(is)^n \;\left(\frac{\mathcal{D}_1}{2\mathcal{D}_2}\right)^n \mathcal{D}_{n}} \nonumber\\
    &=& \lim_{\epsilon \to 0} \sum_{q,p}\frac{2\mathcal{D}_2}{2\pi i\mathcal{D}_1 }\int_{-\infty}^{\infty} ds \; \frac{2s}{s^2+(\frac{2\mathcal{D}_2\epsilon}{\mathcal{D}_1})^2} \bigg( -i\frac{\partial}{\partial s} \bigg)\;  e^ { \frac{\mathcal{D}_1^2}{4\mathcal{D}_2}\left(2is - s^2 - \frac{i\mathcal{D}_1\mathcal{D}_3}{2\mathcal{D}_2^2} s^3+\cdots\right)}.\label{eq:wignerabs1}
\end{eqnarray}
In order to analyze this expression further, we now need to evaluate the diagrams $\mathcal{D}_n$ explicitly. Note that all the $\mathcal{D}_n$ can be expressed in terms of various information theoretic invariants built out of the reduced density matrix $\rho_R(t)$, its higher powers and $|\psi_0\rangle\langle \psi_0|$. Some of the first few invariants which will be relevant for us are (see figure \ref{fig:canno2}):
\beq 
S_0(t) = \langle \psi_0 |\rho_R(t)|\psi_0\rangle,
\eeq 
\beq 
g(t) = \langle \psi_0 | \rho_R^2(t) | \psi_0\rangle,
\eeq 
\beq 
S_2^R(t) = - \log \text{Tr}\,\rho^2_R(t),
\eeq 
where note that $S_2^R$ is the second R\'enyi entropy. 

Since $\rho_R(t)$ evolves in time starting from a pure state, all the amplitudes $\mathcal{D}_n$ are time dependent.  For instance, the first two amplitudes $\mathcal{D}_1$ and $\mathcal{D}_2$ are given by:
\begin{eqnarray}
    \mathcal{D}_1&=& 
    \begin{tikzpicture}[baseline={([yshift=-.5ex]current bounding box.center)}]
  \coordinate (P) at (0,0);
  \draw (P) circle (10pt);

  \coordinate (P1) at (-0.15,0);
  \coordinate (P2) at (0.15,0);
  \draw[black] (P1) circle (1.7pt);
  \filldraw (P2) circle (1.7pt);

  \draw[black, out=120, in=60, min distance=1.5cm] (P1) to (P2);
  \draw[dotted, out=-120, in=-60, min distance=1.5cm] (P1) to (P2);
\end{tikzpicture} + \;\;
\begin{tikzpicture}[baseline={([yshift=-.5ex]current bounding box.center)}]
  \coordinate (P) at (0,0);
  \draw (P) circle (10pt);

  \coordinate (P1) at (-0.15,0);
  \coordinate (P2) at (0.15,0);
  \draw[black] (P1) circle (1.7pt);
  \filldraw (P2) circle (1.7pt);
  \end{tikzpicture} \nonumber \\
  &=& \frac{1}{D^2} \bigg[(1- S_0) \; \bdelta_{q,0} + D \;  S_0 \; \delta_{q,0}\bigg],\label{eq:D1}
\end{eqnarray}
\begin{eqnarray}
    2 \; \mathcal{D}_2&=&
    \begin{tikzpicture}[baseline={([yshift=-.5ex]current bounding box.center)}]
\coordinate (P) at (0,0);
\coordinate (Q) at (40pt,0);

\draw (P) circle (10pt);
\draw (Q) circle (10pt);

\coordinate (P1) at (-4pt,0);
\coordinate (P2) at (4pt,0);
\coordinate (Q1) at (36pt,0);
\coordinate (Q2) at (44pt,0);

\draw[black] (P1) circle (1.7pt);
\filldraw (P2) circle (1.7pt);
\draw[black] (Q1) circle (1.7pt);
\filldraw (Q2) circle (1.7pt);

\draw[black,out = 60,in = 120, looseness = 1.5] (P1) to (Q2);
\draw[black,out = 60,in = 120, looseness =1.5] (P2) to (Q1);

\draw[dotted,out = -60,in = -120, looseness = 1.5] (P1) to (Q2);
\draw[dotted,out = -60,in = -120, looseness =1.5] (P2) to (Q1); 
\end{tikzpicture} \; +
\begin{tikzpicture}[baseline={([yshift=-.5ex]current bounding box.center)}]
\coordinate (P) at (0,0);
\coordinate (Q) at (40pt,0);

\draw (P) circle (10pt);
\draw (Q) circle (10pt);

\coordinate (P1) at (-4pt,0);
\coordinate (P2) at (4pt,0);
\coordinate (Q1) at (36pt,0);
\coordinate (Q2) at (44pt,0);

\draw[black] (P1) circle (1.7pt);
\filldraw (P2) circle (1.7pt);
\draw[black] (Q1) circle (1.7pt);
\filldraw (Q2) circle (1.7pt);

\draw[black,out = 60,in = 120, looseness = 1.5] (P1) to (Q2);
\draw[black,out = 60,in = 120, looseness =1.5] (P2) to (Q1);

 \draw[dotted, out=-120, in=-60, min distance=1.5cm] (P1) to (P2);
  \draw[dotted, out=-120, in=-60, min distance=1.5cm] (Q1) to (Q2);
\end{tikzpicture} \;+
\begin{tikzpicture}[baseline={([yshift=-.5ex]current bounding box.center)}]
\coordinate (P) at (0,0);
\coordinate (Q) at (40pt,0);

\draw (P) circle (10pt);
\draw (Q) circle (10pt);

\coordinate (P1) at (-4pt,0);
\coordinate (P2) at (4pt,0);
\coordinate (Q1) at (36pt,0);
\coordinate (Q2) at (44pt,0);

\draw[black] (P1) circle (1.7pt);
\filldraw (P2) circle (1.7pt);
\draw[black] (Q1) circle (1.7pt);
\filldraw (Q2) circle (1.7pt);

\draw[dotted,out = -60,in = -120, looseness = 1.5] (P1) to (Q2);
\draw[dotted,out = -60,in = -120, looseness =1.5] (P2) to (Q1);

 \draw[black, out=120, in=60, min distance=1.5cm] (P1) to (P2);
  \draw[black, out=120, in=60, min distance=1.5cm] (Q1) to (Q2);
\end{tikzpicture} \;+ 2 \times 
\begin{tikzpicture}[baseline={([yshift=-.5ex]current bounding box.center)}]
\coordinate (P) at (0,0);
\coordinate (Q) at (40pt,0);

\draw (P) circle (10pt);
\draw (Q) circle (10pt); 

\coordinate (P1) at (-4pt,0);
\coordinate (P2) at (4pt,0);
\coordinate (Q1) at (36pt,0);
\coordinate (Q2) at (44pt,0);

\draw[black] (P1) circle (1.7pt);
\filldraw (P2) circle (1.7pt);
\draw[black] (Q1) circle (1.7pt);
\filldraw (Q2) circle (1.7pt);

\draw[black,out = 60,in = 120, looseness = 1.5] (P2) to (Q1);
\draw[dotted,out = -60,in = -120, looseness = 1.5] (P2) to (Q1);
\end{tikzpicture} \nonumber \\
&=& \frac{1}{D^4} \left( \alpha_0\; \delta_{q,0}+\bar{\alpha}_0 \; \bar{\delta}_{q,0}\right), \label{eq:D2}
\end{eqnarray}
where the quantities $\alpha_{0}$ and $\bar{\alpha}_{0}$ are given by:
\begin{eqnarray}
    \alpha_0 &=&(D-1)\left( e^{-S_2^{R}}-2g+S_0^2 \right)- \left( 1-\frac{1}{D} \right)(1-S_0)^2 ,\\
    \bar{\alpha}_{0}&=&\left( D-2-\frac{1}{D} \right) \left( e^{-S_2^{R}}-2g+S_0^2 \right)-\left( 1-\frac{2}{D} \right)(1-S_0)^2  +2 D\; (g-S_0^2). \label{eq:alpha0prime}
\end{eqnarray}
Similarly one can compute the higher point diagrams. In the above formulas, we have only kept the leading contribution in the Weingarten functions (see the discussion around equation \eqref{eq:Wg}). We can further simplify the above expressions at large $D$ as:
\begin{eqnarray}
    \alpha_0 &\simeq & D\left( e^{-S_2^{R}}-2g+S_0^2 \right)-  (1-S_0)^2 \label{eq:alpha0} \\
    \bar{\alpha}_{0}&\simeq & D \left( e^{-S_2^{R}}-S_0^2 \right)-(1-S_0)^2  . \label{eq:alphaneq0}
\end{eqnarray}
At small times, the terms proportional to $D$ dominate, and so we can afford to drop the $O(1)$ terms. At late times, $e^{S_R^2}$ becomes large, and so we can work in the scaling regime where $D \to \infty,\,e^{S_2^R} \to \infty$ with $\frac{D}{e^{S_2^R}}$ fixed. In this regime, the terms we dropped above are again small in $e^{-S_2^R}$.

\begin{figure}[t]
    \centering
    \includegraphics[scale=0.25]{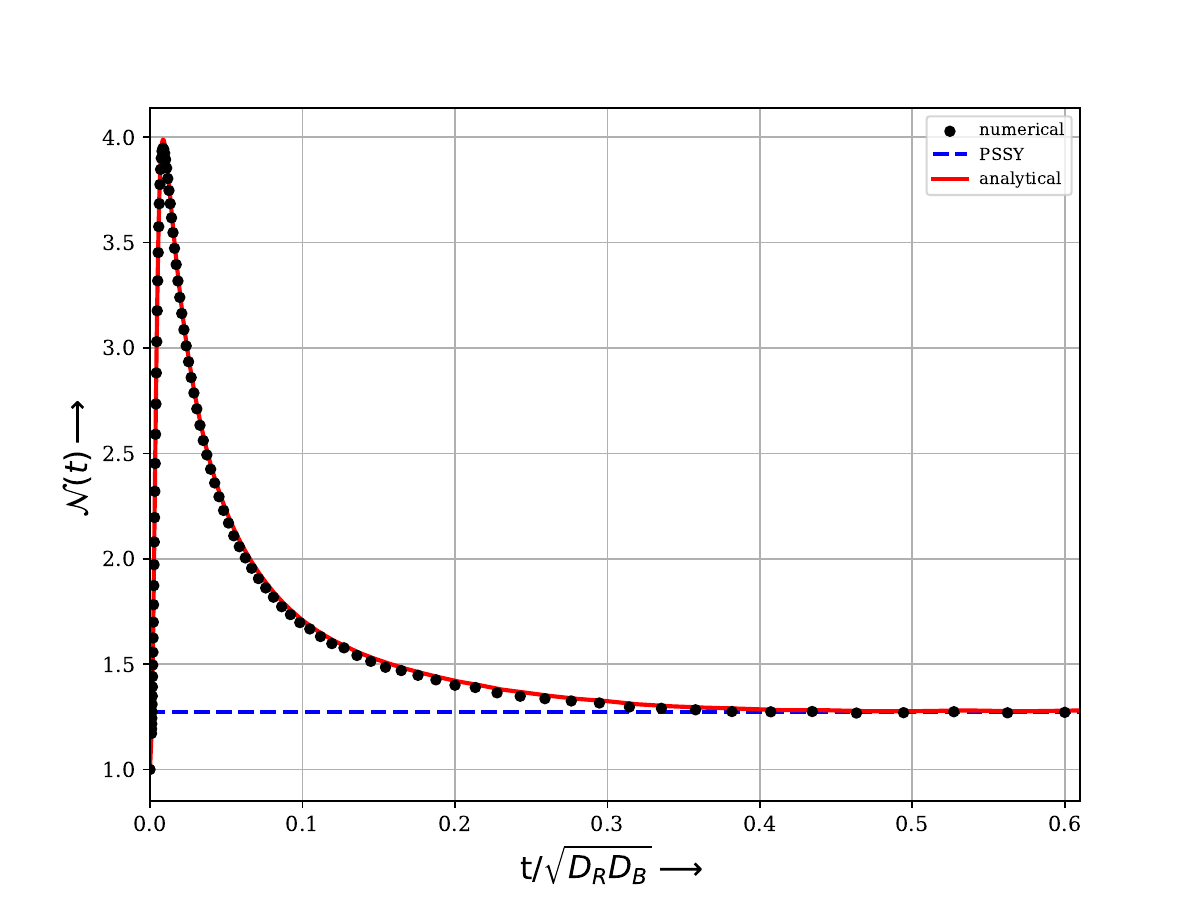}
    \includegraphics[scale=0.25]{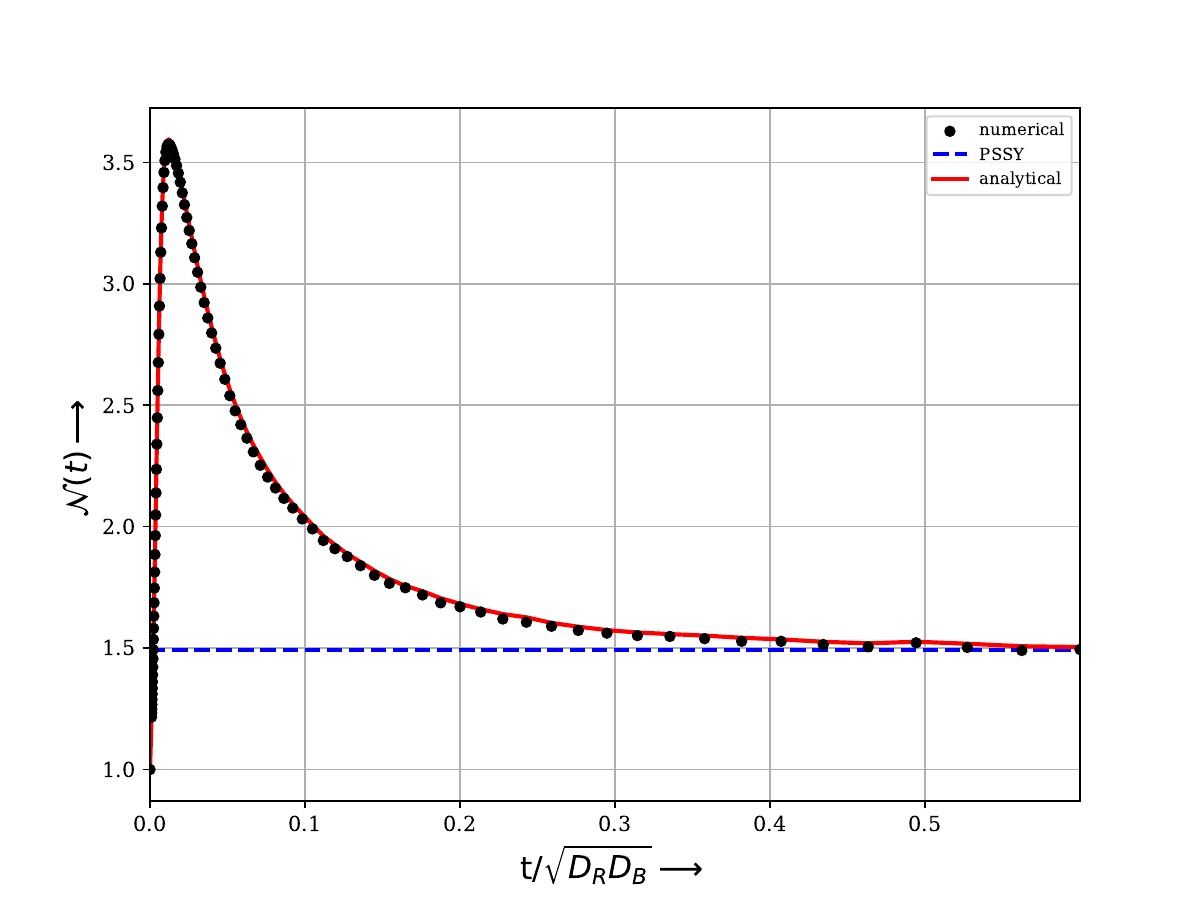}
    \includegraphics[scale=0.25]{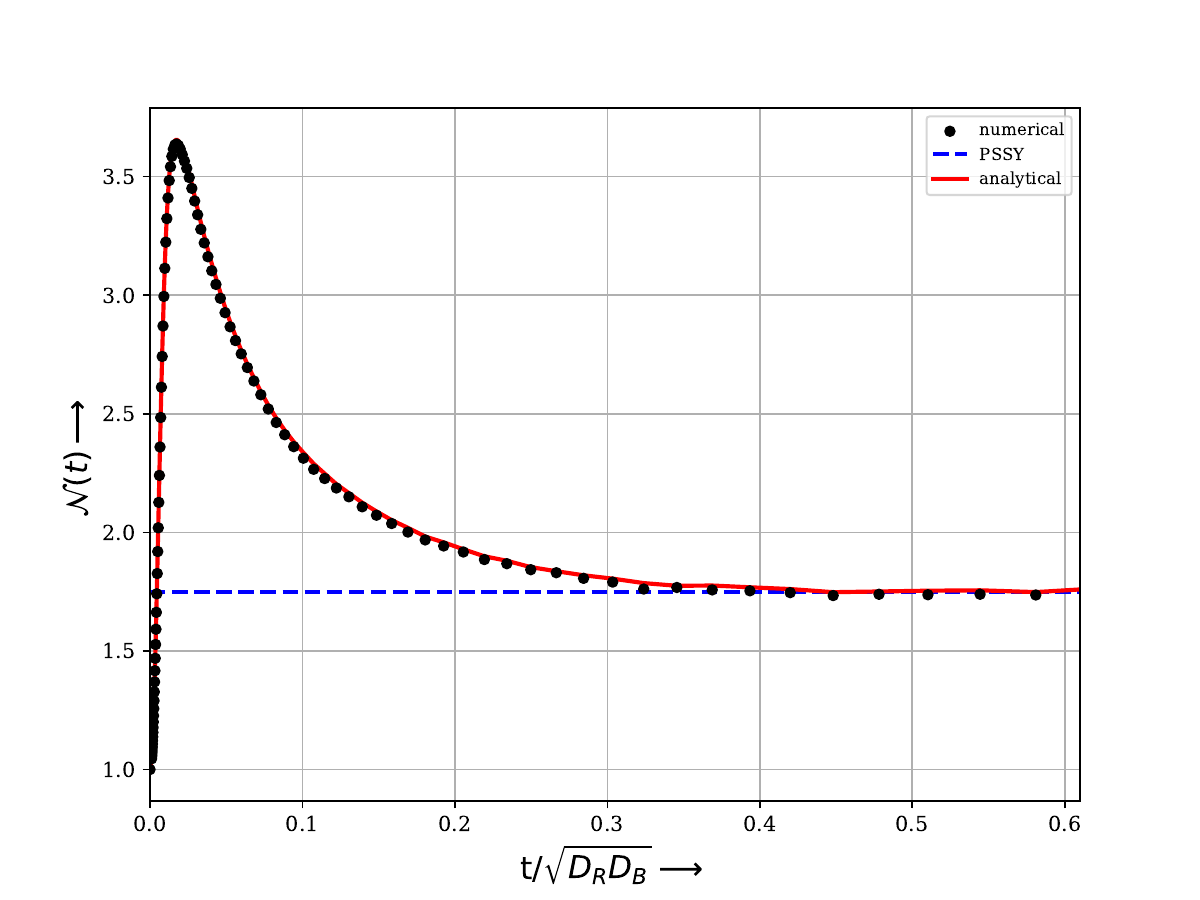}
    
    \caption{Log-log plot of the Wigner negativity for the case where the operators $O_R$ and $O_B$ in equation \eqref{eq:intterms} are chosen from the GUE. The radiation Hilbert space is taken to be $D_R = 101$, while the black hole Hilbert space dimension is taken to be $D_B = 200$ (left), 101 (center) and $50$ (right) respectively. }
    \label{fig:dB=50}
\end{figure}

We will first focus on the case $q\neq 0$, since the negativity coming from $q=0$ can at most contribute some $O(1)$ amount. For $q\neq 0$, the cubic and higher terms in the exponent of equation \eqref{eq:wignerabs1} are suppressed in the large $D$ limit (see Appendix \ref{sec:higherorder} for more details), and so we can resort to the Gaussian approximation for the integral:
\begin{equation} 
\mathcal{N}_{\text{Haar}}\Big|_{q\neq 0} \simeq \lim_{\epsilon \to 0} \sum_{q,p}\frac{2\mathcal{D}_2}{\mathcal{D}_1 }\int_{-\infty}^{\infty} \frac{ds}{2\pi i} \; \frac{2s}{s^2+(\frac{2\mathcal{D}_2\epsilon}{\mathcal{D}_1})^2} \bigg( -i\frac{\partial}{\partial s} \bigg)\;  e^ { \frac{\mathcal{D}_1^2}{4\mathcal{D}_2}\left(2is - s^2\right)}. \label{eq:Gaussian}
\end{equation} 
This is the same integral we encountered in the PSSY model, and gives
\begin{equation}
    \mathcal{N}_{\text{Haar}}\Big|_{q\neq 0} \simeq (1 -S_0) \left(\frac{1}{\sqrt{\pi r_{\bar{0}}}}e^{-r_{\bar{0}}} + \erf(\sqrt{r_{\bar{0}}})\right),
    \end{equation}
where we have defined
\beq 
r_{\bar{0}} = \frac{\mathcal{D}_1^2}{4\mathcal{D}_2} = \frac{(1-S_0)^2}{2\left[D \left( e^{-S_2^{R}}-S_0^2 \right)-(1-S_0)^2  \right]}.
\eeq 
We now consider the $q=0$ case. Here, the Gaussian approximation fails at early times (see Appendix \ref{sec:higherorder}), but becomes valid at later times when the scaling limit $D\to \infty, \,e^{S^2_R}\to \infty, \frac{D}{e^{S^2_R}}$ fixed becomes valid. We will not worry about the failure of the Gaussian approximation at early times too much, as this can at most give an $O(1)$ correction to the negativity. Performing the Gaussian integral as before, we then find
\begin{equation}
     \mathcal{N}_{\text{Haar}}\Big|_{q=0} \simeq S_0 \left(\frac{1}{\sqrt{\pi r_0}}e^{-r_0} + \erf(\sqrt{r_0})\right),
\end{equation}
where $r_0$ is defined as,
\begin{equation}
    r_0=\frac{D^2 S_0^2}{2\left[ D\left( e^{-S_2^R}-2g+S_0^2 \right)-\left(1-S_0\right)^2 \right]}.
\end{equation}
The total negativity is then given by
\begin{equation}
\mathcal{N}_{\text{Haar}}=\mathcal{N}_{\text{Haar}}\Big|_{q=0}+\mathcal{N}_{\text{Haar}}\Big|_{q\neq0}.
\end{equation}
Finally, the ensemble averaged negativity can be obtained from the Haar averaged negativity using equation \eqref{negativity}:
\begin{equation*}
    \overline{\mathcal{N}_{\mathcal{B}}(t)}=\int dH_{BR}\; \mu(H_{BR})\, \mathcal{N}_{\text{Haar}}.
\end{equation*}
In figure \ref{fig:dB=50} we compare our analytical formula for the Wigner negativity with numerical calculations and find a good match. These figures show an interesting feature of negativity growth -- immediately after the black hole is coupled to the radiation bath, the negativity shows a sharp spike and becomes $O(\sqrt{D})$. This instantaneous rise in the negativity happens due to the interactions between the black hole and the radiation bath, and can essentially be thought of as a ``negativity shock''. But soon after, the negativity starts to decay -- this happens because of \emph{decoherence}. When the black hole dimension is much larger than that of the radiation bath, the black hole serves as a good environment and the bath decoheres to a classical density matrix. On the other hand, when the black hole is comparable or much smaller than the radiation bath, then the radiation system does not decohere completely, but only partially. 

It is instructive to study our analytical formula in limiting cases. When the black hole dimension is taken to be small, then the above formula reduces to the one derived in \cite{Basu:2025mmm} for the time evolution of Wigner negativity for pure states. Another interesting case is $t \to \infty$. At very late times,  we expect that $S_0 \sim \tfrac{1}{D}$, $g \sim \tfrac{1}{D^2}$, 
and $e^{-S_2^{R}} \sim e^{-S^R_{2,\text{eq.}}} \sim \tfrac{1}{D}+\tfrac{1}{e^{S_0}}$. In this limit, we get
\begin{equation}
    \mathcal{N}_{\text{Haar}} \simeq  \left(\frac{1}{\sqrt{\pi r}}e^{-r} + \erf(\sqrt{r})\right),\;\; r = \frac{1}{2(D e^{-S^R_{2,\text{eq.}}}-1)}.
\end{equation}
This formula agrees with the answer for the Wigner negativity we obtained in the PSSY model in the microcanonical ensemble, and also with the negativity of Haar random states \cite{White:2020hgn}. Thus, the gravitational result appears as the late time equilibrium value for the negativity in our dynamical model of black hole evaporation. This also explains why two-boundary wormholes were the only relevant wormholes in the gravity calculation -- the Haar average over the Wigner negativity only depends on the second R\'enyi entropy, and the corresponding replica wormholes are two-boundary wormholes.

\section{Stabilizer complexity and the Python's lunch}\label{sec:python}
\begin{figure}[t]
  \centering

  \subfloat[]{\label{fig:QESwithoutPL}%
\begin{minipage}[t]{0.45\textwidth}%
\centering
    \includegraphics[scale=1.58]{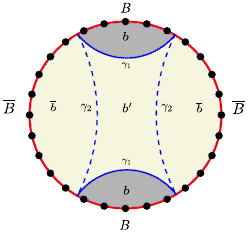}
\end{minipage}}
  \hspace{0.24cm}
  \subfloat[]{\label{fig:QESwithPL}%
\begin{minipage}[t]{0.45\textwidth}%
\centering
    \includegraphics[scale=1.71]{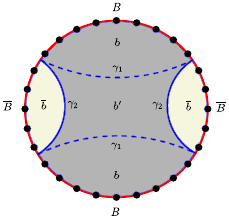}
\end{minipage}}

  \caption{A spatial slice of $\mathrm{AdS}_3$ is shown, where the black dots on the boundary schematically represent the boundary degrees of freedom. Two competing extremal surfaces, $\gamma_1$ and $\gamma_2$, enclose a region $b'$ between them. (a) When $A(\gamma_1) < A(\gamma_2)$, the entanglement wedge of $B$ does not include $b'$. (b) When $A(\gamma_1) > A(\gamma_2)$, the entanglement wedge of $B$ includes $b'$. In this case, the region $b'$ constitutes a python's lunch.}

  \label{fig:PLsetup}
\end{figure}

In the context of AdS/CFT, the entanglement wedge reconstruction paradigm states that bulk degrees of freedom in the entanglement wedge of a given boundary subregion $B$ are encoded in and can be reconstructed in terms of operators in $B$. The basic framework underlying this is the theory of quantum error correction \cite{Almheiri:2014lwa, Dong:2016eik, Harlow:2016vwg, Faulkner:2017vdd,  Petzmap}.  When there exist multiple quantum extremal surfaces (QES) satisfying the homology condition with respect to the same boundary subregion, the entanglement wedge reconstruction paradigm gets more interesting. It was shown in \cite{Engelhardt:2021mue} that the reconstruction of bulk operators up to the outermost (i.e., closest to the boundary subregion) QES can be accomplished by ``simple'' operators in $B$, essentially combining ideas from HKLL \cite{Hamilton:2006az} together with forward and backward time evolution. However, for operators behind the outermost QES (but still within the entanglement wedge), no such simple reconstruction exists. In fact, based on the analogy between the time-reflection symmetric slice in AdS and tensor networks, it was conjectured in \cite{Brown:2019rox} that reconstruction of operators behind the outermost QES should be exponentially complex. This conjecture is known as the  Python's lunch conjecture, referring to the characteristic bulge in the entanglement wedge geometry. Part of the difficulty in proving the python's lunch conjecture is the lack of computable measures of complexity. Here, we will make some remarks about the python's lunch conjecture from the perspective of Wigner negativity. For the most part, we will confine our attention to the simplified setting of time reflection symmetric states, and where the entanglement wedge of the boundary subregion $B$ has two locally minimal surfaces $\gamma_1$ and $\gamma_2$ (see figure \ref{fig:PLsetup} for an example). 

Recall that in our calculation of Wigner negativity in the PSSY model, we observed that the negativity of the radiation state remains small before the Page time, but begins to grow exponentially past the Page time. One way to interpret this result is from the perspective of the Python’s lunch conjecture. As the black hole evaporates past the Page time, the entanglement wedge of the radiation bath forms an island inside the black hole interior. This island region constitutes a python’s lunch from the radiation system perspective, as it lies behind a sub-dominant QES, namely the empty surface. We could interpret the fact that the Wigner negativity becomes exponentially large beyond the Page time as a manifestation of the formation of a python's lunch region in the entanglement wedge.

If this interpretation is correct, then the same phenomenon should be present in more general holographic states. Consider a time-reflection symmetric, bi-partite holographic state $\psi_{B,\overline{B}}$ for complementary boundary subregions $B$ and $\overline{B}$ such that there are two minimal surfaces in the bulk geometry homologous to $B$ (see figure \ref{fig:PLsetup} for an example). Let $\gamma_{\text{out}}$ be the outermost minimal surface and let $\gamma_{\text{min.}}$ be the minimal surface with the smaller entropy. We would like to compute the Wigner negativity of the reduced density matrix on $B$. One complication is that in general the boundary Hilbert space is infinite dimensional, so it is not immediately clear how to apply our previous finite-dimensional results in this setting. We will proceed by truncating the Hilbert space of $B$ to a finite but large prime dimensional ``microcanonical'' subspace (more on this below). A second complication has to do with the choice of computational basis. While the Wigner negativity of the reduced density matrix on $B$ will depend on a choice of basis, we can get a natural basis-independent quantity by averaging over the choice of basis with the Haar measure. This calculation is essentially identical to the late-time limit of the calculation carried out in the previous section, and the answer is given by:
\beq \label{eq:negHaar}
\mathcal{N}_{\text{Haar}} = \left(\frac{1}{\sqrt{\pi r}}e^{-r} + \erf(\sqrt{r})\right),
\eeq
where 
\beq \label{eq:wrongr}
r = \frac{1}{2(D e^{-S^B_{2}}-1)},
\eeq
with $S^B_{2}$ being the second R\'enyi entropy of the reduced density matrix on $B$. An irritating feature of this formula is that it involves the second R\'enyi entropy, which does not have a simple geometric characterization in terms of the original geometry dual to the state $\psi_{B,\overline{B}}$. Nevertheless, we will argue below that the correct formula for the negativity of a general holographic state should involve areas of extremal surfaces in the original geometry. 

As a first pass, we will derive a lower bound on the negativity which only depends on areas of minimal surfaces in the original geometry. We proceed by making the following two assumptions: firstly, we will assume that the area of the outermost minimal surface calculates some kind of a \emph{coarse-grained} entropy for the state $\rho_B$. The precise definition of this coarse-grained entropy is not important (see \cite{Engelhardt:2018kcs, Chandra:2022fwi} for more details), but the only thing we will assume is that the choice of UV regulator in truncating the Hilbert space of $B$ is consistent with this interpretation, in that
\beq 
\frac{1}{4G_N}A(\gamma_{\text{out}}) \leq \log D.
\eeq 
Secondly, we will assume that the von Neumann entropy of the truncated state $\rho_B$ is approximately (i.e., at leading order in $\frac{1}{G_N}$) equal to the area of the minimal area extremal surface $\gamma_{\text{min.}}$. In this case, we have
\beq 
e^{-S_2^B} \geq e^{-S_{vN}} \sim e^{-\frac{1}{4G_N}A(\gamma_{\text{min.}})}.
\eeq 
These two inequalities imply that
\beq 
D e^{-S^B_2} \geq e^{\Delta},\;\; \Delta =  \frac{1}{4G_N}\left(A(\gamma_{\text{out}})-A(\gamma_{\text{min.}})\right).
\eeq 
Note that when the gap $\Delta$ is large, the formula for the Haar averaged Wigner negativity reduces to: 
\beq 
\mathcal{N}_{\text{Haar}} \sim \sqrt{\frac{2}{\pi} D e^{-S^B_2}} \geq \sqrt{\frac{2}{\pi}} e^{\frac{\Delta}{2}}.
\eeq 
Thus, the negativity becomes large, and scales at least exponentially in the gap between the areas of the outermost and minimal extremal surfaces. This line of reasoning leads us to expect that holographic states with a python's lunch in their entanglement wedge will have a large stabilizer complexity for the corresponding reduced density matrix on the boundary subregion $B$ with respect to a generic choice of computational basis. Of course, for this result to be genuinely interesting, we must also be able to show that when the gap is small, the negativity becomes small. More ambitiously, we would want to derive a formula for the negativity, and not just a lower bound. 

We will now sketch out a proposal for a general formula for the Wigner negativity in holographic states. First, we have to be clear on what Hilbert space truncation we will use. Our proposal is that we should write the full state $\psi_{B,\bar{B}}$ in terms of \emph{random tensor networks} (RTNs) \cite{Hayden:2016cfa, Chandra:2023dgq, Geng:2025efs, Bao:2025plr}, or equivalently, \emph{fixed-area} states \cite{Dong:2018seb, Akers:2018fow}, and keep only the RTN/fixed-area state where the probability distribution over RTNs/areas is peaked. Roughly speaking, this can be thought of as truncating the boundary Hilbert space to a ``micro-canonical'' window corresponding to fixed macroscopic variables (such as energy, area etc.). Note that while fixed-area states are defined from a bulk point of view as approximate eigenstates of the area operators, they behave quite similarly to RTNs (see Appendix E of \cite{Penington:2019kki}); in particular, they have a flat entanglement spectrum, and hence R\'enyi entropies independent of the R\'enyi parameter to leading order in $\frac{1}{G_N}$. This analogy between fixed-area/random tensor network states suggests that for such states it is reasonable to make the identifications:
\beq 
D \sim e^{\frac{1}{4G_N}A(\gamma_{\text{out}})}, \;\;\; e^{S^B_2} \sim e^{S^B_{vN}} \sim e^{\frac{1}{4G_N}A(\gamma_{\text{min.}})}.
\eeq 
Thus, the Wigner negativity for such states should be given by 
\beq \label{eq:rightr}
\mathcal{N} = \left(\frac{1}{\sqrt{\pi r}}e^{-r} + \erf(\sqrt{r})\right),\;\;r = \exp\left[\frac{1}{4G_N}(A(\gamma_{\text{out}})-A(\gamma_{\text{min}}))\right].
\eeq
In fact, for such RTNs, the negativity is automatically independent of choice of computational basis, and one does not need explicit Haar averaging (see \cite{Zhang:2024fyp} and appendix \ref{sec:RTN} for more details). 

Of course, general holographic states are superpositions of fixed area states/RTNs \cite{Dong:2018seb, Akers:2018fow, Caputa:2020fbc}:
\beq
|\psi_{B\overline{B}}\rangle = \sum_{A_1,A_2} \sqrt{p(A_1,A_2)} |A_1, A_2\rangle,
\eeq 
where $A_1$ and $A_2$ are the areas of the two competing extremal surfaces. The probability distribution $p(A_1,A_2)$ is sharply peaked around values of $A_1$ and $A_2$ given by the bulk geometry dual to $\psi_{B,\overline{B}}$. As mentioned above, our proposal is that the negativity, at leading order in $\frac{1}{G_N}$, should simply be given by the formula \eqref{eq:rightr} applied to the fixed-area state on which $p(A_1,A_2)$ is peaked. The motivation behind this proposal is as follows: in defining the Wigner negativity by averaging over choice of computational basis, we must decide what to average over. Equation \eqref{eq:wrongr} follows from a naive Haar average over the entire Hilbert space, which does not make sense for infinite dimensional Hilbert spaces. On the other hand, it seems more physically motivated to Haar average over the microstates within each fixed $(A_1,A_2)$ sector of the Hilbert space (or from the point of view of RTNs, within a fixed microcanonical sector with fixed macroscopic variables), while treating $(A_1,A_2)$ as labels for classical superselection sectors (at leading order in $\frac{1}{G_N}$). Alternatively, we can regard this as a particular choice of UV truncation to make sense of the Wigner negativity, i.e., we compute the negativity of the fixed-area state/random tensor network which best approximates the original state $\psi_{B,\overline{B}}$. Thus, we are led to propose that the Wigner negativity of a general holographic state should be given by equation \eqref{eq:rightr} at leading order in $\frac{1}{G_N}$. As an immediate consequence of our proposal, the presence of a python's lunch region in the entanglement wedge results in an exponentially large stabilizer complexity for the corresponding boundary state:
\beq
\mathcal{N} \sim \sqrt{\frac{2}{\pi}}\exp\left[\frac{1}{2}\frac{(A(\gamma_{\text{out}})-A(\gamma_{\text{min}}))}{4G_N} \right].
\eeq 

Interestingly, our proposed formula for the negativity only seems to involve the outermost and the minimal extremal surfaces, but not the bulge surface which plays a crucial role in the python's lunch conjecture. This is not directly a problem, because the python's lunch conjecture is a statement about the complexity of bulk reconstruction within the python's lunch region, and not about the stabilizer complexity of the boundary density matrix. It would be interesting to use our techniques to explicitly calculate the stabilizer complexity of the Petz recovery channel that implements bulk reconstruction in the python's lunch \cite{Petzmap, Penington:2019kki, Parrikar:2024zbb}; we leave this to future work.

\section*{Acknowledgments}
We would like to thank Abhijit Gadde, Shiraz Minwalla, Shiroman Prakash, Sandip Trivedi and Spenta Wadia for helpful discussions and Vijay Balasubramanian and Alex May for comments on an earlier version of the draft. We are grateful to Suman Nandi for his significant help with numerical computations. We acknowledge support from the Department of
Atomic Energy, Government of India, under project identification number RTI 4002, and from the Infosys Endowment for the study of the Quantum Structure of Spacetime.

\section*{Data Availability}

The data and code supporting the findings of this article are available in Ref.~\cite{BasuGithubData}.

\appendix

\section{Evaluating the Integral} \label{sec:integral}

In this section, we will evaluate the integral obtained in equation (\ref{eq:integral}):
\beq 
I = \lim_{\eps \rightarrow 0}   \int_{-\infty}^{\infty} \frac{\d t}{2\pi i} \frac{-2 it}{t^2 + \eps^2} \frac{\d}{\d t} \exp \left[-r \left( -2i t + t^2 \right)\right],\;\;\; r= \frac{e^{S_2}}{2D}.
\eeq 
By completing squares in the exponent, we can write this integral as:
\beq 
I = \lim_{\eps \rightarrow 0}  e^{-r} \int_{-\infty}^{\infty} \frac{\d t}{2\pi i} \frac{-2 it}{t^2 + \eps^2} \frac{\d}{\d t} \exp \left[-r (t-i)^2\right],\;\;\; r= \frac{e^{S_2}}{2D}.
\eeq 
We pick up the pole at $t=i \epsilon$ and then deform the contour in the upper half plane to the contour $\gamma = \mathbb{R} + i $.\footnote{The contour segments at infinity are exponentially suppressed and can be dropped.} This gives:
\beqn 
I &=& 2r +  e^{-r} \int_{-\infty}^{\infty} \frac{\d x}{2\pi i} \frac{-2 i}{(x+i)} \frac{\d}{\d x}e^{-rx^2}\nonumber\\
&=& 2r +  e^{-r} \int_{-\infty}^{\infty} \frac{\d x}{2\pi i} \frac{-2 i}{(x+i)^2} e^{-rx^2},
\eeqn 
where in the second line we have integrated by parts. We can simplify the remaining integral in the following way: 
\beqn 
I &=&  2r +  \frac{1}{\sqrt{4\pi r}} e^{-r} \int_{-\infty}^{\infty} \frac{\d x}{2\pi i} \frac{-2 i}{(x+i)^2}\int_{-\infty}^{\infty} du\, e^{-\frac{1}{4r}u^2 - i u x} \nonumber\\
&=& 2r +  \frac{1}{\sqrt{4\pi r}} e^{-r}\int_{-\infty}^{\infty} du\,e^{-\frac{1}{4r}u^2 } \int_{-\infty}^{\infty} \frac{\d x}{2\pi i} \frac{-2 i}{(x+i)^2} e^{- i u x}.
\eeqn 
Using
\beq
\int_{-\infty}^{\infty} \frac{\d x}{2\pi i} \frac{-2 i}{(x+i)^2} e^{- i u x} = 2u \theta(u) e^{-u},
\eeq 
we get
\beqn 
I &=& 2r +  \frac{1}{\sqrt{4\pi r}} e^{-r}\int_{0}^{\infty} du\,2ue^{-\frac{1}{4r}u^2-u }\nonumber\\
&=& 2r +  \frac{1}{\sqrt{\pi r}} \int_{0}^{\infty} du\,u\,e^{-\frac{1}{4r}(u+2r)^2 }\nonumber\\
&=& 2r +  \frac{1}{\sqrt{\pi r}} \int_{2r}^{\infty} dy\,y\,e^{-\frac{1}{4r}y^2 }-\frac{2r}{\sqrt{\pi r}} \int_{2r}^{\infty} dy\,\,e^{-\frac{1}{4r}y^2 }.
\eeqn 
The integral in the second term above is elementary and is given by
\beq 
\frac{1}{\sqrt{\pi r}} \int_{2r}^{\infty} dy\,y\,e^{-\frac{1}{4r}y^2 }=\sqrt{\frac{r}{\pi}} e^{-r}.
\eeq
The integral in the third term can be written as
\beq
-\frac{2r}{\sqrt{\pi r}} \int_{0}^{\infty} dy\,\,e^{-\frac{1}{4r}y^2 } + \frac{2r}{\sqrt{\pi r}} \int_{0}^{2r} dy\,\,e^{-\frac{1}{4r}y^2 } .
\eeq 
The first of these terms cancels the pole contribution in $I$, while the second term can be written in terms of the error function. This leads to equation \eqref{eq:neg}.


\section{Remarks on fluctuations in the PSSY model}\label{sec:numerics}

We wish to argue that the variance for the Wigner negativity in the PSSY model is suppressed both in $D$ and in $e^{S_0}$. To quantify this, we consider the quantity $\overline{\mathcal{N}^2} - (\overline{\mathcal{N}})^2$, which can be evaluated using the gravitational path integral. For a rough estimate of its order of magnitude, we will use 
the replica trick introduced earlier, namely 
$|W| = \lim_{n \to 1/2} W^{2n}$. This leads us to the following expression:
\begin{equation}
    \overline{\mathcal{N}^2} - \left(\overline{\mathcal{N}}\right)^2=\sum_{q,p}\sum_{q',p'}\lim_{m\to \frac{1}{2}}\; \lim_{n \to \frac{1}{2}}\left[ \overline{W^{2m}(q,p)\;W^{2n}(q',p')}-\overline{W^{2m}(q,p)}\times \overline{W^{2n}(q',p')} \right]. \label{eq:var1}
\end{equation}
One can impose the boundary conditions associated with each term in the above expression and then fill in the bulk by connecting the various boundaries in all possible ways. We observe that only those diagrams appearing in $\overline{W^{2m}(q,p) W^{2n}(q',p')}$ which connect across---that is, involve contractions between both $A(q,p)$ and $A(q',p')$ within at least one of their irreducible components---contribute to the variance. Importantly, any such diagram can only get enhanced provided $q=q'$ and $p=p'$. Thus, these diagrams must necessarily be suppressed in comparison with those which do not connect across. 

Suppose there are $\mathcal{C}(m,n)$ such diagrams. Each diagram may be regarded as a product of $\alpha_1$ one--boundary contractions, $\alpha_2$ two--boundary contractions, and so on up to $\alpha_{2m+2n}$ $(2m+2n)$--boundary contractions, subject to the constraint 
$\sum_i i \alpha_i = 2m+2n$. Such a diagram contributes at order 
\[
(e^{S_0})^{\sum_{i=1}^{2m+2n}\alpha_i}\, (D)^{\sum_{i=1}^{m+n}\alpha_{2i}} .
\]
The fully disconnected configuration 
$\{\alpha_1 = 2m+2n, \quad \alpha_i = 0 \ \text{for}\ i\geq 2\}$ is excluded since all diagrams are required to connect across. One then finds upon inspection that the maximal power of $e^{S_0}$ can be $2m+2n-1$, 
while the maximal power of $D$ can be $m+n$. Now recall, that there are overall prefactors $\left(\tfrac{1}{D^2 e^{S_0}}\right)^{2m} \left(\tfrac{1}{D^2 e^{S_0}}\right)^{2n}$ originating from the definition of the Wigner function. Taking these into account, we may therefore write,
\begin{equation}
    \overline{\mathcal{N}^2} - \left(\overline{\mathcal{N}}\right)^2=\sum_{q,p}\sum_{q',p'}\lim_{m\to \frac{1}{2}}\; \lim_{n \to \frac{1}{2}}\;\delta_{q,q'}\delta_{p,p'}\times \frac{1}{D^{4m+4n}} \times \mathcal{P}(D,e^{S_0})
\end{equation}
Here, $\mathcal{P}(D, e^{S_0})$ denotes a polynomial in $\tfrac{D}{e^{S_0}}$ and $\tfrac{1}{e^{S_0}}$, containing in total $\mathcal{C}(m,n)$ terms. It turns out that when $\tfrac{D}{e^{S_0}} < 1$, we have $\mathcal{P}(D,e^{S_0}) \leq \mathcal{C}(m,n)\,\frac{D}{e^{S_0}}$, where $\mathcal{C}(m,n)$ does not scale with $D$ or $e^{S_0}$. 
Consequently, the entire sum over diagrams is approximately of order $\tfrac{1}{D^{4m+4n-1} e^{S_0}}$. Upon analytically continuing both $m$ and $n$ to $1/2$ and subsequently summing over $q,q'$ and $p,p'$, one finds that the variance scales as $\tfrac{1}{D e^{S_0}}$. On the other hand, when $\tfrac{D}{e^{S_0}} > 1$, one instead finds $\mathcal{P}(D,e^{S_0}) \leq \mathcal{C}(m,n)\, \bigl(\tfrac{D}{e^{S_0}}\bigr)^{n+m}$, so that the entire sum is approximately of order 
$\tfrac{1}{D^{3n+3m} (e^{S_0})^{n+m}}$. In this case as well, after the continuation $m,n\to 1/2$, the variance scales as 
$\tfrac{1}{D e^{S_0}}$. Thus, regardless of the relative size of $D$ and $e^{S_0}$, the variance is suppressed as $O\!\bigl(\tfrac{1}{D e^{S_0}}\bigr)$.

\section{Details of the calculation of $\mathcal{I}(q,p;z)$}
\label{sec:Iqpz}

In this Appendix, we present the details of the derivation of \eqref{eq:Haarfinal'} quoted in the main text. Recall that we had,
\begin{eqnarray}
   \mathcal{I}(q,p;z)&=& \exp \bigg[ iz \;
\begin{tikzpicture}[baseline={([yshift=-.5ex]current bounding box.center)}]
\coordinate (P) at (0,0);

\draw (P) circle (10pt); 

\node[below] at (0,-10pt) {};

\coordinate (P1) at (-0.15,0);
\coordinate (P2) at (0.15,0);

\draw[black] (P1) circle (1.7pt);
\filldraw (P2) circle (1.7pt);


\end{tikzpicture}\; \bigg] \;
\underbrace{\frac{1}{\text{Vol}_{U(D-1)}}\int dU  \exp \left\{iz 
    \bigg[ \; 
\begin{tikzpicture}[baseline={([yshift=-.5ex]current bounding box.center)}]
\coordinate (P) at (0,0);

\draw (P) circle (10pt); 

\node[below] at (0,-10pt) {};

\coordinate (P1) at (-0.15,0);
\coordinate (P2) at (0.15,0);

\draw[black] (P1) circle (1.7pt);
\filldraw (P2) circle (1.7pt);

\draw[-, thick] (P1) to (-0.7,0.7);
\draw[dotted, thick] (P1) to (-0.7,-0.7);
\end{tikzpicture} \;+
 \begin{tikzpicture}[baseline={([yshift=-.5ex]current bounding box.center)}]
\coordinate (P) at (0,0);

\draw (P) circle (10pt); 

\node[below] at (0,-10pt) {};

\coordinate (P1) at (-0.15,0);
\coordinate (P2) at (0.15,0);

\draw[black] (P1) circle (1.7pt);
\filldraw (P2) circle (1.7pt);

\draw[-, thick] (P2) to (0.7,0.7);
\draw[dotted, thick] (P2) to (0.7,-0.7);
\end{tikzpicture} \; +   
\begin{tikzpicture}[baseline={([yshift=-.5ex]current bounding box.center)}]
\coordinate (P) at (0,0);

\draw (P) circle (10pt); 

\node[below] at (0,-10pt) {};

\coordinate (P1) at (-0.15,0);
\coordinate (P2) at (0.15,0);

\draw[black] (P1) circle (1.7pt);
\filldraw (P2) circle (1.7pt);

\draw[-, thick] (P1) to (-0.7,0.7);
\draw[dotted, thick] (P1) to (-0.7,-0.7);
\draw[-, thick] (P2) to (0.7,0.7);
\draw[dotted, thick] (P2) to (0.7,-0.7);

\end{tikzpicture} \; \bigg]\right\}}_{\mathcal{J}}\nonumber. \label{eq:resum}
\end{eqnarray}
Our task here is to systematically compute the integral $\mathcal{J}$. From \eqref{eq:HaarInt1'}, we have 
\begin{eqnarray}
   \mathcal{J} &=&  \sum_{n=0}^{\infty} \frac{(iz)^n}{n!} \;
   \frac{1}{\text{Vol}_{U(D-1)}} \int dU \underbrace{\exp \bigg[ iz 
   \begin{tikzpicture}[baseline={([yshift=-.5ex]current bounding box.center)}]
\coordinate (P) at (0,0);

\draw (P) circle (10pt); 

\node[below] at (0,-10pt) {};

\coordinate (P1) at (-0.15,0);
\coordinate (P2) at (0.15,0);

\draw[black] (P1) circle (1.7pt);
\filldraw (P2) circle (1.7pt);

\draw[-, thick] (P1) to (-0.7,0.7);
\draw[dotted, thick] (P1) to (-0.7,-0.7);
\draw[-, thick] (P2) to (0.7,0.7);
\draw[dotted, thick] (P2) to (0.7,-0.7);

\end{tikzpicture}
    \bigg]}_{A} \; 
   \underbrace{\bigg[
   \begin{tikzpicture}[baseline={([yshift=-.5ex]current bounding box.center)}]
\coordinate (P) at (0,0);

\draw (P) circle (10pt); 

\node[below] at (0,-10pt) {};

\coordinate (P1) at (-0.15,0);
\coordinate (P2) at (0.15,0);

\draw[black] (P1) circle (1.7pt);
\filldraw (P2) circle (1.7pt);

\draw[-, thick] (P1) to (-0.7,0.7);
\draw[dotted, thick] (P1) to (-0.7,-0.7);
\end{tikzpicture} \;+
   \begin{tikzpicture}[baseline={([yshift=-.5ex]current bounding box.center)}]
\coordinate (P) at (0,0);

\draw (P) circle (10pt); 

\node[below] at (0,-10pt) {};

\coordinate (P1) at (-0.15,0);
\coordinate (P2) at (0.15,0);

\draw[black] (P1) circle (1.7pt);
\filldraw (P2) circle (1.7pt);

\draw[-, thick] (P2) to (0.7,0.7);
\draw[dotted, thick] (P2) to (0.7,-0.7);
\end{tikzpicture} 
   \bigg]^n }_{B} . \label{eq:HaarInt1}
\end{eqnarray}
Let's now focus on the $n$-th term in the above sum in equation \eqref{eq:HaarInt1}. Firstly, note that for $n$ odd, the integral vanishes. Further, when $n$ is even, out of all the terms in the binomial expansion of $B$, only those contribute which have the same power of $U$ and $U^*$, because the Haar integral is non-zero only when we have an equal number $U$s and $U^*$s in the integrand. So, from the expansion of $B$, we pick only those terms which have an equal number of $U$s and $U^*$s. For $n=2p$, we can make the replacement:
\begin{equation}
    \bigg[ \blobleft+\blobright \bigg]^{2p}\to\frac{(2p)!}{(p!)^2}\; \bigg( \blobleft \; \; \blobright \bigg)^p,
\end{equation}
and so equation \eqref{eq:HaarInt1} can be written as
\begin{equation}
    \mathcal{J}=\sum_{p=0}^{\infty}\frac{(iz)^{2p}}{(p!)^2} \; \frac{1}{\text{Vol}_{U(D-1)}}\int dU\; \exp \bigg[iz  \blobtwo \bigg] \; \bigg( \blobleft \;\; \blobright \bigg)^p. \label{eq:IntJ}
\end{equation}

We can now perform the $U$ integral. In doing so, we must further expand out the exponential inside the integral. Any given term in the expansion now looks like:
\begin{equation}
    \bigg[ \underbrace{\blobtwo}_{A'} \bigg]^m \; \bigg( \underbrace{\blobleft}_{B_1} \; \; \; \underbrace{\blobright}_{B_2} \bigg)^p. \label{eq:genericterm}
\end{equation}
At this stage, we will need to use the following general formula for Haar integration:
\begin{multline}
    \frac{1}{\text{Vol}_{U(D-1)}}\int dU \; U_{i_1 j_1} U_{i_2 j_2}....U_{i_n j_n}\; U^*_{i_1'j_1'} U^*_{i_2' j_2'}....U^*_{i'_nj'_n}=\sum_{\sigma, \tau\; \in S_n} \text{Wg}(\sigma\tau^{-1},D-1)\;\; \delta_{i_1 i'_{\sigma(1)}} \delta_{i_2 i'_{\sigma(2)}}...\delta_{i_n i'_{\sigma(n)}}\;\\
    \delta_{j_1 j'_{\tau(1)}} \delta_{j_2 j'_{\tau(2)}}...\delta_{j_n j'_{\tau(n)}} .\label{eq:Haargeneral}
\end{multline}
In our diagrammatic approach, each term in equation \eqref{eq:Haargeneral} can be represented by an appropriately contracted diagram. In general, it is not necessary that the solid and dotted lines get contracted in the same way, i.e., the permutations $\sigma$ and $\tau$ in $S_n$ need not be the same. Consider the example of the following contraction which appears as one of the terms in equation \eqref{eq:Haargeneral} for $n=5$:
\begin{equation*}
    \text{Wg}(\sigma\tau^{-1})\; \delta_{i_1,i'_3}\delta_{i_2,i_1'} \delta_{i_3,i_2'} \delta_{i_4,i'_5}\delta_{i_5, i_4'} \; \delta_{j_1,j_2'} \delta_{j_2,j_1'} \delta_{j_3,j_3'} \delta_{j_4,j_4'} \delta_{j_5,j_5'}
\end{equation*}
with $\sigma=(132)(45),\; \tau=(12)(3)(4)(5)$. We can represent this term diagrammatically as:
\begin{equation*}
    \begin{tikzpicture}[baseline={([yshift=-.5ex]current bounding box.center)}]
\coordinate (P) at (0,0);
\coordinate (Q) at (30pt,0);
\coordinate (R) at (60pt,0);

\draw (P) circle (10pt);
\draw (Q) circle (10pt); 
\draw (R) circle (10pt);

\coordinate (P1) at (-4pt,0);
\coordinate (P2) at (4pt,0);
\coordinate (Q1) at (26pt,0);D
\coordinate (Q2) at (34pt,0);
\coordinate (R1) at (56pt,0);3
\coordinate (R2) at (64pt,0);

\draw[black] (P1) circle (1.7pt);
\filldraw (P2) circle (1.7pt);
\draw[black] (Q1) circle (1.7pt);
\filldraw (Q2) circle (1.7pt);
\draw[black] (R1) circle (1.7pt);
\filldraw (R2) circle (1.7pt);

\draw[black,out = 60,in = 120, looseness = 1.5] (P1) to (R2);
\draw[black,out = 60,in = 120, looseness =1.5] (P2) to (Q1);
\draw[black,out = 60,in = 120, looseness =1.5] (Q2) to (R1);

\draw[dotted,out = -60,in = -120, looseness = 2.0] (P1) to (Q2);
\draw[dotted,out = -60,in = -120, looseness =2.0] (P2) to (Q1); 
\draw[dotted,out = -120,in = -60, min distance =1.5cm] (R1) to (R2); 

\coordinate (S) at (90pt,0);
\coordinate (T) at (120pt,0);

\draw (S) circle (10pt);
\draw (T) circle (10pt);

\coordinate (S1) at (86pt,0);
\coordinate (S2) at (94pt,0);
\coordinate (T1) at (116pt,0);
\coordinate (T2) at (124pt,0);

\draw[black] (S1) circle (1.7pt);
\filldraw (S2) circle (1.7pt);
\draw[black] (T1) circle (1.7pt);
\filldraw (T2) circle (1.7pt);

\draw[black,out = 60,in = 120, looseness = 1.5] (S1) to (T2);
\draw[black,out = 60,in = 120, looseness =1.5] (S2) to (T1);

 \draw[dotted, out=-120, in=-60, min distance=1.5cm] (S1) to (S2);
  \draw[dotted, out=-120, in=-60, min distance=1.5cm] (T1) to (T2);
\end{tikzpicture}.
\end{equation*}
From the diagram it appears as if this term factorizes into two irreducible pieces: one $3-$blob contraction and one $2-$blob contraction. One might think that for $n$ blobs, any diagram can be built out of various such irreducible pieces: $n_1$ $1-$blob contractions, $n_2$ $2-$blob contraction, $n_3$ $3-$blob contractions and so on, such that $\sum_{p=1}^n p\;n_p=n$. But this is \textit{not} quite true in general since the Weingarten functions don't factorize accordingly.
In our example above, the first irreducible $3-$blob diagram by itself would be proportional to $\text{Wg}(21,D-1)$ and the $2-$blob piece would be proportional to $\text{Wg}(2,D-1)$. But for finite $D$, 
\begin{equation*}
    \text{Wg}(2^2 1, D-1)\neq \text{Wg}(21,D-1) \; \times \; \text{Wg}(2,D-1).
\end{equation*}
However, things simplify at large $D$, where the Weingarten functions have the asymptotic behavior:
\begin{equation}
    \text{Wg}(\sigma,D)=\frac{1}{D^{n+|\sigma|}}  \prod_{i}(-1)^{|C_i|-1} \mathcal{C}_{|C_i|-1}\;+\mathcal{O}\bigg(\frac{1}{D^{n+|\sigma|+2}}\bigg), \label{eq:Wg}
\end{equation}
where $|\sigma|$ is the smallest number of transpositions (pairwise exchanges) that $\sigma$ is composed of, $\{C_i\}$ denote the various cycles in the permutation $\sigma$, $|C_i|$ is the length of the cycle $C_i$, and $\mathcal{C}_n = \frac{(2n)!}{n!(n+1)!}$ are the Catalan numbers. In the large-$D$ limit, one can check that the Weingarten functions indeed factorize, up to $O(1/D^2)$ corrections. 

To make progress, in our subsequent discussion we will work in the large $D$ limit. In this case, the Haar average over $n$ $(U,U^*)$ pairs can be decomposed into sums of products of various irreducible (i.e., fully connected) pieces with appropriate multiplicity. A word of caution before we proceed: note that the Weingarten functions factorize in the large $D$ limit, up to $O(1/D^2)$ corrections. If we were only interested in computing R\'enyi Wigner negativities, i.e. $\overline{W_{\mathcal{B}}^{2n}}$, these corrections are unimportant (for any fixed R\'enyi index $n$) and can be safely ignored in the $D \to \infty$ limit. But in the integral representation method, we are summing over all moments, and so one might worry that the $O(1/D^2)$ terms could sum up to give a non-trivial contribution to the Wigner negativity. Recall that in the PSSY calculation, we encountered an analogous situation. There we were able to explicitly re-sum and check that the sub-leading corrections are not important. In the present calculation, we will simply assume that the $O(1/D^2)$ corrections can be ignored, without any further justification. We will later compare our analytic results with numerical calculations and find good agreement, thus justifying the above assumption a posteriori. 

Let us now go back to equation \eqref{eq:IntJ}. Using the notation in \eqref{eq:genericterm}, we can schematically express this equation as, 
\begin{equation}
    \mathcal{J}=\sum_{p=0}^{\infty} \sum_{m=0}^{\infty}\frac{(iz)^{2p}}{(p!)^2}\frac{(iz)^m}{m!}\;\;\overline{(A')^m (B_1^p\;B_2^p)}, \label{eq:SumJ1}
\end{equation}
where the overline indicates Haar-averaging. $\overline{(A')^m (B_1^p B_2^p)}$ can be expressed as a sum over all possible contracted diagrams, where each diagram can be understood as a product of irreducible (i.e.\ fully contracted) pieces, as argued above. In the present case, we proceed as follows. For arbitrary $m$ and $p$, the $A'$ blobs can either contract among themselves or be sandwiched between the $B_1$ and $B_2$ blobs. Suppose $n$ of the $A'$ blobs are contracted with the $B_1, B_2$ blobs, while $n'$ of them contract only among themselves, with $n+n' = m$. Then any generic term in $\overline{(A')^m (B_1^p B_2^p)}$ can be written as a product of two factors: one involving only the $n'$ number of $A'$ blobs, and the other consisting of the $n$ $A'$ blobs that contract with the $p$ $(B_1, B_2)$ blobs. For now, we leave the first part aside and focus on the second.  

Observe that any fully connected diagram containing a $B_1$ blob must also contain a $B_2$ blob. Thus, $(B_1, B_2)$ always appear in pairs in any fully contracted diagram, and a generic fully contracted diagram may contain multiple such pairs. Hence, the second part decomposes into a product of irreducible components in the following manner: $p_1$ irreducible pieces each containing one $(B_1, B_2)$ pair, $p_2$ irreducible pieces each containing two $(B_1, B_2)$ pairs, and so on, up to $p_p$ irreducible pieces each containing $p$ number of pairs, subject to the constraint $\sum_i p_i = p$. For a given set of $p$ $B_1$’s and $p$ $B_2$’s, we can first build the possible pairings in $p!$ ways. Then any particular configuration $\{p_i\}$ can be realised in $\frac{p!}{p_1! p_2! \cdots p_p!} \, \frac{1}{(1!)^{p_1} (2!)^{p_2} \cdots (p!)^{p_p}}$
different ways. Thus, the combinatorial factor associated with distributing $p$ pairs of $(B_1, B_2)$ into various irreducible components is given by,
\begin{equation*}
    \frac{(p!)^2}{\prod_{i=1}^{p} p_i!\;(i!)^{p_i}}.
\end{equation*}
We now address the matter of distributing the $n$ copies of $A'$ blobs among the irreducible structures introduced above. For a fixed configuration $\{p_i\}$, this proceeds as follows. Consider first the $p_1$ irreducible pieces of the first kind (each containing a single $(B_1,B_2)$ pair). Suppose that in the $j$-th such piece, $n_{j}^{(1)}$ copies of $A'$ are contracted, with $j=1,2,\ldots,p_1$. Here the superscript indicates the type of irreducible component, while the subscript labels the individual copy within that type. We denote the entire $j$-th piece of the first kind by $\mathcal{D}_{n_{j}^{(1)}}^{(1)}$. In a similar manner, for the $p_2$ irreducible pieces of the second kind (each involving two $(B_1,B_2)$ pairs), we may contract $n_{j}^{(2)}$ copies of $A'$ into the $j$-th such piece, denoting the resulting structure by $\mathcal{D}_{n_{j}^{(2)}}^{(2)}$. The same prescription applies to higher kinds of irreducible components. Note that each $\mathcal{D}_{n_{j}^{(i)}}^{(i)}$ represents not a single object, but rather a sum of all possible fully connected diagrams involving $i$ number of $(B_1, B_2)$ pairs and $n_{j}^{(i)}$ number of $A'$ blobs. In the above way of distributing $A'$ blobs, the consistency condition is that the total number of $A'$ insertions equals $n$, namely  
$\sum_{i=1}^{p} \sum_{j=1}^{p_i} n_{j}^{(i)} \;=\; n$.
Thus, for any fixed choice of $\{p_i\}$, the above distribution of $n$ $A'$ blobs among the irreducible components can be found in 
\begin{equation*}
    \frac{n!}{\prod_{i=1}^{p}\prod_{j=1}^{p_i} n_{j}^{(i)}! }
\end{equation*}
number of ways. For each fixed configuration $\{p_i\}$, one must sum over all possible distributions $\{n_j^{(i)}\}$ of the $A'$ blobs across the corresponding irreducible components. Finally, one should sum over every admissible configuration $\{p_i\}$.

We now turn to the factor arising in any generic term of $\overline{(A')^m (B_1^p B_2^p)}$ which involves only the $n'$ number of $A'$ blobs. These blobs can be selected in $\tfrac{m!}{n'!\,n!}$ distinct ways. Importantly, the $n'$ $A'$ blobs are \emph{not} necessarily fully connected among themselves; rather, they may form all possible factorizable configurations. We will discuss this below. For the moment, we collectively denote the entire set of such configurations as
\begin{equation*}
        \mathcal{D}_{n'}^{(A')}:=\bigg( \underbrace{\blobtwo .... ...\blobtwo}_{n'}\bigg)_{\text{contr}}.
\end{equation*}

One can now readily write down an expression for $\overline{(A')^m\;(B_1 ^pB_2^p)}$ as a sum of products of irreducible diagrams:
\begin{equation}
    \overline{(A')^m\;(B_1 ^pB_2^p)}=\sum_{\{p_i\}}' \sum_{\{n_j^{(i)},\;n'\}}' \frac{(p!)^2}{\prod_{i=1}^{p} p_i!\;(i!)^{p_i}} \times \frac{m!}{n'! \times \prod_{i=1}^{p}\prod_{j=1}^{p_i} n_{j}^{(i)}!} \;\prod_{i=1}^{p}\prod_{j=1}^{p_i} \mathcal{D}_{n_{j}^{(i)}}^{(i)}\times \mathcal{D}_{n'}^{(A')} .
\end{equation}
Here, the primes on the summations indicate that these sums are constrained, subject to the conditions outlined above. Substituting this expression into equation \eqref{eq:SumJ1}, the first observation is that the overall summation over $m$ effectively decouples the sums over $\{n_{j}^{(i)},n'\}$. In other words, one can independently sum over $\{n_{j}^{(i)}\}$ and $n'$, with each ranging from $0$ to $\infty$. Consequently, one may write,
\begin{eqnarray}
    \mathcal{J}&=&\sum_{p=0}^{\infty}\sum_{\{p_i\}}'\frac{(iz)^{2p}}{\prod_{i=1}^{p} p_i!\;(i!)^{p_i}} \times \prod_{k=1}^{p}\left( \sum_{n^{(k)}=0}^{\infty} \frac{(iz)^{n^{(k)}}}{n^{(k)}!} \mathcal{D}_{n^{(k)}}^{(k)} \right)^{p_k} \times \left( \sum_{n'=0}^{\infty} \frac{(iz)^{n'}}{n'!}\mathcal{D}_{n'}^{(A')}  \right) \nonumber \\
    &=& \mathcal{J}_0 \;\; \sum_{p=0}^{\infty}\sum_{\{p_i\}}'\frac{(iz)^{2p}}{\prod_{i=1}^{p} p_i!\;(i!)^{p_i}} \times \prod_{k=1}^{p}\left( \sum_{n^{(k)}=0}^{\infty} \frac{(iz)^{n^{(k)}}}{n^{(k)}!} \mathcal{D}_{n^{(k)}}^{(k)} \right)^{p_k}, \label{eq:sumJ2}
\end{eqnarray}
where we have defined $\mathcal{J}_0$ as, 
\begin{equation}
    \mathcal{J}_0=\sum_{n'=0}^{\infty} \frac{(iz)^{n'}}{n'!}\mathcal{D}_{n'}^{(A')}. \label{eq:sumJ0}
\end{equation}
We will treat this piece separately later. To obtain the first line of equation \eqref{eq:sumJ2}, we have used the constraints 
$\sum_{i=1}^{p}\sum_{j=1}^{p_i} n_j^{(i)} = n$ and $n + n' = m$. We then distributed the factor $(iz)^m$ among the various groups according to the configuration $\{n_j^{(i)}, n'\}$. Now we further observe that the sums over $\{p_i\}$ also decouple for the same reason as before. In this step, we redistribute the factor $(iz)^{2p}$ in accordance with the constraint 
$\sum_i i p_i = p$, and thereby rewrite equation \eqref{eq:sumJ2} as,
\begin{eqnarray}
    \mathcal{J}&=&\mathcal{J}_0 \; \sum_{p_1=0}^{\infty}\frac{1}{p_1!} \left( \frac{(iz)^2 \mathcal{S}_1}{1!} \right)^{p_1}\times \sum_{p_2=0}^{\infty}\frac{1}{p_2!} \left( \frac{(iz)^4 \mathcal{S}_2}{2!} \right)^{p_2}\times \cdots \cdots \nonumber \\
    &=&\mathcal{J}_0\; \;e^{\frac{(iz)^2 \mathcal{S}_1}{1!}} \; e^{\frac{(iz)^4 \mathcal{S}_2}{2!}}\; e^{\frac{(iz)^6 \mathcal{S}_3}{3!}} \cdots \cdots \nonumber \\
    &=& \mathcal{J}_0 \; \exp \left[ \sum_{k=1}^{\infty} \frac{(iz)^{2k} \mathcal{S}_k}{k!} \right], \label{eq:sumJ3}
\end{eqnarray}
where we have defined $\mathcal{S}_k$ as
\begin{equation}
    \mathcal{S}_k= \sum_{n^{(k)}=0}^{\infty} \frac{(iz)^{n^{(k)}}}{n^{(k)}!} \mathcal{D}_{n^{(k)}}^{(k)}. \label{eq;sumSk}
\end{equation}

Recall that $\mathcal{D}_{n^{(k)}}^{(k)}$ denotes the collection of irreducible diagrams containing $k$ pairs of $(B_1, B_2)$ together with $n^{(k)}$ insertions of $A'$ blobs. Each such diagram therefore corresponds to a $(2k+n^{(k)})$-body contraction. Our goal is to reorganize the sum appearing in the exponent of equation \eqref{eq:sumJ3} as a sum over $n$-body contractions, $\mathcal{D}'_n$. This can be expressed as follows:
\begin{equation}
    \mathcal{J}=\mathcal{J}_0 \; \exp \left[ \sum_{n=2}^{\infty} \;(iz)^n \; \mathcal{D}'_{n} \right]. \label{eq:sumJfinal}
\end{equation}
where the explicit expressions for $\mathcal{D}_{n}'$ for the first few values of $n$ are given by:
\begin{eqnarray}
    \mathcal{D}'_{2}&=&\frac{1}{1!0!} \mathcal{D}^{(1)}_{0} \\
    \mathcal{D}'_{3}&=&\frac{1}{1! 1!} \mathcal{D}_{1}^{(1)} \\
    \mathcal{D}'_{4}&=&\frac{1}{1! 2!} \mathcal{D}_{2}^{(1)}+\frac{1}{2!0!} \mathcal{D}_{0}^{(2)} \\
    \mathcal{D}'_{5}&=&\frac{1}{1! 3!} \mathcal{D}_{3}^{(1)}+\frac{1}{2!1!} \mathcal{D}_{1}^{(2)} ,\\
\end{eqnarray}
or diagrmmatically: 
\begin{eqnarray}
        \mathcal{D}_{2}'&=& 
        \begin{tikzpicture}[baseline={([yshift=-.5ex]current bounding box.center)}]
\coordinate (P) at (0,0);
\coordinate (Q) at (40pt,0);

\draw (P) circle (10pt);
\draw (Q) circle (10pt); 

\coordinate (P1) at (-4pt,0);
\coordinate (P2) at (4pt,0);
\coordinate (Q1) at (36pt,0);
\coordinate (Q2) at (44pt,0);

\draw[black] (P1) circle (1.7pt);
\filldraw (P2) circle (1.7pt);
\draw[black] (Q1) circle (1.7pt);
\filldraw (Q2) circle (1.7pt);

\draw[black,out = 60,in = 120, looseness = 1.5] (P2) to (Q1);
\draw[dotted,out = -60,in = -120, looseness = 1.5] (P2) to (Q1);
\end{tikzpicture} \\
\mathcal{D}_{3}' &=&
\begin{tikzpicture}[baseline={([yshift=-.5ex]current bounding box.center)}]
\coordinate (P) at (0,0);
\coordinate (Q) at (30pt,0);
\coordinate (R) at (60pt,0);

\draw (P) circle (10pt);
\draw (Q) circle (10pt); 
\draw (R) circle (10pt);

\coordinate (P1) at (-4pt,0);
\coordinate (P2) at (4pt,0);
\coordinate (Q1) at (26pt,0);
\coordinate (Q2) at (34pt,0);
\coordinate (R1) at (56pt,0);
\coordinate (R2) at (64pt,0);

\draw[black] (P1) circle (1.7pt);
\filldraw (P2) circle (1.7pt);
\draw[black] (Q1) circle (1.7pt);
\filldraw (Q2) circle (1.7pt);
\draw[black] (R1) circle (1.7pt);
\filldraw (R2) circle (1.7pt);

\draw[black,out = 60,in = 120, looseness = 2.0] (P2) to (R1);
\draw[black,out = 120,in = 60, min distance = 1.2cm] (Q1) to (Q2);

\draw[dotted,out = -120,in = -60, min distance = 1.0cm] (P2) to (Q1);
\draw[dotted,out = -120,in = -60, min distance = 1.0cm] (Q2) to (R1); 
\end{tikzpicture}  \label{eq:D3prime}
\end{eqnarray}
and so on. Now, all that remains is to compute $\mathcal{J}_0$. Recall that $\mathcal{D}_{n'}^{(A')}$ is the set of all possible kinds of contracted diagrams one can build out of $n'$ number of $A'$ blobs. Any such diagram can be obtained by putting together $p_1$ number of $1-$blob contractions, $p_2$ number of $2-$blob contractions,...., $p_{n'}$ number of $n'-$blob contractions with the constraint: $\sum_{i=1}^{n'} i\;p_i=n'$. Here all the $i-$blob contractions are assumed to be fully connected. We will denote them with $\mathcal{D}_{i,0}$. Any given configuration $\{ p_i \}$ can be obtained in $\frac{n'!}{\prod_{i=1}^{n'} (i!)^{p_i}\; p_i!}$ number of ways. If we sum over all possible configurations of $\{ p_i\}$ subject to the constraint mentioned above, we get the entire set $\mathcal{D}_{n'}^{(A')}$:
\begin{equation}
    \mathcal{D}_{n'}^{(A')}=\sum_{\{p_i\}} ' \frac{n'!}{(1!)^{p_1} (2!)^{p_2}....(n'!)^{p_{n'}}}\; \frac{1}{p_1! p_2!....p_{n'}!}\times \mathcal{D}_{1,0}^{p_1}\; \mathcal{D}_{2,0}^{p_2}....\; \mathcal{D}_{n',0}^{p_{n'}} .\label{eq:allconn0}
\end{equation}
Here we draw the first few connected diagrams $\mathcal{D}_{i,0}$.
\begin{eqnarray}
    \mathcal{D}_{1,0}&= &
    \begin{tikzpicture}[baseline={([yshift=-.5ex]current bounding box.center)}]
  \coordinate (P) at (0,0);
  \draw (P) circle (10pt);

  \coordinate (P1) at (-0.15,0);
  \coordinate (P2) at (0.15,0);
  \draw[black] (P1) circle (1.7pt);
  \filldraw (P2) circle (1.7pt);

  \draw[black, out=120, in=60, min distance=1.5cm] (P1) to (P2);
  \draw[dotted, out=-120, in=-60, min distance=1.5cm] (P1) to (P2);
\end{tikzpicture}  \\
\mathcal{D}_{2,0}&=&
\begin{tikzpicture}[baseline={([yshift=-.5ex]current bounding box.center)}]
\coordinate (P) at (0,0);
\coordinate (Q) at (40pt,0);

\draw (P) circle (10pt);
\draw (Q) circle (10pt);

\coordinate (P1) at (-4pt,0);
\coordinate (P2) at (4pt,0);
\coordinate (Q1) at (36pt,0);
\coordinate (Q2) at (44pt,0);

\draw[black] (P1) circle (1.7pt);
\filldraw (P2) circle (1.7pt);
\draw[black] (Q1) circle (1.7pt);
\filldraw (Q2) circle (1.7pt);

\draw[black,out = 60,in = 120, looseness = 1.5] (P1) to (Q2);
\draw[black,out = 60,in = 120, looseness =1.5] (P2) to (Q1);

\draw[dotted,out = -60,in = -120, looseness = 1.5] (P1) to (Q2);
\draw[dotted,out = -60,in = -120, looseness =1.5] (P2) to (Q1); 
\end{tikzpicture} \; +
\begin{tikzpicture}[baseline={([yshift=-.5ex]current bounding box.center)}]
\coordinate (P) at (0,0);
\coordinate (Q) at (40pt,0);

\draw (P) circle (10pt);
\draw (Q) circle (10pt);

\coordinate (P1) at (-4pt,0);
\coordinate (P2) at (4pt,0);
\coordinate (Q1) at (36pt,0);
\coordinate (Q2) at (44pt,0);

\draw[black] (P1) circle (1.7pt);
\filldraw (P2) circle (1.7pt);
\draw[black] (Q1) circle (1.7pt);
\filldraw (Q2) circle (1.7pt);

\draw[black,out = 60,in = 120, looseness = 1.5] (P1) to (Q2);
\draw[black,out = 60,in = 120, looseness =1.5] (P2) to (Q1);

 \draw[dotted, out=-120, in=-60, min distance=1.5cm] (P1) to (P2);
  \draw[dotted, out=-120, in=-60, min distance=1.5cm] (Q1) to (Q2);
\end{tikzpicture} \;+
\begin{tikzpicture}[baseline={([yshift=-.5ex]current bounding box.center)}]
\coordinate (P) at (0,0);
\coordinate (Q) at (40pt,0);

\draw (P) circle (10pt);
\draw (Q) circle (10pt);

\coordinate (P1) at (-4pt,0);
\coordinate (P2) at (4pt,0);
\coordinate (Q1) at (36pt,0);
\coordinate (Q2) at (44pt,0);

\draw[black] (P1) circle (1.7pt);
\filldraw (P2) circle (1.7pt);
\draw[black] (Q1) circle (1.7pt);
\filldraw (Q2) circle (1.7pt);

\draw[dotted,out = -60,in = -120, looseness = 1.5] (P1) to (Q2);
\draw[dotted,out = -60,in = -120, looseness =1.5] (P2) to (Q1);

 \draw[black, out=120, in=60, min distance=1.5cm] (P1) to (P2);
  \draw[black, out=120, in=60, min distance=1.5cm] (Q1) to (Q2);
\end{tikzpicture} \\
\mathcal{D}_{3,0}&=& 
\begin{tikzpicture}[baseline={([yshift=-.5ex]current bounding box.center)}]
\coordinate (P) at (0,0);
\coordinate (Q) at (30pt,0);
\coordinate (R) at (60pt,0);

\draw (P) circle (10pt);
\draw (Q) circle (10pt); 
\draw (R) circle (10pt);

\coordinate (P1) at (-4pt,0);
\coordinate (P2) at (4pt,0);
\coordinate (Q1) at (26pt,0);
\coordinate (Q2) at (34pt,0);
\coordinate (R1) at (56pt,0);
\coordinate (R2) at (64pt,0);

\draw[black] (P1) circle (1.7pt);
\filldraw (P2) circle (1.7pt);
\draw[black] (Q1) circle (1.7pt);
\filldraw (Q2) circle (1.7pt);
\draw[black] (R1) circle (1.7pt);
\filldraw (R2) circle (1.7pt);

\draw[black,out = 60,in = 120, looseness = 1.5] (P1) to (R2);
\draw[black,out = 60,in = 120, looseness =1.5] (P2) to (Q1);
\draw[black,out = 60,in = 120, looseness =1.5] (Q2) to (R1);

\draw[dotted,out = -60,in = -120, looseness = 1.5] (P1) to (R2);
\draw[dotted,out = -60,in = -120, looseness =1.5] (P2) to (Q1); 
\draw[dotted,out = -60,in = -120, looseness =1.5] (Q2) to (R1); 
\end{tikzpicture} \;+
\begin{tikzpicture}[baseline={([yshift=-.5ex]current bounding box.center)}]
\coordinate (P) at (0,0);
\coordinate (Q) at (30pt,0);
\coordinate (R) at (60pt,0);

\draw (P) circle (10pt);
\draw (Q) circle (10pt); 
\draw (R) circle (10pt);

\coordinate (P1) at (-4pt,0);
\coordinate (P2) at (4pt,0);
\coordinate (Q1) at (26pt,0);
\coordinate (Q2) at (34pt,0);
\coordinate (R1) at (56pt,0);
\coordinate (R2) at (64pt,0);

\draw[black] (P1) circle (1.7pt);
\filldraw (P2) circle (1.7pt);
\draw[black] (Q1) circle (1.7pt);
\filldraw (Q2) circle (1.7pt);
\draw[black] (R1) circle (1.7pt);
\filldraw (R2) circle (1.7pt);

\draw[black,out = 60,in = 120, looseness = 1.5] (P1) to (R2);
\draw[black,out = 60,in = 120, looseness =1.5] (P2) to (Q1);
\draw[black,out = 60,in = 120, looseness =1.5] (Q2) to (R1);

\draw[dotted, out=-120, in=-60, min distance=1.5cm] (P1) to (P2);
\draw[dotted, out=-120, in=-60, min distance=1.5cm] (Q1) to (Q2);
\draw[dotted, out=-120, in=-60, min distance=1.5cm] (R1) to (R2);

\end{tikzpicture} +
\begin{tikzpicture}[baseline={([yshift=-.5ex]current bounding box.center)}]
\coordinate (P) at (0,0);
\coordinate (Q) at (30pt,0);
\coordinate (R) at (60pt,0);

\draw (P) circle (10pt);
\draw (Q) circle (10pt); 
\draw (R) circle (10pt);

\coordinate (P1) at (-4pt,0);
\coordinate (P2) at (4pt,0);
\coordinate (Q1) at (26pt,0);
\coordinate (Q2) at (34pt,0);
\coordinate (R1) at (56pt,0);
\coordinate (R2) at (64pt,0);

\draw[black] (P1) circle (1.7pt);
\filldraw (P2) circle (1.7pt);
\draw[black] (Q1) circle (1.7pt);
\filldraw (Q2) circle (1.7pt);
\draw[black] (R1) circle (1.7pt);
\filldraw (R2) circle (1.7pt);

\draw[dotted,out = -60,in = -120, looseness = 1.5] (P1) to (R2);
\draw[dotted,out = -60,in = -120, looseness =1.5] (P2) to (Q1);
\draw[dotted,out = -60,in = -120, looseness =1.5] (Q2) to (R1);

\draw[black, out=120, in=60, min distance=1.5cm] (P1) to (P2);
\draw[black, out=120, in=60, min distance=1.5cm] (Q1) to (Q2);
\draw[black, out=120, in=60, min distance=1.5cm] (R1) to (R2);

\end{tikzpicture} \nonumber \\
&+& 3 \times
\begin{tikzpicture}[baseline={([yshift=-.5ex]current bounding box.center)}]
\coordinate (P) at (0,0);
\coordinate (Q) at (30pt,0);
\coordinate (R) at (60pt,0);

\draw (P) circle (10pt);
\draw (Q) circle (10pt); 
\draw (R) circle (10pt);

\coordinate (P1) at (-4pt,0);
\coordinate (P2) at (4pt,0);
\coordinate (Q1) at (26pt,0);D
\coordinate (Q2) at (34pt,0);
\coordinate (R1) at (56pt,0);3
\coordinate (R2) at (64pt,0);

\draw[black] (P1) circle (1.7pt);
\filldraw (P2) circle (1.7pt);
\draw[black] (Q1) circle (1.7pt);
\filldraw (Q2) circle (1.7pt);
\draw[black] (R1) circle (1.7pt);
\filldraw (R2) circle (1.7pt);

\draw[black,out = 60,in = 120, looseness = 1.5] (P1) to (R2);
\draw[black,out = 60,in = 120, looseness =1.5] (P2) to (Q1);
\draw[black,out = 60,in = 120, looseness =1.5] (Q2) to (R1);

\draw[dotted,out = -60,in = -120, looseness = 2.0] (P1) to (Q2);
\draw[dotted,out = -60,in = -120, looseness =2.0] (P2) to (Q1); 
\draw[dotted,out = -120,in = -60, min distance =1.5cm] (R1) to (R2); 
\end{tikzpicture} + 6 \times
\begin{tikzpicture}[baseline={([yshift=-.5ex]current bounding box.center)}]
\coordinate (P) at (0,0);
\coordinate (Q) at (30pt,0);
\coordinate (R) at (60pt,0);

\draw (P) circle (10pt);
\draw (Q) circle (10pt); 
\draw (R) circle (10pt);

\coordinate (P1) at (-4pt,0);
\coordinate (P2) at (4pt,0);
\coordinate (Q1) at (26pt,0);D
\coordinate (Q2) at (34pt,0);
\coordinate (R1) at (56pt,0);3
\coordinate (R2) at (64pt,0);

\draw[black] (P1) circle (1.7pt);
\filldraw (P2) circle (1.7pt);
\draw[black] (Q1) circle (1.7pt);
\filldraw (Q2) circle (1.7pt);
\draw[black] (R1) circle (1.7pt);
\filldraw (R2) circle (1.7pt);

\draw[black,out = 60,in = 120, looseness = 2.0] (P1) to (Q2);
\draw[black,out = 60,in = 120, looseness =2.0] (P2) to (Q1);
\draw[black,out = 120,in = 60, min distance=1.5cm] (R1) to (R2);

\draw[dotted,out = -60,in = -120, looseness = 1.5] (Q1) to (R2);
\draw[dotted,out = -60,in = -120, looseness =1.5] (Q2) to (R1); 
\draw[dotted,out = -120,in = -60, min distance=1.5cm] (P1) to (P2);
\end{tikzpicture} + 3 \times 
\begin{tikzpicture}[baseline={([yshift=-.5ex]current bounding box.center)}]
\coordinate (P) at (0,0);
\coordinate (Q) at (30pt,0);
\coordinate (R) at (60pt,0);

\draw (P) circle (10pt);
\draw (Q) circle (10pt); 
\draw (R) circle (10pt);

\coordinate (P1) at (-4pt,0);
\coordinate (P2) at (4pt,0);
\coordinate (Q1) at (26pt,0);D
\coordinate (Q2) at (34pt,0);
\coordinate (R1) at (56pt,0);3
\coordinate (R2) at (64pt,0);

\draw[black] (P1) circle (1.7pt);
\filldraw (P2) circle (1.7pt);
\draw[black] (Q1) circle (1.7pt);
\filldraw (Q2) circle (1.7pt);
\draw[black] (R1) circle (1.7pt);
\filldraw (R2) circle (1.7pt);

\draw[dotted,out = -60,in = -120, looseness = 1.5] (P1) to (R2);
\draw[dotted,out = -60,in = -120, looseness =1.5] (P2) to (Q1);
\draw[dotted,out = -60,in = -120, looseness =1.5] (Q2) to (R1);

\draw[black,out = 60,in = 120, looseness = 2.0] (P1) to (Q2);
\draw[black,out = 60,in = 120, looseness =2.0] (P2) to (Q1); 
\draw[black,out = 120,in = 60, min distance =1.5cm] (R1) to (R2); 
\end{tikzpicture} \label{eq:D3,0}
\end{eqnarray}

Now we notice again that in \eqref{eq:sumJ0}, we are summing over all possible values of $n'$. Hence the restricted $p-$sums in \eqref{eq:allconn0} can be performed independently. And we get,
\begin{eqnarray}
    \mathcal{J}_0&=&\sum_{p_1=0}^{\infty}\frac{1}{p_1!}\bigg(\frac{(iz)}{1!}\mathcal{D}_{1,0}\bigg)^{p_1} \; \sum_{p_2=0}^{\infty}\frac{1}{p_2!}\bigg(\frac{(iz)^2}{2!}\mathcal{D}_{2,0}\bigg)^{p_2}\cdots\nonumber \\
    &=& \exp \bigg( \sum_{p=1}^{\infty} \frac{(iz)^p}{p!} \; \mathcal{D}_{p,0} \bigg) .\label{eq:I0final}
\end{eqnarray}
We have computed all the pieces we require to write down a closed form expression for $\overline{e^{it\;W_{\mathcal{U.B}}}}$. From \eqref{eq:resum} and using \eqref{eq:sumJfinal}, \eqref{eq:I0final} we finally get:
\begin{equation} 
   \mathcal{I}(q,p;z) =  \exp \bigg( \sum_{n=1}^{\infty} \;(iz)^n \; \mathcal{D}_{n}\bigg), \label{eq:Haarfinal}
\end{equation}

\section{Higher order terms}
\label{sec:higherorder}

While performing the Gaussian approximation in Eq. \eqref{eq:Gaussian}, we argued that higher-order contributions -- cubic, quartic, and beyond -- are suppressed. In this appendix, we explain why this happens. Specifically, for $q \neq 0$, we demonstrate that the ratio $\frac{\mathcal{D}_3 \mathcal{D}_1}{\mathcal{D}_2^2}$ appearing in Eq.\eqref{eq:wignerabs1} is indeed suppressed at early as well as late times. Building on this analysis, we will then identify a general pattern indicating that higher-order terms, such as the quartic contribution, are similarly suppressed. Next, we will consider the $q=0$ case and show how in this setting, the Gaussian approximation can fail.

\noindent\underline{\textbf{$q\neq0$ case}:}\quad We begin by evaluating $\mathcal{D}_3$. From equation \eqref{eq:HigherDia}, it follows that $\mathcal{D}_3 = \frac{\mathcal{D}_{3,0}}{3!} + \mathcal{D}_3'$. Using \eqref{eq:D3,0} and \eqref{eq:D3prime}, we find that the diagrams contributing for the case $q \neq 0$ are given by,
\begin{eqnarray}
    \mathcal{D}_3&=& \frac{1}{3!}\times
    \begin{tikzpicture}[baseline={([yshift=-.5ex]current bounding box.center)}]
\coordinate (P) at (0,0);
\coordinate (Q) at (30pt,0);
\coordinate (R) at (60pt,0);

\draw (P) circle (10pt);
\draw (Q) circle (10pt); 
\draw (R) circle (10pt);

\coordinate (P1) at (-4pt,0);
\coordinate (P2) at (4pt,0);
\coordinate (Q1) at (26pt,0);
\coordinate (Q2) at (34pt,0);
\coordinate (R1) at (56pt,0);
\coordinate (R2) at (64pt,0);

\draw[black] (P1) circle (1.7pt);
\filldraw (P2) circle (1.7pt);
\draw[black] (Q1) circle (1.7pt);
\filldraw (Q2) circle (1.7pt);
\draw[black] (R1) circle (1.7pt);
\filldraw (R2) circle (1.7pt);

\draw[black,out = 60,in = 120, looseness = 1.5] (P1) to (R2);
\draw[black,out = 60,in = 120, looseness =1.5] (P2) to (Q1);
\draw[black,out = 60,in = 120, looseness =1.5] (Q2) to (R1);

\draw[dotted,out = -60,in = -120, looseness = 1.5] (P1) to (R2);
\draw[dotted,out = -60,in = -120, looseness =1.5] (P2) to (Q1); 
\draw[dotted,out = -60,in = -120, looseness =1.5] (Q2) to (R1); 
\end{tikzpicture} \;+  \frac{1}{3!}\times
\begin{tikzpicture}[baseline={([yshift=-.5ex]current bounding box.center)}]
\coordinate (P) at (0,0);
\coordinate (Q) at (30pt,0);
\coordinate (R) at (60pt,0);

\draw (P) circle (10pt);
\draw (Q) circle (10pt); 
\draw (R) circle (10pt);

\coordinate (P1) at (-4pt,0);
\coordinate (P2) at (4pt,0);
\coordinate (Q1) at (26pt,0);
\coordinate (Q2) at (34pt,0);
\coordinate (R1) at (56pt,0);
\coordinate (R2) at (64pt,0);

\draw[black] (P1) circle (1.7pt);
\filldraw (P2) circle (1.7pt);
\draw[black] (Q1) circle (1.7pt);
\filldraw (Q2) circle (1.7pt);
\draw[black] (R1) circle (1.7pt);
\filldraw (R2) circle (1.7pt);

\draw[black,out = 60,in = 120, looseness = 1.5] (P1) to (R2);
\draw[black,out = 60,in = 120, looseness =1.5] (P2) to (Q1);
\draw[black,out = 60,in = 120, looseness =1.5] (Q2) to (R1);

\draw[dotted, out=-120, in=-60, min distance=1.5cm] (P1) to (P2);
\draw[dotted, out=-120, in=-60, min distance=1.5cm] (Q1) to (Q2);
\draw[dotted, out=-120, in=-60, min distance=1.5cm] (R1) to (R2);

\end{tikzpicture} +  \frac{1}{3!}\times
\begin{tikzpicture}[baseline={([yshift=-.5ex]current bounding box.center)}]
\coordinate (P) at (0,0);
\coordinate (Q) at (30pt,0);
\coordinate (R) at (60pt,0);

\draw (P) circle (10pt);
\draw (Q) circle (10pt); 
\draw (R) circle (10pt);

\coordinate (P1) at (-4pt,0);
\coordinate (P2) at (4pt,0);
\coordinate (Q1) at (26pt,0);
\coordinate (Q2) at (34pt,0);
\coordinate (R1) at (56pt,0);
\coordinate (R2) at (64pt,0);

\draw[black] (P1) circle (1.7pt);
\filldraw (P2) circle (1.7pt);
\draw[black] (Q1) circle (1.7pt);
\filldraw (Q2) circle (1.7pt);
\draw[black] (R1) circle (1.7pt);
\filldraw (R2) circle (1.7pt);

\draw[dotted,out = -60,in = -120, looseness = 1.5] (P1) to (R2);
\draw[dotted,out = -60,in = -120, looseness =1.5] (P2) to (Q1);
\draw[dotted,out = -60,in = -120, looseness =1.5] (Q2) to (R1);

\draw[black, out=120, in=60, min distance=1.5cm] (P1) to (P2);
\draw[black, out=120, in=60, min distance=1.5cm] (Q1) to (Q2);
\draw[black, out=120, in=60, min distance=1.5cm] (R1) to (R2);

\end{tikzpicture} \nonumber \\
&+& \frac{1}{2} \times
\begin{tikzpicture}[baseline={([yshift=-.5ex]current bounding box.center)}]
\coordinate (P) at (0,0);
\coordinate (Q) at (30pt,0);
\coordinate (R) at (60pt,0);

\draw (P) circle (10pt);
\draw (Q) circle (10pt); 
\draw (R) circle (10pt);

\coordinate (P1) at (-4pt,0);
\coordinate (P2) at (4pt,0);
\coordinate (Q1) at (26pt,0);D
\coordinate (Q2) at (34pt,0);
\coordinate (R1) at (56pt,0);3
\coordinate (R2) at (64pt,0);

\draw[black] (P1) circle (1.7pt);
\filldraw (P2) circle (1.7pt);
\draw[black] (Q1) circle (1.7pt);
\filldraw (Q2) circle (1.7pt);
\draw[black] (R1) circle (1.7pt);
\filldraw (R2) circle (1.7pt);

\draw[black,out = 60,in = 120, looseness = 1.5] (P1) to (R2);
\draw[black,out = 60,in = 120, looseness =1.5] (P2) to (Q1);
\draw[black,out = 60,in = 120, looseness =1.5] (Q2) to (R1);

\draw[dotted,out = -60,in = -120, looseness = 2.0] (P1) to (Q2);
\draw[dotted,out = -60,in = -120, looseness =2.0] (P2) to (Q1); 
\draw[dotted,out = -120,in = -60, min distance =1.5cm] (R1) to (R2); 
\end{tikzpicture} + 
\begin{tikzpicture}[baseline={([yshift=-.5ex]current bounding box.center)}]
\coordinate (P) at (0,0);
\coordinate (Q) at (30pt,0);
\coordinate (R) at (60pt,0);

\draw (P) circle (10pt);
\draw (Q) circle (10pt); 
\draw (R) circle (10pt);

\coordinate (P1) at (-4pt,0);
\coordinate (P2) at (4pt,0);
\coordinate (Q1) at (26pt,0);D
\coordinate (Q2) at (34pt,0);
\coordinate (R1) at (56pt,0);3
\coordinate (R2) at (64pt,0);

\draw[black] (P1) circle (1.7pt);
\filldraw (P2) circle (1.7pt);
\draw[black] (Q1) circle (1.7pt);
\filldraw (Q2) circle (1.7pt);
\draw[black] (R1) circle (1.7pt);
\filldraw (R2) circle (1.7pt);

\draw[black,out = 60,in = 120, looseness = 2.0] (P1) to (Q2);
\draw[black,out = 60,in = 120, looseness =2.0] (P2) to (Q1);
\draw[black,out = 120,in = 60, min distance=1.5cm] (R1) to (R2);

\draw[dotted,out = -60,in = -120, looseness = 1.5] (Q1) to (R2);
\draw[dotted,out = -60,in = -120, looseness =1.5] (Q2) to (R1); 
\draw[dotted,out = -120,in = -60, min distance=1.5cm] (P1) to (P2);
\end{tikzpicture} + \frac{1}{2} \times 
\begin{tikzpicture}[baseline={([yshift=-.5ex]current bounding box.center)}]
\coordinate (P) at (0,0);
\coordinate (Q) at (30pt,0);
\coordinate (R) at (60pt,0);

\draw (P) circle (10pt);
\draw (Q) circle (10pt); 
\draw (R) circle (10pt);

\coordinate (P1) at (-4pt,0);
\coordinate (P2) at (4pt,0);
\coordinate (Q1) at (26pt,0);D
\coordinate (Q2) at (34pt,0);
\coordinate (R1) at (56pt,0);3
\coordinate (R2) at (64pt,0);

\draw[black] (P1) circle (1.7pt);
\filldraw (P2) circle (1.7pt);
\draw[black] (Q1) circle (1.7pt);
\filldraw (Q2) circle (1.7pt);
\draw[black] (R1) circle (1.7pt);
\filldraw (R2) circle (1.7pt);

\draw[dotted,out = -60,in = -120, looseness = 1.5] (P1) to (R2);
\draw[dotted,out = -60,in = -120, looseness =1.5] (P2) to (Q1);
\draw[dotted,out = -60,in = -120, looseness =1.5] (Q2) to (R1);

\draw[black,out = 60,in = 120, looseness = 2.0] (P1) to (Q2);
\draw[black,out = 60,in = 120, looseness =2.0] (P2) to (Q1); 
\draw[black,out = 120,in = 60, min distance =1.5cm] (R1) to (R2); 
\end{tikzpicture} \nonumber \\
&+& \begin{tikzpicture}[baseline={([yshift=-.5ex]current bounding box.center)}]
\coordinate (P) at (0,0);
\coordinate (Q) at (30pt,0);
\coordinate (R) at (60pt,0);

\draw (P) circle (10pt);
\draw (Q) circle (10pt); 
\draw (R) circle (10pt);

\coordinate (P1) at (-4pt,0);
\coordinate (P2) at (4pt,0);
\coordinate (Q1) at (26pt,0);
\coordinate (Q2) at (34pt,0);
\coordinate (R1) at (56pt,0);
\coordinate (R2) at (64pt,0);

\draw[black] (P1) circle (1.7pt);
\filldraw (P2) circle (1.7pt);
\draw[black] (Q1) circle (1.7pt);
\filldraw (Q2) circle (1.7pt);
\draw[black] (R1) circle (1.7pt);
\filldraw (R2) circle (1.7pt);

\draw[black,out = 60,in = 120, looseness = 2.0] (P2) to (R1);
\draw[black,out = 120,in = 60, min distance = 1.2cm] (Q1) to (Q2);

\draw[dotted,out = -120,in = -60, min distance = 1.0cm] (P2) to (Q1);
\draw[dotted,out = -120,in = -60, min distance = 1.0cm] (Q2) to (R1); 
\end{tikzpicture}  
\end{eqnarray}
We now proceed to evaluate these diagrams. For this purpose, it is convenient to introduce two projectors: 
$ P_{0} = |\psi_{0}\rangle \langle \psi_{0}|$
and $ P_{\perp} = \mathbb{1} - P_{0}$. Expressed in terms of these projectors, and retaining only the leading contributions of the relevant Weingarten functions, we obtain
\begin{multline}
    \mathcal{D}_3 \simeq \frac{1}{D^6} \Bigg[ -\frac{1}{3}\text{Tr}\left( (\rho_R P_{\perp})^3 \right)+ \frac{1}{3D^2}\left(\text{Tr} (\rho_R P_{\perp})\right)^3 \\
   + \frac{3}{2D} \text{Tr}\left( (\rho_R P_{\perp})^2 \right) \; \text{Tr} \left( \rho_R P_{\perp} \right)-\text{Tr}\left( (\rho_R P_{\perp})^2\;\rho_R P_0 \right)  \Bigg].
\end{multline}
Note that $\rho_R(t)$ is time dependent. We have already evaluated $\mathcal{D}_2$ for $q\neq0$ in equation \eqref{eq:alphaneq0}. For convenience, we reproduce it here:
\begin{equation}
    2\mathcal{D}_2\simeq\frac{1}{D^4} \left(D\; \text{Tr}\left( (\rho_R P_{\perp})^2 \right) - \left(\text{Tr} (\rho_R P_{\perp})\right)^2+2D\;\text{Tr}\left( \rho_R P_{\perp}\rho_R P_0 \right) \right).
\end{equation}
And for $\mathcal{D}_1$ we have,
\begin{equation}
    \mathcal{D}_1= \frac{1}{D^2} \left( 1- \langle \psi_0|\rho_R|\psi_0 \rangle \right) .
\end{equation}
Now, we can write:
\begin{eqnarray}
    \text{Tr}\left( (\rho_R P_{\perp})^3 \right)&=&\text{Tr}\left( (\rho_R)^3 \right)-3\; \langle \psi_0|\rho_R ^3|\psi_0\rangle + 3\; \langle \psi_0|\rho_R |\psi_0\rangle \; \langle \psi_0|\rho_R^2|\psi_0\rangle-\langle\psi_0|\rho_R|\psi_0\rangle^3 \nonumber \\
    \text{Tr}\left( (\rho_R P_{\perp})^2 \right)&=& \text{Tr}\left( (\rho_R)^2 \right)- 2 \; \langle \psi_0|\rho_R^2|\psi_0\rangle +\langle \psi_0|\rho_R | \psi_0\rangle^2 \nonumber \\
    \text{Tr}\left( \rho_R P_{\perp} \right) &=&1-\langle\psi_0|\rho_R| \psi_0 \rangle \\
    \text{Tr}\left( (\rho_R P_{\perp})^2\;\rho_R P_0 \right) &=& \langle \psi_0|\rho_R^3|\psi_0\rangle-2 \; \langle \psi_0|\rho_R^2|\psi_0\rangle \langle \psi_0|\rho_R|\psi_0\rangle + \langle \psi_0|\rho_R|\psi_0\rangle^3 \nonumber\\
    \text{Tr}\left( \rho_R P_{\perp}\;\rho_R P_0 \right)&=& \langle \psi_0|\rho_R^2|\psi_0\rangle- \langle \psi_0|\rho_R|\psi_0\rangle^2.\nonumber
\end{eqnarray}
We now redefine all these invariants by scaling out a suitable factor of $e^{-S_2^R}$:
\begin{eqnarray}
    \text{Tr}\left( (\rho_R)^3 \right) &=& e^{-2S_2^R} \hat{S}_3 ,\\
    \langle \psi_0|\rho_R^3|\psi_0\rangle &=& e^{-3 S_2^R} \hat{h}, \\
    \langle \psi_0|\rho_R^2|\psi_0\rangle &=& e^{-2 S_2^R} \hat{g} ,\\
    \langle \psi_0|\rho_R|\psi_0\rangle &=& e^{- S_2^R} \hat{S}_0.
\end{eqnarray}
All these hatted quantities remain $\mathcal{O}(1)$ at all times. With these redefinitions, the quantities $\mathcal{D}_3$, $\mathcal{D}_2$, and $\mathcal{D}_1$ can be expressed as follows:
\begin{multline}
    \mathcal{D}_3\simeq\frac{1}{D^6} \; e^{-3S_2^R} \Bigg[ -\frac{1}{3} \left(e^{S_2^R}\hat{S}_3-3\hat{h}+3 \hat{g}\hat{S}_0-\hat{S}_0^3 \right)+\frac{1}{3D^2} \left( e^{S_2^R}-\hat{S}_0 \right)^3 \\
    +\frac{3}{2D}\left( e^{S_2^R}-2\hat{g}+\hat{S}_0^2 \right) \left( e^{S_2^R}-\hat{S}_0 \right)-\left( \hat{h}-2\hat{g} \hat{S}_0+\hat{S}_0^3\right) \Bigg] \label{eq:D3redef}
\end{multline}
\begin{eqnarray}
    \mathcal{D}_2 &\simeq & \frac{1}{D^4} \frac{1}{2!} \; e^{-2S_2^R} \left[ D\left( e^{S_2^R}-2\hat{g}+\hat{S}_0^2 \right)+2D\left( \hat{g}-\hat{S}_0\right) -\left( e^{S_2^R}-\hat{S}_0 \right)^2 \right] \label{eq:D2redef} \\
    \mathcal{D}_1 &=& \frac{1}{D^2} e^{-S_2^R} \left[ e^{S_2^R}-\hat{S}_0\right] \label{eq:D1redef}
\end{eqnarray}
At small times, $e^{S_2^R}$ is of order unity. Consequently, at early times, the quantities inside the square brackets in equations \eqref{eq:D3redef} and \eqref{eq:D1redef} are also of order one, while the quantity inside the square bracket in equation \eqref{eq:D2redef} is of order $D$. As a result, we obtain
\[
\mathcal{D}_3 \simeq \frac{1}{D^6} e^{-3S_2^R} \times \mathcal{O}(1), 
\qquad 
\mathcal{D}_2 \simeq \frac{1}{D^4} e^{-2S_2^R} \times D \times \mathcal{O}(1), 
\qquad 
\mathcal{D}_1 \simeq \frac{1}{D^2} e^{-S_2^R} \times \mathcal{O}(1).
\]
Thus, we find that
\begin{equation}
     \frac{\mathcal{D}_3 \mathcal{D}_1}{\mathcal{D}_2^2} \simeq \mathcal{O}\!\left(\tfrac{1}{D^2}\right)  \qquad ,\qquad \cdots\;\;(\text{small times}).
\end{equation}
As time progresses, $e^{S_2^R}$ continues to grow. At late times, $e^{S_2^R}$ becomes large and we consider the limit where $D \to \infty$, $e^{S_2^R} \to \infty$, with the ratio $\tfrac{D}{e^{S_2^R}} = \hat{r}$ held fixed. In this regime, we find from \eqref{eq:D3redef}, \eqref{eq:D2redef} and \eqref{eq:D1redef},
\begin{eqnarray}
    \mathcal{D}_3 &\simeq & \frac{1}{D^6} \times e^{-2S_2^R}\times \left( -\frac{1}{3}\hat{S}_3+\frac{1}{3 \;\hat{r}^2}+\frac{3}{2 \;\hat{r}} \right) \\
    \mathcal{D}_2 &\simeq& \frac{1}{D^4} \times \left(\hat{r}-1 \right) \\
    \mathcal{D}_1 &\simeq& \frac{1}{D^2} 
\end{eqnarray}
Hence at late times, we get,
\begin{equation}
     \frac{\mathcal{D}_3 \mathcal{D}_1}{\mathcal{D}_2^2} \simeq e^{-2S_2^R} \times f_1(\hat{r})  \qquad ,\qquad \cdots (\text{late times}).
\end{equation}
where $f_1(\hat{r})$ is $\mathcal{O}(1)$. Thus, we show that the cubic term is suppressed for $q\neq0$, either by powers of $D$ at early times or by powers of $e^{S_2^R}$ at late times. At this point, the trend becomes evident. The quartic term in \eqref{eq:wignerabs1} appears with the prefactor $\frac{\mathcal{D}_4 \mathcal{D}_1^2}{\mathcal{D}_2^3}$. A similar evaluation of the relevant diagrams contributing to $\mathcal{D}_4$ shows that, for small times, $\mathcal{D}_4 \simeq \tfrac{1}{D^8} e^{-4S_2^R} \times D \times \mathcal{O}(1)$, 
while at late times $\mathcal{D}_4 \simeq \tfrac{1}{D^8} e^{-2S_2^R} \times f_2(\hat{r})$. As a consequence, one can readily verify that at early times the ratio $\tfrac{\mathcal{D}_4 \mathcal{D}_1^2}{\mathcal{D}_2^3}$ is of order $\tfrac{1}{D^2}$. On the other hand, at late times this ratio is of order $\tfrac{1}{e^{2S_2^R}}$. Thus, one can systematically check that the higher-order terms appearing in the exponent in \eqref{eq:wignerabs1} are indeed suppressed.

\noindent\underline{\textbf{$q=0$ case:}} \quad We now turn to the case $q=0$. The structure of the calculation is, to a large extent, similar to the $q \neq 0$ case, so we will only sketch the steps here while highlighting the key differences that arise. For $\mathcal{D}_3$, we have the following diagrams:
\begin{eqnarray}
    \mathcal{D}_3&=&\frac{1}{3!}\left(
    \begin{tikzpicture}[baseline={([yshift=-.5ex]current bounding box.center)}]
\coordinate (P) at (0,0);
\coordinate (Q) at (30pt,0);
\coordinate (R) at (60pt,0);

\draw (P) circle (10pt);
\draw (Q) circle (10pt); 
\draw (R) circle (10pt);

\coordinate (P1) at (-4pt,0);
\coordinate (P2) at (4pt,0);
\coordinate (Q1) at (26pt,0);
\coordinate (Q2) at (34pt,0);
\coordinate (R1) at (56pt,0);
\coordinate (R2) at (64pt,0);

\draw[black] (P1) circle (1.7pt);
\filldraw (P2) circle (1.7pt);
\draw[black] (Q1) circle (1.7pt);
\filldraw (Q2) circle (1.7pt);
\draw[black] (R1) circle (1.7pt);
\filldraw (R2) circle (1.7pt);

\draw[black,out = 60,in = 120, looseness = 1.5] (P1) to (R2);
\draw[black,out = 60,in = 120, looseness =1.5] (P2) to (Q1);
\draw[black,out = 60,in = 120, looseness =1.5] (Q2) to (R1);

\draw[dotted,out = -60,in = -120, looseness = 1.5] (P1) to (R2);
\draw[dotted,out = -60,in = -120, looseness =1.5] (P2) to (Q1); 
\draw[dotted,out = -60,in = -120, looseness =1.5] (Q2) to (R1); 
\end{tikzpicture} +  
\begin{tikzpicture}[baseline={([yshift=-.5ex]current bounding box.center)}]
\coordinate (P) at (0,0);
\coordinate (Q) at (30pt,0);
\coordinate (R) at (60pt,0);

\draw (P) circle (10pt);
\draw (Q) circle (10pt); 
\draw (R) circle (10pt);

\coordinate (P1) at (-4pt,0);
\coordinate (P2) at (4pt,0);
\coordinate (Q1) at (26pt,0);
\coordinate (Q2) at (34pt,0);
\coordinate (R1) at (56pt,0);
\coordinate (R2) at (64pt,0);

\draw[black] (P1) circle (1.7pt);
\filldraw (P2) circle (1.7pt);
\draw[black] (Q1) circle (1.7pt);
\filldraw (Q2) circle (1.7pt);
\draw[black] (R1) circle (1.7pt);
\filldraw (R2) circle (1.7pt);

\draw[black,out = 60,in = 120, looseness = 1.5] (P1) to (R2);
\draw[black,out = 60,in = 120, looseness =1.5] (P2) to (Q1);
\draw[black,out = 60,in = 120, looseness =1.5] (Q2) to (R1);

\draw[dotted, out=-120, in=-60, min distance=1.5cm] (P1) to (P2);
\draw[dotted, out=-120, in=-60, min distance=1.5cm] (Q1) to (Q2);
\draw[dotted, out=-120, in=-60, min distance=1.5cm] (R1) to (R2);

\end{tikzpicture} + 
\begin{tikzpicture}[baseline={([yshift=-.5ex]current bounding box.center)}]
\coordinate (P) at (0,0);
\coordinate (Q) at (30pt,0);
\coordinate (R) at (60pt,0);

\draw (P) circle (10pt);
\draw (Q) circle (10pt); 
\draw (R) circle (10pt);

\coordinate (P1) at (-4pt,0);
\coordinate (P2) at (4pt,0);
\coordinate (Q1) at (26pt,0);D
\coordinate (Q2) at (34pt,0);
\coordinate (R1) at (56pt,0);3
\coordinate (R2) at (64pt,0);

\draw[black] (P1) circle (1.7pt);
\filldraw (P2) circle (1.7pt);
\draw[black] (Q1) circle (1.7pt);
\filldraw (Q2) circle (1.7pt);
\draw[black] (R1) circle (1.7pt);
\filldraw (R2) circle (1.7pt);

\draw[black,out = 60,in = 120, looseness = 1.5] (P1) to (R2);
\draw[black,out = 60,in = 120, looseness =1.5] (P2) to (Q1);
\draw[black,out = 60,in = 120, looseness =1.5] (Q2) to (R1);

\draw[dotted,out = -60,in = -120, looseness = 2.0] (P1) to (Q2);
\draw[dotted,out = -60,in = -120, looseness =2.0] (P2) to (Q1); 
\draw[dotted,out = -120,in = -60, min distance =1.5cm] (R1) to (R2); 
\end{tikzpicture}
    \right) \nonumber \\
    &=& \frac{1}{D^6} \frac{1}{3!} \left( \text{Tr}\left( (\rho_R P_{\perp})^3 \right) +\frac{2}{D^2} \left( \text{Tr}(\rho_R P_{\perp}) \right)^3-\frac{3}{D} \text{Tr}\left( (\rho_R P_{\perp})^2\right)\; \text{Tr} (\rho_R P_{\perp}) \right)
\end{eqnarray}
Using the same redefinitions in terms of the hatted quantities introduced above for $q\neq 0$ case, we can then express the result as,
\begin{multline}
    \mathcal{D}_3=\frac{1}{D^6}\frac{1}{3!}\; e^{-3S_2^R} \Bigg[ \left(e^{S_2^R}\hat{S}_3-3\hat{h}+3 \hat{g}\hat{S}_0-\hat{S}_0^3 \right)+\frac{2}{D^2} \left( e^{S_2^R}-\hat{S}_0 \right)^3 \\
    -\frac{3}{D}\left( e^{S_2^R}-2\hat{g}+\hat{S}_0^2 \right) \left( e^{S_2^R}-\hat{S}_0 \right) \bigg]
\end{multline}
Similarly, for $\mathcal{D}_2$, we obtain,
\begin{eqnarray}
    2\mathcal{D}_2&=& \frac{1}{D^4}\left(D\; \text{Tr}\left( (\rho_R P_{\perp})^2 \right) - \left(\text{Tr} (\rho_R P_{\perp})\right)^2 \right) \nonumber \\
    &=& \frac{1}{D^4}\; e^{-2S_2^R}\left[ D\left( e^{S_2^R}-2\hat{g}+\hat{S}_0^2 \right)-\left( e^{S_2^R}-\hat{S}_0\right)^2 \right]
\end{eqnarray}
The key point is that, in this case, $\mathcal{D}_1$ is enhanced; that is,
\begin{eqnarray}
    \mathcal{D}_1&=&\frac{1}{D}\langle \psi_0|\rho_R|\psi_0\rangle \nonumber \\
    &=&\frac{1}{D}\; e^{-S_2^R} \hat{S}_0
\end{eqnarray}
At small times, one readily finds 
\begin{equation}
\mathcal{D}_3 \simeq \frac{1}{D^6} e^{-3S_2^R} \times \mathcal{O}(1), 
\qquad 
\mathcal{D}_2 \simeq \frac{1}{D^4} e^{-2S_2^R} \times D \times \mathcal{O}(1)
\end{equation}
Consequently, the ratio 
$\frac{\mathcal{D}_3 \, \mathcal{D}_1}{\mathcal{D}_2^2}$
is of order $\tfrac{1}{D}$ at early times. 
At late times, however, $e^{S_2^R}$ becomes large, and one finds 
\begin{equation}
\mathcal{D}_3 \simeq \frac{1}{D^6} e^{-2S_2^R} \times f_3(\hat{r}), 
\qquad 
\mathcal{D}_2 \simeq \frac{1}{D^4} (\hat{r}-1), 
\qquad 
\mathcal{D}_1 \simeq \frac{1}{D^2} \, \hat{r}\,\hat{S}_0 ,
\end{equation}
where $f_3(\hat{r})$ is again of order one. Thus, at late times the ratio $\frac{\mathcal{D}_3 \, \mathcal{D}_1}{\mathcal{D}_2^2}$ is suppressed as $\mathcal{O}(e^{-2S_2^R})$. So far everything seems fine. But the next nontrivial (quartic) term in \eqref{eq:wignerabs1} with the coefficient
$\frac{\mathcal{D}_4\,\mathcal{D}_1^2}{\mathcal{D}_2^3}$ turns out to be problematic. 
The discussion above already indicates its form of $\mathcal{D}_4$ at early and late times.  At early times one finds
$\mathcal{D}_4 \simeq \frac{1}{D^8} e^{-4S_2^R}\times D \times \mathcal{O}(1)$. At late times $\mathcal{D}_4$ instead scales as
$\mathcal{D}_4 \simeq \frac{1}{D^8} e^{-2S_2^R}\times f_4(\hat{r})$, with $f_4(\hat{r})=\mathcal{O}(1)$. It is then straightforward to verify that, at early times, the ratio $\frac{\mathcal{D}_4\,\mathcal{D}_1^2}{ \mathcal{D}_2^3}$ is of order 1, while at late times the same ratio is suppressed as $e^{-2S_2^R}$. Furthermore, the pattern is as follows: at early times, the coefficients of $s^n$ in \eqref{eq:wignerabs1} with $n=3,5,7,\dots$ are suppressed as $1/D$, whereas the coefficients with $n=4,6,8,\dots$ are of order 1. At late times, all terms beyond $s^2$ are suppressed by powers of $e^{S_2^R}$. Thus, the Gaussian approximation appears to break down for $q=0$ at early times, but remains valid at late times where the scaling limit is valid.

\section{Wigner negativity for random tensor networks} \label{sec:RTN}

In this section, we will compute Wigner negativity in the simplest instance of a random tensor network -- a bipartite state with a single random tensor -- quite similar to Page's setup. Consider a bipartite system $\mathcal{H}_{B} \otimes \mathcal{H}_{\overline{B}}$ with Hilbert space dimensions $d_B$ and $d_{\overline{B}}$, respectively. We prepare a random pure state,
\begin{equation}
|\Psi\rangle = U \;|0\rangle_{\overline{B}}\otimes |0\rangle_B ,
\end{equation}
where $U$ is a Haar-random unitary acting on the full Hilbert space. The seed state $|0\rangle_{\overline{B}}\otimes |0\rangle_B$ is chosen to be the one where both subsystems are in first basis state in their respective computational bases. The corresponding circuit diagram is shown below.
\[
|\Psi\rangle \; = \;
\begin{tikzpicture}[thick, font=\large, baseline={(current bounding box.center)}]

    \node[draw, minimum width=1.6cm, minimum height=1.0cm] (U) at (0,0) {$U$}; 

    \coordinate (topleft) at ($(U.north west)+(0.5cm,0)$);
    \coordinate (topright) at ($(U.north east)+(-0.5cm,0)$);
    \draw (topleft) -- ++(0,1.0) node[above] {$\overline{B}$};
    \draw (topright) -- ++(0,1.0) node[above] {$B$};

    \coordinate (bottomleft) at ($(U.south west)+(0.5cm,0)$);
    \coordinate (bottomright) at ($(U.south east)+(-0.5cm,0)$);
    \draw (bottomleft) -- ++(0,-1.0) node[below] {$|0\rangle$};
    \draw (bottomright) -- ++(0,-1.0) node[below] {$|0\rangle$};

\end{tikzpicture}
\]

For the state under consideration, we are interested in evaluating the Wigner negativity of subsystem $B$. 
To establish a connection with fixed-area states in quantum gravity, we can identify, $d_B = \exp(A_1 / 4G_N)$ and $d_{\overline{B}} = \exp(A_2 / 4G_N)$, 
where $A_1$ and $A_2$ denote the areas of the two competing extremal surfaces as we have discussed earlier. In the present setting, the negativity will be expressed as a function of $d_B$ and $d_{\overline{B}}$. In what follows, we focus on the regime in which both $d_B$ and $d_{\overline{B}}$ are large, while their ratio $d_B / d_{\overline{B}}$ is held fixed.

We expand the state $|\Psi\rangle$ in the computational basis $\{\,|i\rangle_B \otimes |j\rangle_{\overline{B}}\,\}$ where $i$ runs from $0$ to $d_B-1$ and $j$ runs from $0$ to $d_{\overline{B}}-1$. The reduced density matrix of subsystem $B$ is then obtained by tracing over $\overline{B}$, yielding,
\begin{equation}
    \langle l |\rho_B|k\rangle=\sum_{i,i'} U_{il\;;\;00}\;U^*_{i'k\;;\;00}\;\delta_{ii'} \label{eq:RTNrho}
\end{equation}
Subsequently, the Wigner function corresponding to this density matrix is given by,
\begin{equation*}
    W(q,p)=\frac{1}{d_B}\sum_{k,l=0}^{d_B-1}A_{kl}\;\langle l|\rho_B|k\rangle
\end{equation*}
where, $A_{kl}$ is the phase point operator: $A_{kl}=e^{\frac{2i\pi p}{d_B}(k-l)} \widehat{\delta}_{2q,k+l}$. 

Our goal is to compute the Wigner negativity for the state $\rho_B$, or more precisely the Haar--averaged Wigner negativity (where this time the Haar average is over the random tensor). As in previous sections, in order to handle the absolute value of the Wigner function, we employ the following integral representation:
\begin{align}
    \overline{\mathcal{N}}=\sum_{q,p}\overline{|W(q,p)|} \qquad , \qquad \overline{|W|}=\lim_{\epsilon \to 0}\frac{1}{2 \pi i}\int_{-\infty}^{\infty} dz\;\frac{2z}{z^2+\epsilon^2}\;\left( -i\frac{\partial}{\partial z}\right) \; \overline{e^{izW}}. \label{eq:NegPL}
\end{align}
Here, the overline denotes Haar averaging over the random tensor. 
The calculation proceeds in a manner similar to that outlined in the main text, so we will only sketch the essential steps here. 
We expand out $e^{iz W}$ and evaluate $\overline{W^n}$ by computing the relevant Haar integrals. 

First, using \eqref{eq:RTNrho}, we express $W^n$ in terms of the Haar-random unitaries,
\begin{multline}
    W^n=\frac{1}{d_B^n}\sum_{\substack{k_1,l_1,....,k_n,l_n\\i_1,i_1',......,i_n,i_n'}} \;A_{k_1l_1}A_{k_2 l_2}..... A_{k_n l_n} \;\; \delta_{i_1 i_1'} \delta_{i_2 i_2'} ...... \delta_{i_n i_n'} \times\\
    U_{i_1 l_1;00} U_{i_2 l_2;00}.....U_{i_n l_n;00} \;U_{i_1' k_1;00}^* \;U_{i_2' k_2;00}^* .....\;U_{i_n' k_n;00}^* 
\end{multline}
The above expression can be conveniently recast in a compact diagrammatic form. We show this below.

\newcommand{\mydiagramcircle}[2][1]{%
\begin{tikzpicture}
\begin{scope}[scale=#1, transform shape, every node/.style={font=\large}]

\node[draw, circle, minimum size=2.3cm] (center) at (0,0) {};

\node[circle,draw,thick,inner sep=1.5pt] (l)  at ([xshift=-0.6cm, yshift=0.5cm]center) {};
\node at ([xshift=0.22cm]l.east) {$\ell_{#2}$};  

\node[circle,draw,thick,inner sep=1.5pt] (k)  at ([xshift=0.6cm, yshift=0.5cm]center) {};
\node at ([xshift=-0.22cm]k.west) {$k_{#2}$};     

\node[circle,draw,thick,inner sep=1.5pt] (i)  at ([xshift=-0.6cm, yshift=-0.5cm]center) {};
\node at ([xshift=0.22cm]i.east) {$i_{#2}$};     

\node[circle,draw,thick,inner sep=1.5pt] (ip) at ([xshift=0.6cm, yshift=-0.5cm]center) {};
\node at ([xshift=-0.22cm]ip.west) {$i'_{#2}$};  

\node[draw,minimum width=0.6cm,minimum height=0.5cm] (A)   at (0,1.8) {A};
\node[draw,minimum width=0.6cm,minimum height=0.5cm] (one) at (0,-1.8) {$\mathds{1}$};

\draw[blue, line width=0.8pt] (l.north) to[out=90,in=190,looseness=1.1] (A.west);
\draw[blue, line width=0.8pt] (k.north) to[out=90,in=-10,looseness=1.1] (A.east);

\draw[black, line width=0.8pt] (i.south)  to[out=-90,in=190,looseness=1.1] (one.west);
\draw[black, line width=0.8pt] (ip.south) to[out=-90,in=-10,looseness=1.1] (one.east);

\draw[blue, line width=0.8pt] (l.west)  -- ++(-1.3, 1.3);  
\draw[blue, line width=0.8pt] (k.east)  -- ++(1.3, 1.3);   
\draw[black, line width=0.8pt] (i.west)  -- ++(-1.3,-1.3);  
\draw[black, line width=0.8pt] (ip.east) -- ++(1.3,-1.3);   

\end{scope}
\end{tikzpicture}%
}

\begin{equation}
W^n =
\raisebox{-0.5\height}{\mydiagramcircle[0.6]{1}} \times
\hspace{0.6cm}
\raisebox{-0.5\height}{\mydiagramcircle[0.6]{2}} \times
\cdots \cdots \times
\raisebox{-0.5\height}{\mydiagramcircle[0.6]{n}} \label{eq:PLW^n}
\end{equation}
where each piece denotes,

\begin{equation}
    \raisebox{-0.5\height}{\mydiagramcircle[0.8]{}}:=\; \frac{1}{d_B} \sum_{k,l;i,i'} \; A_{k\ell} \;U_{i\ell;00} U^*_{i' k;00}\; \delta_{ii'}
\end{equation}
In the diagram, the diagonal legs represent the row indices of the unitaries $U$ and $U^*$. 
The column indices are suppressed, as they are always fixed to zero. The indices $k,\ell$ correspond to the subsystem $B$ and are depicted by blue lines, 
while the indices $i,i'$ correspond to the subsystem $\overline{B}$ and are represented by black lines. 
The symbol $A$ denotes the phase-point operator, and $\mathds{1}$ corresponds to $\delta_{ii'}$. 
This diagrammatic representation is particularly convenient for visualizing the effect of averaging over the Haar random unitaries. We now perform the Haar integrals using the following formula:
\begin{multline}
    \int dU \;U_{i_1 l_1;00} U_{i_2 l_2;00}.....U_{i_n l_n;00} \;U_{i_1' k_1;00}^* \;U_{i_2' k_2;00}^* .....\;U_{i_n' k_n;00}^* =\sum_{\sigma \in S_n} \delta_{i_1 i'_{\sigma(1)}} \delta_{k_1 l_{\sigma(1)}} \; \delta_{i_2 i'_{\sigma(2)}} \delta_{k_2 l_{\sigma(2)}} \cdots \\
    \cdots \delta_{i_n i'_{\sigma(n)}} \delta_{k_n l_{\sigma(n)}} \times\left( \sum_{\tau} \text{Wg}(\tau, d)\right) \\
    \approx \frac{1}{d_B^n d_{\overline{B}}^n} \sum_{\sigma \in S_n} \delta_{i_1 i'_{\sigma(1)}} \delta_{k_1 l_{\sigma(1)}} \; \delta_{i_2 i'_{\sigma(2)}} \delta_{k_2 l_{\sigma(2)}} \cdots \cdots \delta_{i_n i'_{\sigma(n)}} \delta_{k_n l_{\sigma(n)}} \label{eq:HaarAvgPL}
\end{multline}
Here, $d = d_B d_{\overline{B}}$, and the second line follows in the limit where $d_B$ and $d_{\overline{B}}$ are both large, i.e., $d$ is large. Using this result, we can compute $\overline{W^n}$ by contracting the free legs in \eqref{eq:PLW^n} according to all elements of the permutation group $S_n$. Importantly, \eqref{eq:HaarAvgPL} dictates that the upper and lower legs are contracted identically. As an illustration, consider the case $n=5$. The quantity $\overline{W^5}$ admits $5!$ possible contractions, one of which is shown below. This contraction corresponds to the permutation $\sigma = (123)(45)$. The blue lines, associated with subsystem $B$, form loops that produce the factor $\mathrm{Tr}(A^3)\,\mathrm{Tr}(A^2)$, while the black lines, associated with subsystem $\overline{B}$, yield $\mathrm{Tr}(\mathds{1})^2$. Now for any $n$, 
\begin{equation}
    \Tr(A^n) = \begin{cases}
        1  \quad  \text{if} \; n \; \text{is} \; \text{odd}, \\ 
        d_B  \quad  \text{if} \; n \; \text{is} \;\text{even}.
    \end{cases} \label{eq:TrPL}
\end{equation}
and $\mathrm{Tr}(\mathds{1}) = d_{\overline{B}}$.
\newcommand{\onediskdiagram}[2][1]{%
\begin{tikzpicture}[scale=#1, transform shape, baseline={(current bounding box.center)}]
\begin{scope}[every node/.style={font=\large}]
\node[draw, circle, minimum size=2.0cm] (center) at (0,0) {#2};

\node[circle,draw,thick,inner sep=1.5pt] (l)  at ([xshift=-0.5cm, yshift=0.45cm]center) {};
\node[circle,draw,thick,inner sep=1.5pt] (k)  at ([xshift=0.5cm, yshift=0.45cm]center) {};
\node[circle,draw,thick,inner sep=1.5pt] (i)  at ([xshift=-0.5cm, yshift=-0.45cm]center) {};
\node[circle,draw,thick,inner sep=1.5pt] (ip) at ([xshift=0.5cm, yshift=-0.45cm]center) {};

\node[draw,minimum width=0.6cm,minimum height=0.5cm] (A)   at (0,1.8) {A};
\node[draw,minimum width=0.6cm,minimum height=0.5cm] (one) at (0,-1.8) {$\mathds{1}$};

\draw[blue, line width=0.8pt] (l.north) to[out=90,in=190,looseness=1.1] (A.west);
\draw[blue, line width=0.8pt] (k.north) to[out=90,in=-10,looseness=1.1] (A.east);
\draw[black, line width=0.8pt] (i.south)  to[out=-90,in=190,looseness=1.1] (one.west);
\draw[black, line width=0.8pt] (ip.south) to[out=-90,in=-10,looseness=1.1] (one.east);
\end{scope}
\end{tikzpicture}%
}

\newcommand{\mybigdiagram}[1][0.6]{%
\begin{tikzpicture}[scale=#1, transform shape, every node/.style={font=\large}, baseline={(current bounding box.center)}]
\node (D1) at (0,0) {\onediskdiagram[0.7]{1}};
\node (D2) at (3.0,0) {\onediskdiagram[0.7]{2}};
\node (D3) at (6.0,0) {\onediskdiagram[0.7]{3}};
\node (D4) at (9.0,0) {\onediskdiagram[0.7]{4}};
\node (D5) at (12.0,0){\onediskdiagram[0.7]{5}};

\coordinate (D1k) at ($(D1)+(0.35,0.315)$);
\coordinate (D2l) at ($(D2)+(-0.35,0.315)$);
\coordinate (D2k) at ($(D2)+(0.35,0.315)$);
\coordinate (D3l) at ($(D3)+(-0.35,0.315)$);
\coordinate (D3k) at ($(D3)+(0.35,0.315)$);
\coordinate (D1l) at ($(D1)+(-0.35,0.315)$);

\coordinate (D4k) at ($(D4)+(0.35,0.315)$);
\coordinate (D5l) at ($(D5)+(-0.35,0.315)$);
\coordinate (D5k) at ($(D5)+(0.35,0.315)$);
\coordinate (D4l) at ($(D4)+(-0.35,0.315)$);

\coordinate (D1ip) at ($(D1)+(0.35,-0.315)$);
\coordinate (D2i)  at ($(D2)+(-0.35,-0.315)$);
\coordinate (D2ip) at ($(D2)+(0.35,-0.315)$);
\coordinate (D3i)  at ($(D3)+(-0.35,-0.315)$);
\coordinate (D3ip) at ($(D3)+(0.35,-0.315)$);
\coordinate (D1i)  at ($(D1)+(-0.35,-0.315)$);

\coordinate (D4ip) at ($(D4)+(0.35,-0.315)$);
\coordinate (D5i)  at ($(D5)+(-0.35,-0.315)$);
\coordinate (D5ip) at ($(D5)+(0.35,-0.315)$);
\coordinate (D4i)  at ($(D4)+(-0.35,-0.315)$);

\draw[blue, thick] (D1k) to[out=60, in=120, looseness=1.5] (D2l);
\draw[blue, thick] (D2k) to[out=60, in=120, looseness=1.5] (D3l);
\draw[blue, thick] (D3k) to[out=55, in=125, looseness=1.5] (D1l);
\draw[blue, thick] (D4k) to[out=60, in=120, looseness=1.5] (D5l);
\draw[blue, thick] (D5k) to[out=55, in=125, looseness=2.5] (D4l);

\draw[black, thick] (D1ip) to[out=-60, in=-120, looseness=1.5] (D2i);
\draw[black, thick] (D2ip) to[out=-60, in=-120, looseness=1.5] (D3i);
\draw[black, thick] (D3ip) to[out=-55, in=-125, looseness=1.5] (D1i);
\draw[black, thick] (D4ip) to[out=-60, in=-120, looseness=1.5] (D5i);
\draw[black, thick] (D5ip) to[out=-55, in=-125, looseness=2.5] (D4i);
\end{tikzpicture}%
}

\begin{eqnarray}
\overline{W^5}\;\; &\supset&
\vcenter{\hbox{\mybigdiagram[0.65]}} \nonumber \\
&=& \frac{1}{d_B^5} \times \frac{1}{d_B^5 d_{\overline{B}}^5}\times \text{Tr}(A^3) \text{Tr}(A^2) \text{Tr}(\mathds{1})^2 \qquad ....... \qquad \sigma=(123) (45)
\end{eqnarray}

This example highlights a key point: the value of any term in $\overline{W^n}$ depends solely on the cycle structure of the corresponding permutation. For $S_n$, \emph{any} generic element can be decomposed into $c_1$ $1$-cycles, $c_2$ $2$-cycles, $\dots$, $c_n$ $n$-cycles, subject to the constraint $\sum_{i=1}^n i\,c_i = n$. The number of permutations with such a cycle structure is $\frac{n!}{c_1! \, c_2! \cdots c_n!} \times \frac{1}{1^{c_1} 2^{c_2} \cdots n^{c_n}} \,$.
Each of these permutations contributes the same value, namely $\big( \mathrm{Tr}(A)\,\mathrm{Tr}(\mathds{1}) \big)^{c_1}
\big( \mathrm{Tr}(A^2)\,\mathrm{Tr}(\mathds{1}) \big)^{c_2} \cdots
\big( \mathrm{Tr}(A^n)\,\mathrm{Tr}(\mathds{1}) \big)^{c_n}$. Now we define
\begin{equation}
D_n := \mathrm{Tr}(A^n)\,\mathrm{Tr}(\mathds{1}) \label{eq:ConnPL}
\end{equation}
These objects $D_n$ may be regarded as the analogues of the fully connected irreducible diagrams introduced in the main text. With this definition, the sum over all permutations in $S_n$ can be reorganized as a sum over distinct cycle structures, subject to the constraint $\sum_i i c_i = n$. Thus, we can write
\begin{equation}
\overline{W^n} \;=\;\frac{1}{\left(d_B^2 d_{\overline{B}}\right)^n} \; \sum_{\{c_i\}}'
\frac{n!}{c_1! \, c_2! \cdots c_n!} \;\;
\prod_{i=1}^n \left(\frac{D_i}{i}\right)^{\,c_i}.
\end{equation}
At this stage, one can easily recognize the pattern. Having encountered similar structures in our earlier discussions, we can immediately write down the corresponding expression for $\overline{e^{izW}}$.
\begin{equation}
    \overline{e^{izW}}=\sum_{n=0}^{\infty}\frac{(iz)^n}{n!}\;\overline{W^n}=\exp \left[ \sum_{n=1}^{\infty} \left( \frac{iz}{d_B^2 d_{\overline{B}}}\right)^n \;\frac{D_n}{n} \right]
\end{equation}
Following the same line of reasoning laid out in the PSSY analysis, we can effectively truncate the exponent in the above expression at $n=2$. Substituting this truncated expression into \eqref{eq:NegPL}, we obtain the ensemble-averaged absolute value of the Wigner function, $|W|$:
\begin{equation}
    \overline{|W(q,p)|}\approx \frac{D_1}{d_B^2 d_{\overline{B}}} \erf \left( \frac{D_1}{\sqrt{2 D_2}} \right)+\frac{1}{d_B^2 d_{\overline{B}}}\sqrt{\frac{2 D_2}{\pi}}\;e^{-\frac{D_1^2}{2D_2}}
\end{equation}
From \eqref{eq:ConnPL} and \eqref{eq:TrPL}, it follows that $D_1 = d_{\overline{B}},\;\; D_2 = d_B d_{\overline{B}}$. Consequently, the averaged negativity is given by,
\begin{equation}
    \overline{\mathcal{N}}=\erf \left( \sqrt{\frac{d_{\overline{B}}}{2d_B}} \right)+\sqrt{\frac{2 d_B}{\pi d_{\overline{B}}}} \;e^{-\frac{d_{\overline{B}}}{2d_B}} \label{eq:PLavgNeg}
\end{equation}

Hence, we obtain an explicit expression for the Wigner negativity as a function of the Hilbert space dimensions of the subsystems $B$ and $\overline{B}$ in the simplest tensor network model. More precisely, the negativity depends on the ratio $d_B/d_{\overline{B}}$. Drawing an analogy with fixed-area states in quantum gravity, this corresponds to a dependence on $\exp\big((A_1 - A_2)/4G_N\big)$, i.e., the difference between the areas of the competing extremal surfaces $\gamma_1$ and $\gamma_2$. Equation \eqref{eq:PLavgNeg} is our final expression for the Wigner negativity in the simplest random tensor network state considered here.  

\bibliographystyle{apsrev4-2}
\bibliography{Reference_wigner}

\end{document}